\documentclass[aps,pra,reprint,superscriptaddress]{revtex4-1}
\usepackage{amsmath}
\usepackage{resizegather}

\usepackage{amssymb}
\newcommand{\tr}{\mathsf{Tr}}
\usepackage{braket}
\usepackage[T1]{fontenc}
\usepackage{dsfont}
\usepackage{physics}
\usepackage{siunitx}
\usepackage{mathtools}
\usepackage{enumitem}
\usepackage{tikz} 
\usetikzlibrary{decorations.pathmorphing,patterns}
\usetikzlibrary{backgrounds}
\usetikzlibrary{decorations.markings}
\usepackage{epstopdf}
\tikzset{
    set arrow inside/.code={\pgfqkeys{/tikz/arrow inside}{#1}},
    set arrow inside={end/.initial=>, opt/.initial=},
    /pgf/decoration/Mark/.style={
        mark/.expanded=at position #1 with
        {
            \noexpand\arrow[\pgfkeysvalueof{/tikz/arrow inside/opt}]{\pgfkeysvalueof{/tikz/arrow inside/end}}
        }
    },
    arrow inside/.style 2 args={
        set arrow inside={#1},
        postaction={
            decorate,decoration={
                markings,Mark/.list={#2}
            }
        }
    },
}
\usetikzlibrary{shapes}
\definecolor{myred}{HTML}{FF0000} 
\usetikzlibrary{external}
\definecolor{QRc}{HTML}{E64F40} 
\definecolor{nonQRc}{HTML}{3C49E3} 

    \usepackage{xcolor}
\usepackage{mdframed}
\usepackage{multienum}

\newcommand{\san}[1]{\mathsf{#1}}

\newcommand{\proj}[1]{\ketbra{#1}}
\DeclareMathOperator{\id}{\mathbb{I}}

\usepackage{epstopdf}
\usepackage{float}

\makeatletter
\newcommand*{\centerfloat}{%
  \parindent \z@
  \leftskip \z@ \@plus 1fil \@minus \textwidth
  \rightskip\leftskip
  \parfillskip \z@skip}
\makeatother

\usepackage{todonotes}
\usepackage{hyperref}

\newcommand{\nstar}{n^{*}}

\begin{document}
\title{Near-term quantum-repeater experiments with nitrogen-vacancy centers: Overcoming the limitations of direct transmission}
\author{Filip Rozp\k{e}dek}
\thanks{These three authors contributed equally; f.d.rozpedek@tudelft.nl}
\affiliation{QuTech, Delft University of Technology, Lorentzweg 1, 2628 CJ Delft, The Netherlands}
\affiliation{Kavli Institute of Nanoscience, Delft University of Technology, Lorentzweg 1, 2628 CJ Delft, The Netherlands}
\author{Raja Yehia}
\thanks{These three authors contributed equally; f.d.rozpedek@tudelft.nl}
\affiliation{QuTech, Delft University of Technology, Lorentzweg 1, 2628 CJ Delft, The Netherlands}
\affiliation{Sorbonne Université, CNRS, Laboratoire d'Informatique de Paris 6, F-75005 Paris, France}
\author{Kenneth Goodenough}
\thanks{These three authors contributed equally; f.d.rozpedek@tudelft.nl}
\author{Maximilian Ruf}
\author{Peter C.~Humphreys}
\author{Ronald Hanson}
\author{Stephanie Wehner}
\affiliation{QuTech, Delft University of Technology, Lorentzweg 1, 2628 CJ Delft, The Netherlands}
\affiliation{Kavli Institute of Nanoscience, Delft University of Technology, Lorentzweg 1, 2628 CJ Delft, The Netherlands}
\author{David Elkouss}
\affiliation{QuTech, Delft University of Technology, Lorentzweg 1, 2628 CJ Delft, The Netherlands}

\begin{abstract}
Quantum channels enable the implementation of communication tasks inaccessible to their classical counterparts. The most famous example is the distribution of secret keys. However, in the absence of quantum repeaters, the rate at which these tasks can be performed is dictated by the losses in the quantum channel. In practice, channel losses have limited the reach of quantum protocols to short distances.
Quantum repeaters have the potential to significantly increase the rates and reach beyond the limits of direct transmission. However, no experimental implementation has overcome the direct transmission threshold. Here, we
propose three quantum repeater schemes and assess their ability to generate secret key when implemented on a setup using nitrogen-vacancy (NV) centers in diamond with near-term experimental parameters. 
We find that one of these schemes - the so-called single-photon scheme, requiring no quantum storage - has the ability to surpass the capacity - the highest secret-key rate achievable with direct transmission - by a factor of 7 for a distance of approximately 9.2 km with near-term parameters, establishing it as a prime candidate for the first experimental realization of a quantum repeater.
\end{abstract}    
    \maketitle


\section{Introduction}
There exist communication tasks for which quantum resources allow for qualitative advantages. Examples of such tasks include clock synchronization~\cite{jozsa2000quantum,krvco2002quantum,preskill2000quantum,Giovanetti_01}, distributed computation~\cite{spiller2006quantum}, anonymous information transmission~\cite{christandl2005quantum,brassard2007anonymous}, and the distribution of secret keys \cite{Bennett_84,ekert1991quantum}.  While some of these tasks have been implemented over short distances, their implementation over long distances remains a formidable challenge.

One of the main hurdles for long-distance quantum communication is the \emph{loss} of photons, whether it is through fiber or free-space. Unfortunately, the no-cloning theorem~\cite{park1970concept} makes the amplification of the transmitted quantum states impossible. For tasks such as the generation of shared secret key or entanglement, this limits the corresponding generation rate to scale at best linearly in the transmissivity $\eta$ of the fiber joining two distant parties~\cite{takeoka2014fundamental,takeoka2014squashed,pirandola2015fundamental}.

Luckily, while quantum mechanics prevents us from overcoming the effects of losses through amplification, it is possible to do so using repeater stations~\cite{briegel1998quantum, sangouard2011quantum, Munro_15}. Formally, we call a quantum repeater a device that allows for a better performance than can be achieved over the direct communication channel alone~\cite{parameterregimes}. This performance is measured differently for different tasks, such as secret-key generation or transmission of quantum information. Consequently, the optimal performance that can be achieved over the direct channel without using repeaters, called the channel capacity, is also different for these two tasks. Here we will assess our proposed repeater schemes for the task of secret-key generation, as it is easier to realize experimentally. Our formal definition of a repeater---as opposed to a relative definition with respect to some setup of reference---endows the demonstration of a quantum repeater with a fundamental meaning that cannot be affected by future technological developments in the field.

However, a successful experimental implementation of a quantum repeater has not yet been demonstrated. This is mainly due to the additional noise introduced by such a quantum repeater. While the implementation of a single quantum repeater does not necessarily imply that that setup can be scaled up to a larger number of repeater nodes (due to the effects of noise and decoherence), the first demonstration of a functioning quantum repeater will form an important step toward practical quantum communication and the quantum internet~\cite{kimble2008quantum}.

A multitude of quantum repeater schemes have been put forward~\cite{dur1999quantum,duan2001long,simon2007quantum,sangouard2008robust, sangouard2011quantum, guha2015rate, jiang2009quantum, munro2010quantum, bernardes2012hybrid, munro2012quantum, muralidharan2014ultrafast, Azuma_15}, each with their own strengths and weaknesses. It should be noted here that most of the earlier repeater proposals aim at overcoming transmission losses using heralded entanglement generation and compensate for noise arising in quantum memories using two-way entanglement distillation. However, some of the schemes, e.g.~, those in Refs.~\cite{jiang2009quantum, munro2010quantum, bernardes2012hybrid} introduce error correction to overcome operational errors while those in Refs.~\cite{munro2012quantum, muralidharan2014ultrafast, Azuma_15} use error correction also for dealing with losses. Although \emph{a priori} it is not clear which of those schemes will perform best with current or near-term experimental parameters, it is clear that operating on large number of qubits in each repeater node, necessary for the implementation of error correction, is a significant experimental challenge. Therefore, it is not expected that the first realizations of quantum repeaters will be based on large error-correction schemes. Hence, here we will focus on simple schemes without encoding, leaving out also entanglement distillation. We will discuss entanglement distillation in the context of our findings in the discussion at the end of the article.

In this work, we propose three such schemes and together with the fourth scheme analyzed before~\cite{luong2015overcoming,parameterregimes}, we assess their performance for generating secret key. We consider their implementation based on nitrogen-vacancy centers in diamond (NV centers), a system which has properties making it an excellent candidate for long-distance quantum communication applications~\cite{humphreys2017deterministic,Abobeih2018,kalb2017entanglement, taminiau2014universal, vandam2017multiplexed, blok2015towards, cramer2015repeated, reiserer2016robust, gao2015coherent, bogdanovic2016design}. 

\begin{figure}
\centerfloat
\includegraphics[clip, trim = 0mm 0mm 0mm 0mm,width=0.47\textwidth]{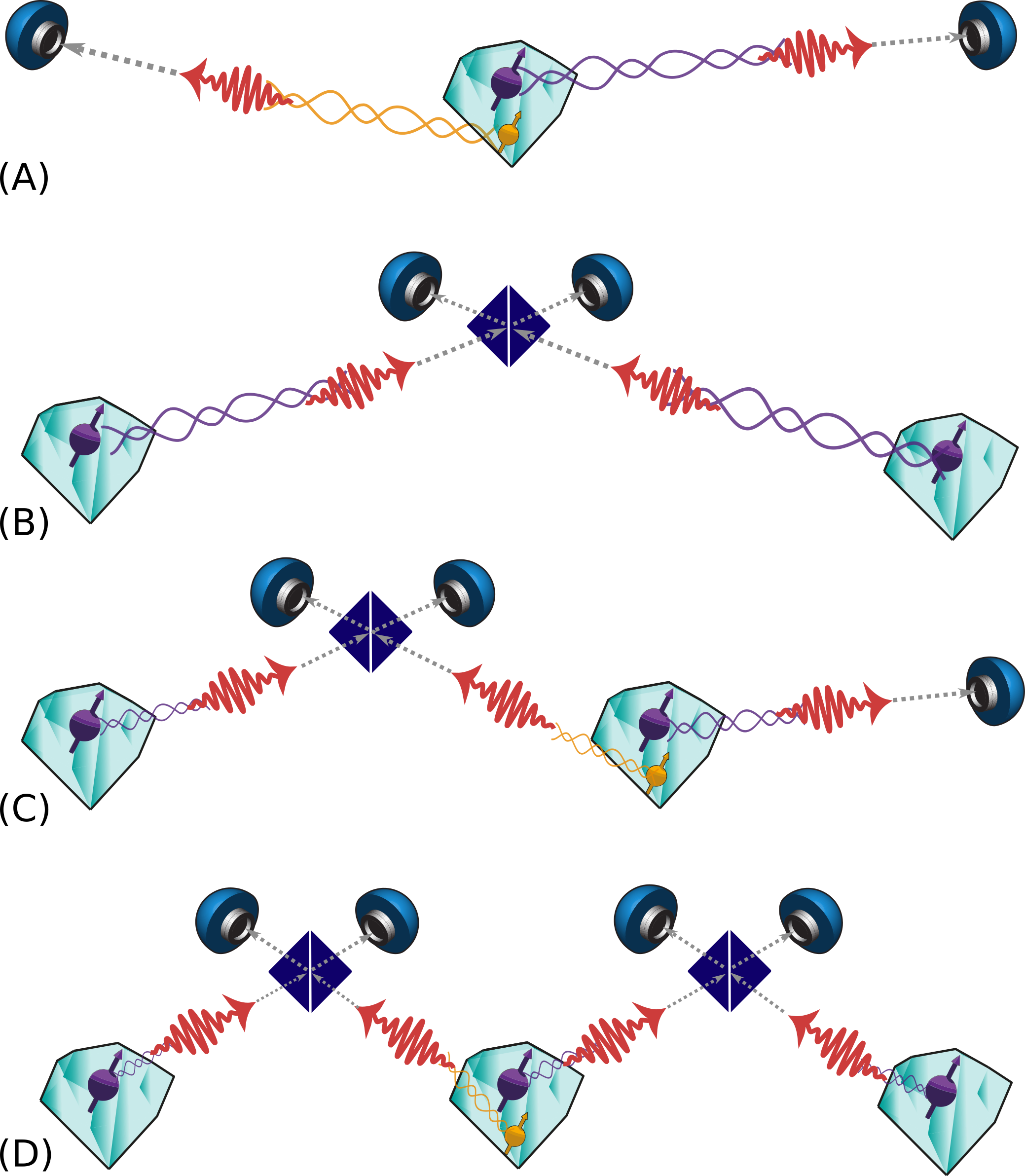}
\caption{Schematic overview of the four quantum repeater schemes assessed in this paper. From top to bottom: the Single Sequential Quantum Repeater (SiSQuaRe) scheme (A), the single-photon scheme (B), the Single-Photon with Additional Detection Setup (SPADS) scheme (C) and the Single-Photon Over Two Links (SPOTL) scheme (D). The purple particles represent NV electron spins capable of emitting photons (red wiggly arrows) while the yellow particles represent carbon $^{13}$C nuclear spins. Dark blue squares depict the beam splitters used to erase the which-way information of the photons, followed by blue photon detectors. For more details on the different proposals, see Sec.~\ref{sec:protocols}.}
\label{fig:overview}
\end{figure}

The four considered schemes are the following: the ``single sequential quantum repeater node'' (first proposed and studied in Ref.~\cite{luong2015overcoming}, then further analyzed in Ref.~\cite{parameterregimes}), the single-photon scheme (proposed originally in the context of remote entanglement generation~\cite{cabrillo1999creation}, also studied in the context of secret-key generation without quantum memories~\cite{lucamarini2018overcoming}), and two schemes which are a combination of the first two. See Fig.~\ref{fig:overview} for a schematic overview of the repeater proposals considered in this work.

We compare the \emph{secret-key rate} of each of these schemes to the highest theoretically achievable secret-key rate using direct transmission, the \emph{secret-key capacity of the pure-loss channel}~\cite{pirandola2015fundamental}. We show that one of these schemes, the \emph{single-photon scheme}, can surpass the secret-key capacity by a factor of 7 for a distance of $\approx9.2$ km with near-term parameters. This shows the viability of this scheme for the first experimental implementation of a quantum repeater.

In Sec.~\ref{sec:protocols}, we discuss and detail the different repeater proposals that will be assessed in this work. In Sec.~\ref{sec:modeling}, we expand on how the different components of the repeater proposals would be implemented experimentally. Sec.~\ref{sec:secretkeyrate} details how to calculate the secret-key rate achieved with the quantum repeater proposals from the modeled components.
In Sec.~\ref{sec:benchmarks} we discuss how to assess the performance of a quantum repeater.
The comparison of the different repeater proposals is performed in Sec.~\ref{sec:results}, which allows us to conclude with our results in Sec.~\ref{sec:conclusions}. The numerical results of this article were produced with a \texttt{PYTHON} and a \texttt{MATHEMATICA} script, which are available upon request.

\section{Quantum repeater schemes}
\label{sec:protocols}
\par{In the following section, we present the quantum repeater schemes that will be assessed in this work. All these schemes use NV-center-based setups which involve memory nodes consisting of an electron spin qubit acting as an optical interface and possibly an additional carbon $^{13}$C nuclear spin qubit acting as a long-lived quantum memory. Specifically, the optical interface of the electron spin allows for the generation of spin-photon entanglement, where the photonic qubits can then be transmitted over large distances. The carbon nuclear spin acts as a long-lived memory, but can be accessed only through the interaction with the electron spin. Here, we briefly go over all the proposed schemes, consider why they are interesting from an experimental perspective, and discuss their advantages and disadvantages.}


\subsection{The Single Sequential Quantum Repeater (SiSQuaRe) scheme}
\begin{figure}
\centerfloat
\includegraphics[clip, trim = 0mm 0mm 0mm 0mm,width=0.47\textwidth]{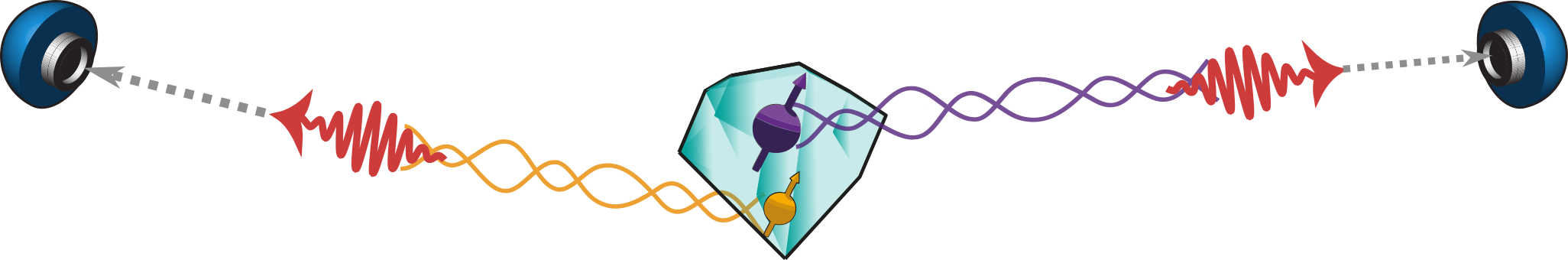}
\caption{Schematic overview of the SiSQuaRe scheme. The NV center in the middle first attempts to generate an entangled photon-electron pair, after which it tries to send the photon through the fiber to Alice. Alice then directly measures the photon, using either a BB84 or a six-state measurement. Then after the state of the electron spin is swapped to the carbon $^{13}$C nuclear spin, the same is attempted on Bob's side. After both Alice and Bob measured a photon, a Bell-state measurement is performed on the two quantum states held by the middle node. Alice and Bob can use their measurement outcomes together with the outcome of the Bell-state measurement to generate a single raw bit of key.}
\label{fig:SiSQuaRe}
\end{figure}

The first scheme that we discuss here was proposed and analyzed in Ref.~\cite{luong2015overcoming} and further studied in Ref.~\cite{parameterregimes}. The scheme involves a node holding two quantum memories in the middle of Alice and Bob (see Fig.~\ref{fig:SiSQuaRe}). This middle node tries to send a photonic qubit, encoded in the time-bin degree of freedom, that is entangled with one of the quantum memories, through a fiber to Alice. This is attempted repeatedly until the photon successfully arrives, after which Alice performs a BB84~\cite{Bennett_84} or a six-state measurement~\cite{bruss1998optimal, bechmann1999incoherent}. By performing such a measurement, the quantum memory will be steered into a specific state depending on the measurement outcome. Now the same is attempted on Bob's side. After Bob has measured a photon, the middle node performs a Bell-state measurement on both quantum memories. Using the classical information of the outcome of the Bell-state measurement, Alice and Bob can generate a single raw bit.
In our model, the middle node has only one photonic interface (corresponding to the NV electron spin), and hence has to send the photon sequentially first to Alice and then to Bob.

While trying to send a photon to Bob, the state stored in the middle node will decohere. A possible way to compensate for the effects of decoherence is to introduce a so-called \emph{cut-off}~\cite{parameterregimes}. The cut-off is a limit on the number of attempts we allow the middle node to try and send a photon to Bob. If the cut-off is reached, the stored state is discarded, and the middle node attempts again to send a photon to Alice. Since the scheme starts from scratch, we are effectively trading off the generation time versus the quality of our state. By optimizing over the cut-off, it is possible to considerably increase the distance over which secret key can be generated~\cite{parameterregimes}.


\subsubsection*{Setup and scheme}

We will now describe the exact procedure of this scheme, when Alice and Bob use a nitrogen-vacancy center in diamond as quantum memories and as a photon source. The scheme that we study is the following:
\begin{enumerate}[label={(\arabic*)}]
\item The quantum repeater attempts to generate an entangled qubit-qubit state between a photon and its electron spin, and sends the photon to Alice through a fiber.
\item The first step is repeated until a photon arrives at Alice's side, after which she performs a BB84 or a six-state measurement. The electron state is swapped to the carbon spin.
\item The quantum repeater attempts to do the same on Bob's side while the state in the carbon spin is kept stored. This state will decohere during the next steps.
\item Repeat until a photon arrives at Bob's side, who will perform a BB84 or a six-state measurement. If the number of attempts $n$ reaches the cut-off $n^{*}$, restart from step 1.
\item The quantum repeater performs a Bell-state measurement and communicates the result to Bob.
\item All the previous steps are repeated until sufficient data have been generated.
\end{enumerate}

\subsection{The single-photon scheme}
\label{subsecsingle-photonscheme}
\par{Cabrillo \emph{et al}.~\cite{cabrillo1999creation} devised a procedure that allows for the heralded generation of entanglement between a separated pair of matter qubits (their proposal discusses specific implementation with single atoms, but the scheme can also be applied to other platforms such as NV centers or quantum dots) using linear optics. For the atomic ensemble platform this scheme also forms a building block of the Duan, Lukin, Cirac, Zoller (DLCZ) quantum repeater scheme~\cite{duan2001long}.
Here we will refer to this scheme as a single-photon scheme as the entanglement generation is heralded by a detection of only a single photon. This requirement of successful transmission of only a single photon from one node makes it possible for this scheme to qualify as a quantum repeater (see below for more details).
 
The basic setup of the single-photon scheme consists of placing a beam splitter and two detectors between Alice and Bob, with both parties simultaneously sending a photonic quantum state toward the beam splitter. The transmitted quantum state is entangled with a quantum memory, and the state space of the photon is spanned by the two states corresponding to the presence and absence of a photon. Immediately after transmitting their photons through the fiber, both Alice and Bob measure their quantum memories in a BB84 or six-state basis (see the discussion of which quantum key distribution protocol is optimal for each scheme in Sec.~\ref{sec:secfrac} and in Sec.~\ref{sec:bb846state}). 
Note that this is equivalent to preparing a specific state of the photonic qubit and therefore is closely linked to the measurement device independent quantum key distribution (MDI QKD)~\cite{lo2012measurement} as discussed in Appendix~\ref{sec:MDI}. However, preparing specific states that involve the superposition of the presence and absence of a photon on its own is generally experimentally challenging. The NV implementation allows us to achieve this task precisely by preparing spin-photon entanglement and then measuring the spin qubit. Afterwards, by conditioning on the click of a single detector only, Alice and Bob can use the information of which detector clicked to generate a single raw bit of key; see Appendix~\ref{sec:SingleClick} and Ref.~\cite{cabrillo1999creation} for more information.

The main motivation of this scheme is that, informally, we only need one photon to travel half the distance between the two parties to get an entangled state. This thus effectively reduces the effects of losses, and in the ideal scenario the secret-key rate would scale with the square root of the total transmissivity $\eta$, as opposed to linear scaling in $\eta$ (which is the optimal scaling without a quantum repeater~\cite{pirandola2016capacities}).}
\par{However, one problem that one faces when implementing this scheme is that the fiber induces a phase shift on the transmitted photons. This shift can change over time, e.g. due to fluctuations in the temperature and vibrations of the fiber. The uncertainty of the phase shift induces dephasing noise on the state, reducing the quality of the state.}

To overcome this problem, a two-photon scheme was proposed by Barrett and Kok~\cite{barrett2005efficient}, which does not place such high requirement on the optical stability of the setup. Specifically, in the Barrett and Kok scheme the problem of optical phase fluctuations is overcome by requiring two consecutive clicks and performing additional spin-flip operations on both of the remote memories. The Barrett and Kok scheme has seen implementation in many experiments~\cite{hensen2015loophole,hensen2016loophole,bernien2013heralded,maunz2007quantum}.
However, the requirement of two consecutive clicks implies that a setup using only the Barrett and Kok scheme with two memory nodes will never be able to satisfy the demands of a quantum repeater. Specifically, the probability of getting two consecutive clicks will not be higher than the transmissivity of the fiber between the two parties and therefore will not surpass the secret-key capacity. 

In the single-photon scheme, on the other hand, the dephasing caused by the unknown optical phase shift is overcome by using active \emph{phase-stabilization} of the fiber to reduce the fluctuations in the induced phase. This technique has been used in the experimental implementations of the single-photon scheme for remote entanglement generation using quantum dots~\cite{stockill2017phase,delteil2016generation}, NV centers~\cite{humphreys2017deterministic} and atomic ensembles~\cite{chou2005cw}. For experimental details relating to NV implementation, we refer the reader to Sec.~\ref{sec:modeling}. This phase-stabilization technique effectively reduces the uncertainty in the phase, allowing us to significantly mitigate the resulting dephasing noise; see Appendix~\ref{appendix:lossesandnoise} for mathematical details.

In contrast to the Barrett and Kok scheme, the single-photon scheme cannot produce a perfect maximally entangled state, even in the case of perfect operations and perfect phase-stabilization. This is because losses in the channel result in a significant probability of having both nodes emitting a photon which can also lead to a single click in one of the detectors, yet the memories will be projected onto a product state. As we discuss below, this noise can be traded versus the probability of success of the scheme by reducing the weight of the photon-presence term in the generated spin-photon entangled state. This is discussed in more detail below and the full analysis is presented in Appendix~\ref{sec:SingleClick}.
\par{The single-photon scheme with phase-stabilization is a promising candidate for a near-term quantum repeater with NV centers. We note here that recently other QKD schemes that use the MDI framework have been proposed. These schemes, similar to our proposal, use single-photon detection events to overcome the linear scaling of the secret-key rate with $\eta$~\cite{lucamarini2018overcoming,tamaki2018information, ma2018phase}. In these proposals, in contrast to our single-photon scheme, no quantum memories are used, but instead Alice and Bob send phase-randomized optical pulses to the middle heralding station.}

\subsubsection*{Setup and scheme}
\par
\begin{figure}
\centerfloat
\includegraphics[clip, trim = 0mm 0mm 0mm 0mm,width=0.47\textwidth]{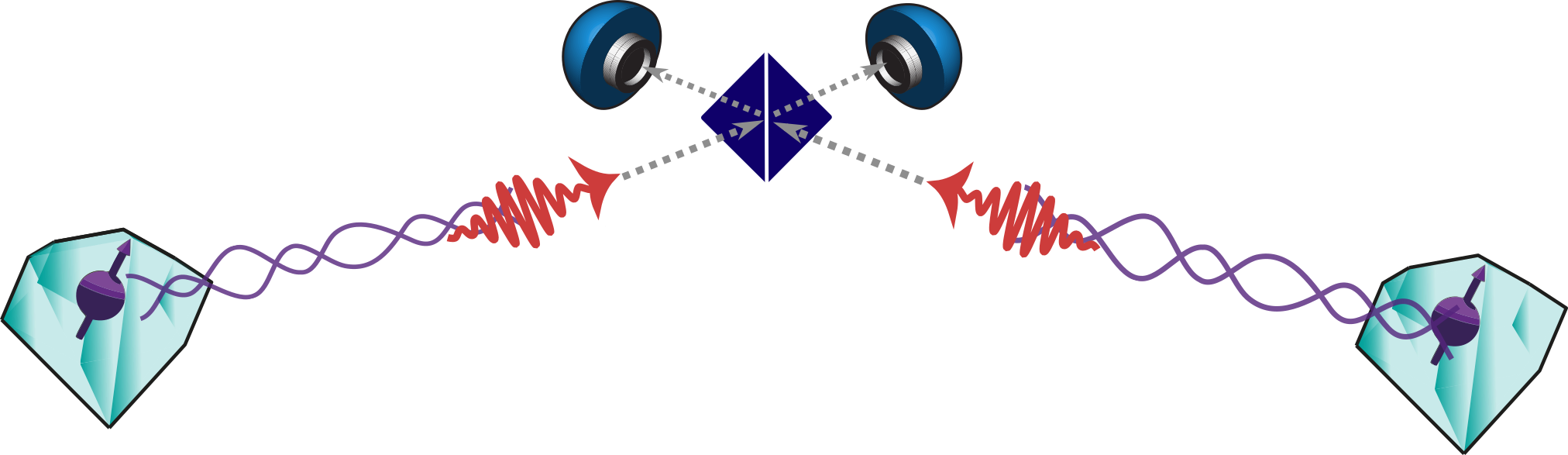}
\caption{Schematic overview of the single-photon scheme. Alice and Bob simultaneously transmit a photonic state from their NV centers toward a balanced beam splitter in the center. This photonic qubit, corresponding to the presence and absence of a photon, is initially entangled with the NV electron spin. If only one of the detectors (which can be seen at the top of the figure) registers a click, Alice and Bob can use the information of which detector clicked to generate a single raw bit of key.}
\label{fig:singleclick}
\end{figure}

{In the setup of the single-photon scheme, Alice and Bob are separated by a fiber where in the center there is a beam splitter with two detectors (see Fig.~\ref{fig:singleclick}). They will both create entanglement between a photonic qubit and a stored spin and send the photonic qubit to the beam splitter. }

Alice and Bob thus perform the following,
\begin{enumerate}[label={(\arabic*)}]
\item Alice and Bob both prepare a state
$\ket{\psi} = \sin\theta\ket{\downarrow}\ket{0} + \cos\theta\ket{\uparrow}\ket{1} $
where $\ket{\downarrow}$($\ket{\uparrow}$) refers to the dark (bright) state of the electron-spin qubit, $\ket{0}$ ($\ket{1}$) indicates the absence (presence) of a photon, and $\theta$ is a tunable parameter.
\item Alice and Bob attempt to both separately send the photonic qubit to the beam splitter.
\item Alice and Bob both perform a six-state measurement on their memories.
\item The previous steps are repeated until only one of the detectors between the parties clicks.
\item The information of which detector clicked gets sent to Alice and Bob for classical correction.
\item All the previous step are repeated until sufficient data have been generated.
\end{enumerate}

The parameter $\theta$ can be chosen by preparing a non-uniform superposition of the dark and bright state of the electron spin $\ket{\psi} = \sin\theta\ket{\downarrow} + \cos\theta\ket{\uparrow}$ via coherent microwave pulses. This is done before applying the optical pulse to the electron which entangles it with the presence and absence of a photon. The parameter $\theta$ can then be tuned in such a way as to maximize the secret-key rate. In the next section, we will briefly expand on some of the issues arising when losses and imperfect detectors are present. We defer the full explanation and calculations until Appendix~\ref{sec:SingleClick}.
 

\subsubsection*{Realistic setup} 
In any realistic implementation of the single-photon scheme, a large number of attempts is needed before a photon detection event is observed. Furthermore, a single detector registering a click does not necessarily mean that the state of the memories is projected onto the maximally entangled state. This is due to multiple reasons, such as losing photons in the fiber or in some other loss process between the emission and detection, arrival of the emitted photons outside of the detection time window and the fact that \emph{dark counts} generate clicks at the detectors. Photon loss in the fiber effectively acts as amplitude damping on the state of the photon when using the state space spanned by the presence and absence of the photon~\cite{pirandola2015fundamental,ivan2011operator}. Dark counts are clicks in the detectors, caused by thermal excitations. These clicks introduce noise, since it is impossible to distinguish between clicks caused by thermal excitations and the photons traveling through the fiber if they arrive in the same time window. All these sources of loss and noise acting on the photonic qubits are discussed in detail in Appendix~\ref{appendix:lossesandnoise}. Finally, we note that we assume here the application of non-number-resolving detectors. This can lead to additional noise in the low-loss regime, since the event in which two photons got emitted cannot be distinguished from the single-photon emission events even if no photons got lost. However, in any realistic loss regime this is not a problem, since the probability of two such photons arriving at the heralding station is quadratically suppressed with respect to events where only one photon arrives. In the realistic regime, almost all the noise coming from the impossibility of distinguishing two-photon from single-photon emission events is the result of photon loss. Namely, if a two-photon emission event occurs and the detector registers a click, then with dominant probability it is due to only a single photon arriving, while the other one being lost. Hence the use of photon-number-resolving detectors would not give any visible benefit with respect to the use of the non-number-resolving ones. For a detailed calculation of the effects of losses and dark counts for the single-photon scheme, see Appendix~\ref{sec:SingleClick}.


\subsection{Single-Photon with Additional Detection Setup (SPADS) scheme}
\begin{figure}
\centerfloat
\includegraphics[clip,width=0.47\textwidth]{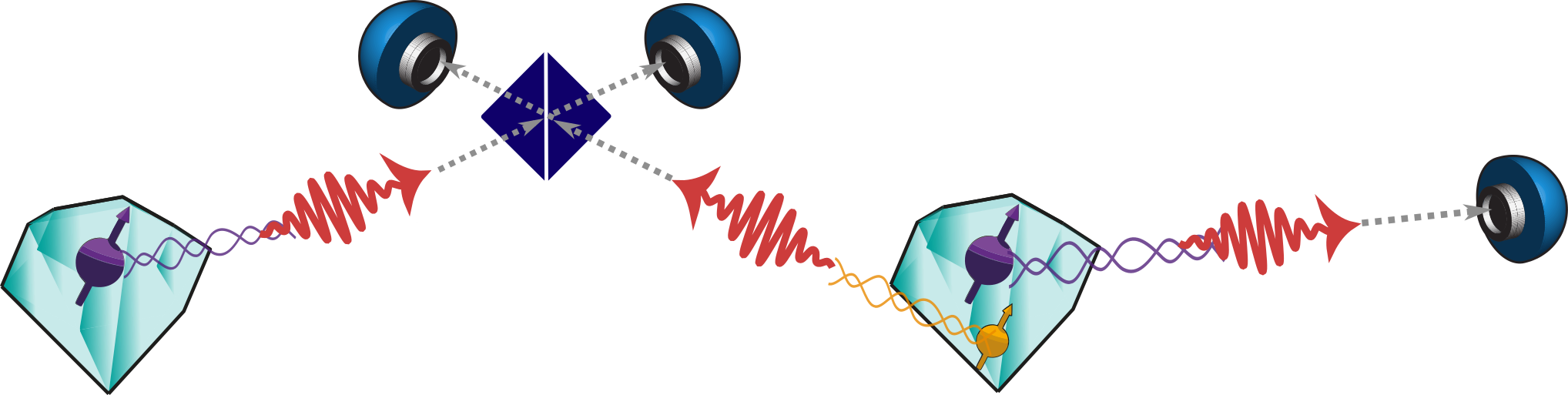}
\caption{Schematic overview of the SPADS scheme. First, the two NV centers run the single-photon scheme, such that Alice measures her electron spin directly after every attempt. After success, the middle node swaps its state to the carbon spin. Then the middle node generates electron-photon entangled pairs where the photonic qubit is encoded in the time-bin degree of freedom and sent to Bob. This is attempted until Bob successfully measures the photon or until the cut-off is reached. If the cut-off is reached, the scheme gets restarted, otherwise the middle node performs an entanglement swapping on its two memories and communicates the classical outcome to Alice and Bob, who can correct their measurement outcomes to obtain a bit of raw key.}
\label{fig:SPADS}
\end{figure}

The third scheme that we consider here is the Single-Photon with Additional Detection Setup (SPADS) scheme, which is effectively a combination of the single-photon scheme and the SiSQuaRe scheme as shown in Fig.~\ref{fig:SPADS}. If the middle node is positioned at two-thirds of the total distance away from Alice, the rate of this setup would scale, ideally, with the cube root of the transmissivity $\eta$.

This scheme runs as follows:
\begin{enumerate}[label={(\arabic*)}]
\item Alice and the repeater run the single-photon scheme until success, however, only Alice performs her spin measurement immediately after each spin-photon entanglement generation attempt. This measurement is either in a six-state or BB84 basis.  
\item The repeater swaps the state of the electron spin onto the carbon spin.
\item The repeater runs the second part of the SiSQuaRe scheme with Bob. This means it generates spin-photon entanglement between an electron and the time-bin encoded photonic qubit. Afterwards, it sends the photonic qubit to Bob. This is repeated until Bob successfully measures his photon in a six-state or BB84 basis or until the cut-off $n^{*}$ is reached, in which case the scheme is restarted with step 1.
\item After Bob has received the photon and communicated this to the repeater, the repeater performs a Bell-state measurement on its two quantum memories and communicates the classical result to Bob.
\item All the previous steps are repeated until sufficient data have been generated.
\end{enumerate}

The motivation for introducing this scheme is twofold. First, we note that by using this scheme we divide the total distance between Alice and Bob into three segments: two segments corresponding to the single-photon subscheme and the third segment over which the time-bin-encoded photons are sent. This gives us one additional independent segment with respect to the single-photon or the SiSQuaRe scheme on their own. Hence, for distances where no cut-off is required, we expect the scaling of the secret-key rate with the transmissivity to be better than the ideal square root scaling of the previous two schemes. Furthermore, dividing the total distance into more segments should also allow us to reach larger distances before dark counts become significant.  When considering the resources necessary to run this scheme, we note that the additional third node needs to be equipped only with a photon detection setup.

Second, we note that the SPADS scheme can also be naturally compared to the scenario in which an NV center is used as a single-photon source for direct transmission between Alice and Bob. Both the setup for the SPADS scheme and such direct transmission involve Alice using an NV for emission and Bob having only a detector setup. Hence, the SPADS scheme corresponds to inserting a new NV node (the repeater) between Alice and Bob without changing their local experimental setups at all. This motivates us to compare the achievable secret-key rate of the SPADS scheme and direct transmission. We perform this comparison on a separate plot in Sec.~\ref{sec:results}.


\subsection{Single-Photon Over Two Links (SPOTL) scheme}
The final scheme that we study here is the Single-Photon Over Two Links (SPOTL) scheme, and it is another combination of the single-photon and SiSQuaRe schemes. A node is placed between Alice and Bob which tries to sequentially generate entanglement with their quantum memories by using the single-photon scheme (see Fig.~$\ref{fig:QR3NV}$). The motivation for this scheme is that, while using relatively simple components and without imposing stricter requirements on the memories than in the previous schemes, its secret-key rate would ideally scale with the fourth root of the transmissivity $\eta$.

\subsubsection*{Setup and scheme}
The setup that we study is the following:
\begin{figure}
\centerfloat
\includegraphics[clip, trim = 0mm 0mm 0mm 0mm,width=0.47\textwidth]{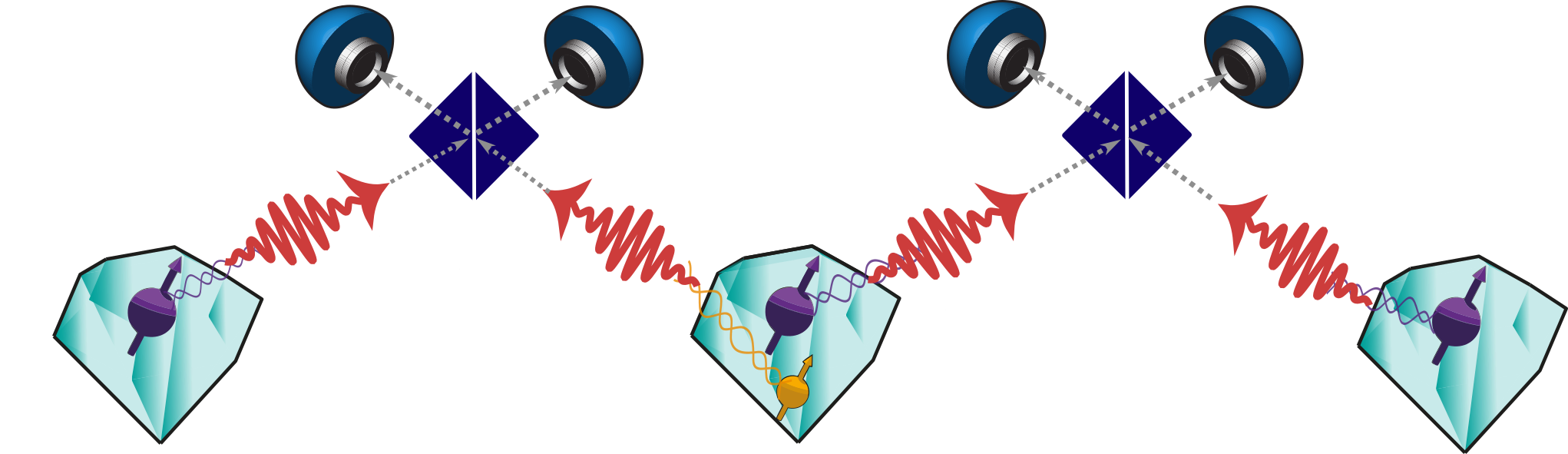}
\caption{Schematic overview of the setup for the SPOTL scheme. This scheme is a combination of the SiSQuaRe and single-photon scheme. Instead of sending photons directly through the fiber as in the SiSQuaRe scheme, entanglement is established between the middle node and Alice and Bob using the single-photon scheme.}
\label{fig:QR3NV}
\end{figure}
\begin{enumerate}[label={(\arabic*)}]
\item  Alice and the repeater run the single-photon scheme until success with the tunable parameter $\theta = \theta_A$. However, only Alice performs her spin measurement immediately after each spin-photon entanglement generation attempt. This measurement is in a six-state basis.
\item  The repeater swaps the state of the electron spin onto the carbon spin.
\item Bob and the repeater run the single-photon scheme until success or until the cut-off $n^{*}$ is reached, in which case the scheme is restarted with step 1. The tunable parameter is set here to $\theta = \theta_B$. Again, only Bob performs his spin measurement immediately after each spin-photon entanglement generation attempt and this measurement is in a six-state basis.
\item The quantum repeater performs a Bell-state measurement and communicates the result to Bob.
\item All the previous steps are repeated until sufficient data have been generated.
\end{enumerate}
We note that for larger distances the optimal cut-off becomes smaller.  Then, since we lose the independence of the attempts on both sides, the scaling of the secret-key rate with distance is expected to drop to $\sqrt{\eta}$, which is the same as for the single-photon scheme. However, the total distance between Alice and Bob is now split into four segments. Alice and Bob thus send photons over only one fourth of the total distance. Thus, this scheme should be able to generate key over much larger distances than the previous ones, as the dark counts will start becoming significant for larger distances only.


\section{NV-implementation}
\label{sec:modeling}
Having proposed different quantum repeater schemes, we now move on to describe their experimental implementation based on nitrogen-vacancy centers in diamond~\cite{Doherty2013}. This defect center is a prime candidate for a repeater node due to its packaged combination of a bright optical interface featuring spin-conserving optical transitions that enable high-fidelity single-shot readout~\cite{robledo2011high} and individually addressable, weakly coupled $^{13}$C memory qubits that can be used to store quantum states in a robust fashion \cite{maurer2012room,kalb2017entanglement}. Moreover, second-long coherence times of an NV electron spin have been achieved recently by means of dynamical decoupling sequences~\cite{Abobeih2018}.

By applying selective optical pulses and coherent microwave rotations, we first generate spin-photon entanglement at an NV center node~\cite{bernien2013heralded}. To generate entanglement between two distant NV electron spins, these emitted photons are then overlapped on a central beam splitter to remove their which-path information. Subsequent detection of a single photon heralds the generation of a spin-spin entangled state~\cite{bernien2013heralded}. For all schemes based on single-photon entanglement generation, we need to employ active phase-stabilization techniques to compensate for phase shifts of the transmitted photons, which will reduce the entangled state fidelity, as introduced in Sec.~\ref{subsecsingle-photonscheme}. These fluctuations arise from both mechanical vibrations and temperature-induced changes in optical path length, as well as phase fluctuations of the lasers used during spin-photon entanglement generation. This problem can be mitigated by using light reflected off the diamond surface to probe the phase of an effectively formed interferometer between the two NV nodes and the central beam splitter, and by feeding the acquired error signal back to a fiber stretcher that changes the relative optical path length~\cite{humphreys2017deterministic}.

The electron spin state can be swapped to a surrounding $^{13}$C nuclear spin to free up the single optical NV interface per node for a subsequent entangling round; a weak (approx.~few kHz), always-on, distance-dependent magnetic hyperfine interaction between the electron and $^{13}$C spin forms the basis of a dynamical decoupling based universal set of nuclear gates that allow for high-fidelity control of individual nuclear spins~\cite{taminiau2014universal, cramer2015repeated, reiserer2016robust,kalb2017entanglement}. Crucially, the so-formed memory can retain coherence for thousands of remote entangling attempts despite stochastic electron spin reset operations, quasi static noise, and microwave control infidelities during the subsequent probabilistic entanglement generation attempts~\cite{reiserer2016robust,Kalb2018} (see Appendix \ref{sec:noisemem} for details).

In the NV node containing both the electron and carbon nuclear spin, it is also possible to perform a deterministic Bell-state measurement on the two spins. Specifically, a combination of two nuclear-electron spin gates and two sequential electron spin state measurements reads out the combined nuclear-electron spin state in the $Z$ and $X$ bases, enabling us to discriminate all four Bell states~\cite{pfaff2014unconditional}.

For an NV center in free space, only $\sim 3 \%$ of photons are emitted in the \textit{zero-phonon line} (ZPL) that can be used for secret-key generation. This poses a key challenge for a repeater implementation, since this means that the probability of successfully detecting an emitted photon is low. Therefore, we consider a setup in which the NV center is embedded in an optical cavity with a high ratio of quality factor $Q$ to mode volume $V$ to enhance this probability via the Purcell effect in the weak coupling regime~\cite{Purcell1946}. This directly translates into a lower optical excited state lifetime that is beneficial to shorten the time window during which we detect ZPL photons after the beam splitter, reducing the impact of dark counts on the entangled state. Additionally, a cavity introduces a preferential mode into which the ZPL photons are emitted that can be picked up efficiently. This leads to a higher expected collection efficiency than the non cavity case~\cite{bogdanovic2016design}. Enhancement of the ZPL has been successfully implemented for different cavity architectures, including photonic crystal cavities~\cite{Englund2010,Wolters2010,VanDerSar2011,Faraon2012,Hausmann2013,Lee2014,Li2015,Riedrich-Moller2015}, microring resonators~\cite{Faraon2011}, whispering gallery mode resonators~\cite{Barclay2011,Gould2016} and open, tunable cavities~\cite{Kaupp2013,Johnson2015,riedel2017deterministic}. However, cavity-assisted entanglement generation has not yet been demonstrated for these systems, limited predominantly by broad optical lines of surface-proximal NV centers. Therefore, we focus on the open, tunable microcavity approach~\cite{Hunger2010}, since it has the potential for incorporating micron-scale diamond slabs inside the cavity, while allowing to keep high $Q/V$ values and providing \emph{in situ} spatial and spectral tunability~\cite{Janitz2015}. In these diamond slabs, an NV center can be microns away from surfaces, potentially allowing to maintain bulklike optical and spin properties as needed for the considered repeater protocols.   

\section{Calculation of the secret-key rate}
\label{sec:secretkeyrate}
\par{With the modeling of each of the components of the different setups in hand, the performance of each setup can be estimated. The performance of a setup is assessed in this paper by its ability to generate secret key between two parties, Alice and Bob. We note here that the ability of a quantum repeater to generate secret key can be measured in two different ways - in its \emph{throughput} and its \emph{secret-key rate}. The throughput is equal to the amount of secret key generated per unit time, while the secret-key rate equals the amount of secret key generated per \emph{channel use}. In this paper, we will focus on the secret-key rate only. This is because it allows us to make concrete information-theoretical statements about our ability to generate secret key. Moreover, we note that the secret-key rate is also more universal in the sense that it can be easily converted into the throughput by multiplying it with the repetition rate of our scheme (number of attempts we can perform in a unit time). It must be also noted here that demonstrating repeater schemes that achieve higher throughput than the currently available QKD systems based on direct transmission will be a great challenge. This is because the sources of photonic states used within those QKD systems operate at the GHz repetition rates, while the performance of the repeater schemes will be limited by many additional factors such as transmission latency and time of local operations at the memory nodes. These issues are not captured by the secret-key rate directly. Nevertheless, as mentioned before, the universality of the secret-key rate allows for the interconversion between the two quantities. We further discuss the differences between the throughput and secret-key rate in Sec.~\ref{sec:discussion}. 

\par{The secret-key rate $R$ is equal to

\begin{align}
R = \frac{Y r}{N_{\textrm{modes}}}\ ,
\end{align}

where $Y$ and $r$ are the yield and secret-key fraction, respectively. The yield $Y$ is defined as the average number of raw bits generated per channel use and the secret-key fraction $r$ is defined as the amount of secret key that can be extracted from a single raw bit (in the limit of asymptotically many rounds). Here $N_{\textrm{modes}}$ is the number of optical modes needed to run the scheme. Time-bin encoding requires two modes while the single-photon scheme uses only one mode. Hence, $N_{\textrm{modes}} = 2$ for all the schemes that use time-bin encoding in at least one of the arms of the setup. For the schemes that use only the single-photon subschemes as their building blocks, we have that $N_{\textrm{modes}} = 1$}.
\par{In the remainder of this section, we will briefly detail how to calculate the yield and secret-key fraction, from which we can estimate the secret-key rate of each scheme.}


\subsection{Yield}
\label{sec:yield}
The yield depends not only on the used scheme but also on the losses in the system. We model the general emission and transmission of photons through fibers from NV centers in diamond as in Fig.~\ref{fig:photonlosses}. That is, with probability $p_{\textrm{ce}}$ spin-photon entanglement is generated and the photon is coupled into a fiber. The photons that successfully got coupled into the fiber might not be useful for quantum information processing since they are not coherent. Thus, we filter out those photons that are not emitted at the zero-phonon line, reducing the number of photons by a further factor of $p_{\textrm{zpl}}$. Then, over the length of the fiber, a photon gets lost with probability $1-\eta_f = 1-e^{-\frac{L}{L_0}}$, where $L_0$ is the attenuation length and $\eta_f$ is the transmissivity. After exiting the fiber, the photon gets registered as a click by the detector with probability $p_{\textrm{det}}$. Finally, the photon gets accepted as a successful click if the click happens within the time window $t_{\textrm{w}}$ of the detector (see Appendix~\ref{appendix:lossesandnoise} for more details).

\begin{centering}
\begin{figure}
\centerfloat
\includegraphics[width = 9cm]{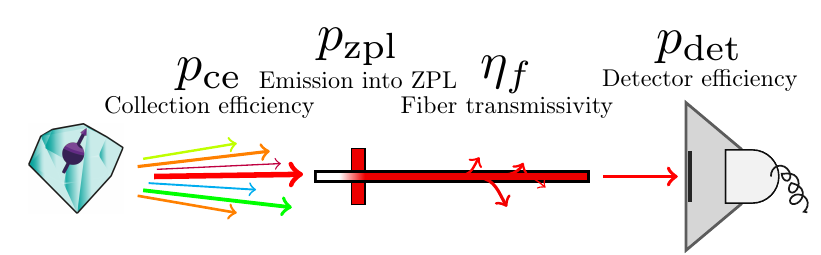}

\caption{The model of photon-loss proccesses occurring in our repeater setups. The parameter $p_\textrm{ce}$ is the photon-collection efficiency, which includes the probability that the photon is successfully coupled into the fiber. Only photons emitted at the zero-phonon line (ZPL) can be used for quantum information processing.  All non-ZPL photons are filtered out, such that a fraction $p_\textrm{zpl}$ of the photons remains. The photons are then transmitted through a fiber with transmissivity $\eta_{f}$. Such successful transmissions are registered by the detector with probability $p_\textrm{det}$. Additionally, a significant fraction of photons can arrive in the detector outside of the detection time window $t_{\textrm{w}}$. Such photons will effectively also get discarded. Here we describe the total efficiency of our apparatus by a single parameter, $p_{\textrm{app}} = p_{\textrm{ce}}p_\textrm{zpl}p_{\textrm{det}}$.}
\label{fig:photonlosses}
\end{figure}
\end{centering}

The yield can then be calculated as the reciprocal of the expected number of channel uses needed to get one single raw bit,

\begin{equation}
Y=\frac{1}{\mathbb{E}[N]}\ ,
\end{equation}

with $N$ being the random variable that models the number of channel uses needed for generating a single raw bit.


\subsubsection*{Yield of the single-photon scheme}
The yield of the single-photon scheme is relatively easy to calculate, since the single condition heralding the success of the scheme is a single click in one of the detectors in the heralding station. Therefore, the yield $Y$ is simply the probability that an individual attempt will result in a single click in one of the detectors. This probability will depend on the losses in the system, dark counts and the angle $\theta$. A full calculation of the yield is given in Appendix~\ref{sec:SingleClick}.


\subsubsection*{Yield of the SiSQuaRe, SPADS, and SPOTL schemes}
The SiSQuaRe, SPADS, and SPOTL schemes require two conditions for the heralding of the successful generation of a raw bit, namely the scheme needs to succeed both on Alice's and Bob's side independently. In this case we are going to take a very conservative perspective and assume the total number of channel uses to be the sum of the required channel uses on Alice's and Bob's side of the memory repeater node,
\begin{equation}
\mathbb{E}[N]=\mathbb{E}[N_{A}+N_{B}]\ .
\end{equation}
Moreover, every time Bob reaches $n^{*}$ attempts, both parties start the scheme over again. The cut-off increases the average number of channel uses, thus decreasing the yield. Denoting by $p_{A}$ and $p_{B}$ the probability that a single attempt of the subscheme on Alice's and Bob's side respectively succeeds, we find (see Appendix \ref{sec:expnumbsum} for the derivation)
\begin{equation}
\mathbb{E}[N_{A} + N_{B}] =  \frac{1}{p_{A}\left(1-\left(1-p_{B}\right)^{n^{*}}\right)}+\frac{1}{p_{B}}\ .
\end{equation}}


\subsection{Secret-key fraction}
\label{sec:secfrac}
The secret-key fraction is the fraction of key that can be extracted from a single raw state. It is a function of the average quantum bit error rates in the $X$, $Y$, and $Z$basis~\cite{scarani2009security,watanabe2007key} (QBER), and depends on the protocol (such as the BB84~\cite{Bennett_84} or six-state protocol~\cite{bruss1998optimal, bechmann1999incoherent}) and classical postprocessing used (such as the advantage distillation post-processing~\cite{watanabe2007key})
.

Here we consider the entanglement-based version of the BB84 and six-state protocols. That is, Alice and Bob both perform measurements on their local qubits which share quantum correlations. We note that both the BB84 and the six-state protocol can in principle be run either in a symmetric or asymmetric way. Symmetric means that the probabilities of performing measurements in all the used bases are the same, while for asymmetric protocols they can be different. We note in the asymptotic regime, which is the regime that we consider here, it is possible to set this probability bias to approach unity and still maintain security~\cite{lo2005efficient}. Unfortunately, for technical reasons, within our model it is not possible to run an asymmetric six-state protocol when time-bin-encoded photons are used~\cite{parameterregimes}.

Moreover, as we mentioned above, it is also possible to apply different types of classical postprocessing of the raw key generated through the BB84 or the six-state protocol. In particular, here we consider two types of post processing: the standard one-way error correction and a more involved two-way error correction protocol called advantage distillation, which can tolerate much more errors. Specifically, here we consider the advantage distillation protocol proposed in Ref.~\cite{watanabe2007key}, as this advantage distillation protocol has high efficiency (in particular, in the scenario of no noise, the efficiency of this protocol equals unity). Hence, in our model we effectively consider two protocols for generating secret key: BB84 with standard one-way error correction and six-state with advantage distillation. We refer the reader to Appendix~\ref{sec:secretkeyfracderiv} for the mathematical expressions for the secret-key fraction for all the considered protocols.

Now we can state explicitly which QKD protocols will be considered for each scheme, which in turn depends on the type of measurements that Alice and Bob perform in that scheme. There are two physical implementations of measurements that Alice and Bob perform, depending on the scheme under consideration. That is, they either measure a quantum state of a spin or of a time-bin encoded photons. Since the fully asymmetric six-state protocol with advantage distillation has higher efficiency than both symmetric and asymmetric BB84 protocol with one-way error correction, we will use this six-state protocol for both the single-photon and SPOTL scheme. The SiSQuaRe and SPADS schemes involve direct measurement on time-bin encoded photons. Hence, for these schemes, we consider the maximum of the amount of key that can be obtained using the fully asymmetric BB84 protocol and the symmetric six-state protocol with advantage distillation (which can tolerate more noise, but has three times lower efficiency than the fully asymmetric BB84 protocol).

To estimate the QBER, we model all the noisy and lossy processes that take place during the protocol run. From this, we calculate the qubit error rates and yield, from which we can retrieve the secret-key fraction. We invite the interested reader to read about the details of these calculations in Appendices~\ref{sec:SingleClick} and~\ref{appendix:QBERsingleclicktimes2}. The derivation of the QBER and the yield for the SiSQuaRe scheme is performed in Ref.~\cite{parameterregimes}. Moreover, in this work we introduce certain refinements to the model which we discuss in Appendix~\ref{sec:SiSQuaReChanges}.
With the QBER in hand, we can calculate the resulting secret-key fraction for the considered protocols as presented in Appendix~\ref{sec:secretkeyfracderiv}. 

\par{We note here that we consider only the secret-key rate in the asymptotic limit, and that we thus do not have to deal with non-asymptotic statistics.
}


\section{Assessing the performance of quantum repeater schemes}
\label{sec:benchmarks}

\par{In this section, we will detail four benchmarks that will be used to assess the performance of quantum repeaters. The usage of such benchmarks for repeater assessment has been done in Refs.~\cite{parameterregimes, luong2015overcoming}, and achieving a rate greater than such benchmarks can be seen as milestones toward the construction of a quantum repeater.}
\par{The considered benchmarks are defined with respect to the efficiencies of processes involving photon loss when emitting photons at NV centers, transmitting them through an optical fiber and detecting them at the end of the fiber as described in Sec.~\ref{sec:yield} and as shown in Fig.~\ref{fig:photonlosses}.}
\par{Having this picture in mind, we can now proceed to present the considered benchmarks. The first three of these benchmarks are inspired by fundamental limits on the maximum achievable secret-key rate if Alice and Bob are connected by quantum channels which model quantum key distribution over optical fiber without the use of a (possible) quantum repeater.} 

\par{\textbf{The first of these benchmarks} we consider here is also the most stringent one, the so-called \emph{capacity of the pure-loss channel}. The capacity of the pure-loss channel is the maximum achievable secret-key rate over a channel modeling a fiber of transmissivity $\eta_f$, and is given by~\cite{pirandola2015fundamental}

\begin{align}
-\log_2\left(1-\eta_f\right)\ . \label{eq:benchmark1}
\end{align}
This is the maximum secret-key rate achievable, meaning that even if Alice and Bob had perfect unbounded quantum computers and memories, they could not generate secret key at a larger rate. If, by using a quantum repeater setup, a higher rate can be achieved than $-\log_2(1-\eta_f)$, we are certain our quantum repeater setup allowed us to do something that would be impossible with direct transmission. Surpassing the secret-key capacity has been widely used as a defining feature of a quantum repeater~\cite{khalique2015practical,krovi_15,pant2017rate,guha2015rate,luong2015overcoming,parameterregimes,takeoka2014fundamental,takeoka2014squashed,pirandola2015fundamental,goodenough2016assessing,kaur2017upper,sharma2017bounding}. Unfortunately, and as could be expected, surpassing the capacity is experimentally challenging. This motivates the introduction of other, easier to surpass, benchmarks. These benchmarks are still based on (upper bounds on) the secret-key capacity of quantum channels which model realistic implementations of quantum communications over fibers.}
\par{\textbf{The second benchmark} is built on the idea of including the losses of the apparatus into the transmissivity of the fiber. The resultant channel with all those losses included we call here \emph{the extended channel}. The benchmark is thus equal to 
\begin{align}
-\log_2\left(1-\eta_f p_{\textrm{app}}\right)\ . \label{eq:benchmark2}
\end{align}
Here $p_{\textrm{app}}$ describes all the intrinsic losses of the devices used, that is, the collection efficiency $p_{\textrm{ce}}$ at the emitting diamond, the probability that the emitted photon is within the zero-phonon line $p_{\textrm{zpl}}$ (which is necessary for generating quantum correlations), and photon detection efficiency $p_{\textrm{det}}$, so that $p_{\textrm{app}} = p_{\textrm{ce}}p_{\textrm{zpl}}p_{\textrm{det}}$.}

\textbf{The third benchmark} we consider is the so-called \emph{thermal channel bound}, which takes into account the effects of dark counts. The secret-key capacity of the thermal channel has been studied extensively~\cite{davis2018energy,goodenough2016assessing,sharma2017bounding,kaur2017upper, pirandola2015fundamental,ottaviani2016secret,laurenza2018finite,laurenza2018tight}. We consider the following bound on the secret-key capacity of the thermal channel,
\begin{align}
-\log_2\left[\left(1-\eta_f p_{\textrm{app}}\right)\left(\eta_f p_{\textrm{app}}\right)^{\overline{n}}\right]-g\left(\overline{n}\right)\ , \label{eq:benchmark3}
\end{align}
if $\overline{n}\leq \frac{\eta_f p_{\textrm{app}}}{1-\eta_f p_{\textrm{app}}}$, and otherwise zero~\cite{pirandola2015fundamental}. Here $\overline{n}$ is the average number of thermal photons per channel use and is equal to $t_{\textrm{w}}$, the time window of the detector, times the average number of dark counts per second; see Ref.~\cite{parameterregimes} for more details. The function $g(x)$ is defined as $g(x) \equiv \left(x+1\right)\log_2\left(x+1\right)-x\log_2\left(x\right)$. We note here that the time window of the detector $t_{\textrm{w}}$ is not fixed in our model but is optimized over for every distance in order to achieve the highest possible secret-key rate. Hence, in this benchmark we fix $t_{\textrm{w}} = 5$ ns which is the shortest duration of the time window that we consider in our secret-key rate optimization.   

Finally, the secret-key rate achieved with \emph{direct transmission using the same devices} can be seen as \textbf{the fourth benchmark}. Specifically, here we mean the secret-key rate achieved when Alice uses her electron spin to generate spin-photon entanglement and sends the time-bin-encoded photon to Bob. She then measures her electron spin while Bob measures the arriving photon. However, to take a conservative view, we will only use this direct transmission benchmark for the SPADS scheme. This is motivated by the fact that for both the SPADS scheme and the direction transmission scheme, the experimental setups on Alice's and Bob's side are the same, ensuring that the two rates can be compared fairly. We note that similarly as in the modeled secret-key rates achievable with our proposed repeater schemes, also for this direct transmission benchmark we optimize over the time window $t_{\textrm{w}}$ for each distance.

The secret-key capacity stated in Eq.~\eqref{eq:benchmark1} is the main benchmark that we consider. Surpassing it establishes the considered scheme as a quantum repeater. The two expressions in Eqs.~\eqref{eq:benchmark2} and \eqref{eq:benchmark3} and the achieved rate with direct transmission are additional benchmarks, which guide the way toward the implementation of a quantum repeater. We define all the considered benchmarks for the channel with the same fiber attenuation length $L_0$ as the channel used for the corresponding achievable secret-key rate.


\section{Numerical results}
\label{sec:results}

We now have a full model of the rate of the presented quantum repeater protocols as a function of the underlying experimental parameters. In this section, we will first state all the parameters required by our model and then present the results and conclusions drawn from the numerical implementation of this model. In particular, in Sec.~\ref{sec:bb846state}, we will first provide a deeper insight into the benefits of using the six-state protocol and advantage distillation in specific schemes. In Sec.~\ref{sec:optimal}, we determine the optimal positioning of the repeater nodes for our schemes and investigate the dependence of the secret-key rate achievable with those schemes on the photon emission angle $\theta$ and the cutoff $n^*$ for the appropriate schemes.
In Sec.~\ref{sec:mainplots}, we then use the insights acquired in the previous section to compare the achievable secret-key rates for all the proposed repeater schemes with the secret-key capacity and other proposed benchmarks. In particular, we show that the single-photon scheme significantly outperforms the secret-key capacity and hence can be used to demonstrate a quantum repeater. Finally, in Sec.~\ref{sec:runtime}, we determine the duration of the experiment that would allow us to demonstrate such a quantum repeater with the single-photon scheme. 

The parameters that we will use are either parameters that have been achieved in an experiment or correspond to expected parameters when the NV center is embedded in an optical Fabry-Perot microcavity. The parameters we will use are listed in Table 1. 
\begin{table*}[]
    \centering
    \begin{ruledtabular}
    \begin{tabular}{l l l}
	    \textbf{Parameter} & \textbf{Notation} & \textbf{Value} \\
         Dephasing of $^{13}$C due to interaction& $a_{0}$ & $1/2000$ per attempt~\cite{reiserer2016robust,Kalb2018} \\
         Dephasing of $^{13}$C with time& $a_{1}$ & $1/3$ per second~\cite{maurer2012room} \\
       Depolarizing of $^{13}$C due to interaction&  $b_{0}$ &  $1/5000$ per attempt ~\cite{reiserer2016robust} \\
         Depolarizing of $^{13}$C with time & $b_{1}$ & $1/3$ per second~\cite{maurer2012room}\\
        Memory-photon entanglement preparation time&  $t_{\textrm{prep}}$  & 6 $\mu$s~\cite{hensen2015loophole}\\
        Depolarizing parameter for the measurement of the electron spin& $F_{m}$ &  $0.95$~\cite{humphreys2017deterministic} \\
         Depolarizing parameter for two qubit gates in quantum memories& $F_{g}$ & $0.98$~\cite{kalb2017entanglement} \\
         Dephasing parameter for the memory-photon state preparation& $F_{\textrm{prep}}$  & $0.99$~\cite{hensen2015loophole}\\
         Collection efficiency & $p_{\textrm{ce}}$ & $0.49$~\cite{hensen2015loophole,bogdanovic2016design} \\
        Emission into the zero-phonon line & $p_{\textrm{zpl}}$ & $0.46$~\cite{riedel2017deterministic}\\
        Detector efficiency& $p_{\textrm{det}}$ & $0.8$~\cite{hensen2015loophole}\\
          Dark count rate & $d$ & $10$ per second~\cite{hensen2015loophole}\\
        Characteristic time of the NV emission& $\tau$ & $6.48$ ns~\cite{riedel2017deterministic,Fox2006}\\
        Detection window offset & $t^{\textrm{offset}}_{\textrm{w}}$ & $1.28$ ns~\cite{hensen2015loophole}\\
          Attenuation length & $L_{0}$ & $0.542$ km~\cite{hensen2015loophole} \\
        Refractive index of the fiber & $n_{\textrm{ri}}$ &  $1.44$~\cite{LaserEncyclopedia} \\
         Optical phase uncertainty of the spin-spin entangled state& $\Delta\phi$& \ang{14.3}~\cite{humphreys2017deterministic} 
    \end{tabular}
    \end{ruledtabular}
    \caption{Parameters used for the nitrogen-vacancy center setups considered in this paper.}
    \label{tab:Parameters}
\end{table*}

To be now more specific, the photon collection efficiency $p_{\textrm{ce}}$ and the probability of emitting into the zero-phonon line $p_{\textrm{zpl}}$ are the two crucial parameters relying on the implementation of the optical cavity. The quoted value of $p_{\textrm{ce}}$ has not been experimentally demonstrated yet, while the value of $p_{\textrm{zpl}}$ has not been demonstrated in the context of quantum communication. All the other independent parameters in the above list that are not related to the setup with a cavity have been demonstrated in experiments relevant for remote entanglement generation. The parameters that have not been discussed in the main text are discussed in the appendixes.


\subsection{Comparing BB84 and six-state advantage distillation protocols}
\label{sec:bb846state}
We first investigate here when the BB84 or six-state advantage distillation protocol performs better. It was shown in Ref.~\cite{parameterregimes} that in the SiSQuaRe scheme there is a trade-off --- for the low-noise regime (small distances) the fully asymmetric BB84 protocol is preferable, while in the high-noise regime (large distances) the problem of noise can be overcome by using a six-state protocol supplemented with advantage distillation. This technique allows us to increase the secret-key fraction at the expense of reducing the yield by a factor of three, since a six-state protocol in which Alice and Bob perform measurements on photonic qubits does not allow for the (fully) asymmetric protocol within our model. Numerically, we find that for the SPADS and SPOTL scheme advantage distillation is \emph{necessary} to generate nonzero secret-key at any distance. This is due to the fact that there is a significant amount of noise in these schemes. Thus, for the SPADS (SPOTL) scheme the (a)symmetric six-state protocol with advantage distillation is optimal.

To provide more insight into the performance of those different QKD schemes for different parameter regimes, we plot the achievable secret-key fraction for the SPADS and SPOTL schemes as a function of the depolarizing parameter due to imperfect electron spin measurement $F_m$ in Fig.~\ref{fig:optprot} (see Appendix~\ref{sec:noisemem} for the discussion of the corresponding noise model). Noise due to imperfect measurements is one of the significant noise sources in our setup, since the SPADS scheme involves three and the SPOTL scheme four single-qubit measurements on the memory qubits. The data have been plotted for a fixed distance of $12.5L_0$, where $L_0 = 0.542$ km is the attenuation length of the fiber. Moreover, since on this plot we aim at maximizing only the secret-key fraction over the tunable parameters, we set the cutoff $n^{*}$ to one and the detection time window $t_{\textrm{w}}$ to 5 ns (the smallest detection time window we use) for both schemes. Furthermore, within the single-photon subscheme the heralding station is always placed exactly in the middle between the two memory nodes. We also consider the positioning of the memory repeater node to be two-thirds away from Alice for the SPADS scheme and in the middle for the SPOTL scheme as discussed in the next section. For the SPOTL scheme we also assume $\theta_A = \theta_B$, which we will justify in the next section.

We see that for the current experimental value of $F_m =0.95$ both schemes can generate key only if the advantage distillation postprocessing is used. As $F_m$ increases, we observe that for the SPADS scheme first the six-state protocol without advantage distillation and then the BB84 protocol start generating key. For the SPOTL scheme the value of $F_m$ at which the six-state protocol without advantage distillation starts generating key is much larger than the corresponding value of $F_m$ for any of the studied protocols for the SPADS scheme. This is because the SPOTL scheme involves more noisy processes than the SPADS scheme. This also provides an approximate quantification of the benefit of using advantage distillation. Specifically, looking at the SPOTL scheme, it can be observed that while at the current experimental value of $F_m =0.95$, advantage distillation allows for generating key, but at a higher value of the depolarizing parameter $F_m =0.97$, still no key can be generated with standard one-way post-processing. Moreover, we see that utilizing advantage distillation for the SPADS scheme allows for the generation of key, even with very noisy measurements when $F_m=0.91$. We also observe two distinct scalings of the secret-key fraction with $F_m$ in the regime where nonzero amount of key is generated. These two scalings depend on whether we use a symmetric or asymmetric protocol. Specifically, for the SPADS scheme the symmetric six-state protocol is used. Therefore, the corresponding two curves have a slope that is approximately three times smaller than the other three curves corresponding to the protocols that run in the fully asymmetric mode. 

\begin{figure}
\centering
\includegraphics[width=0.53\textwidth, trim= 0cm 0cm 0cm 0cm, clip]{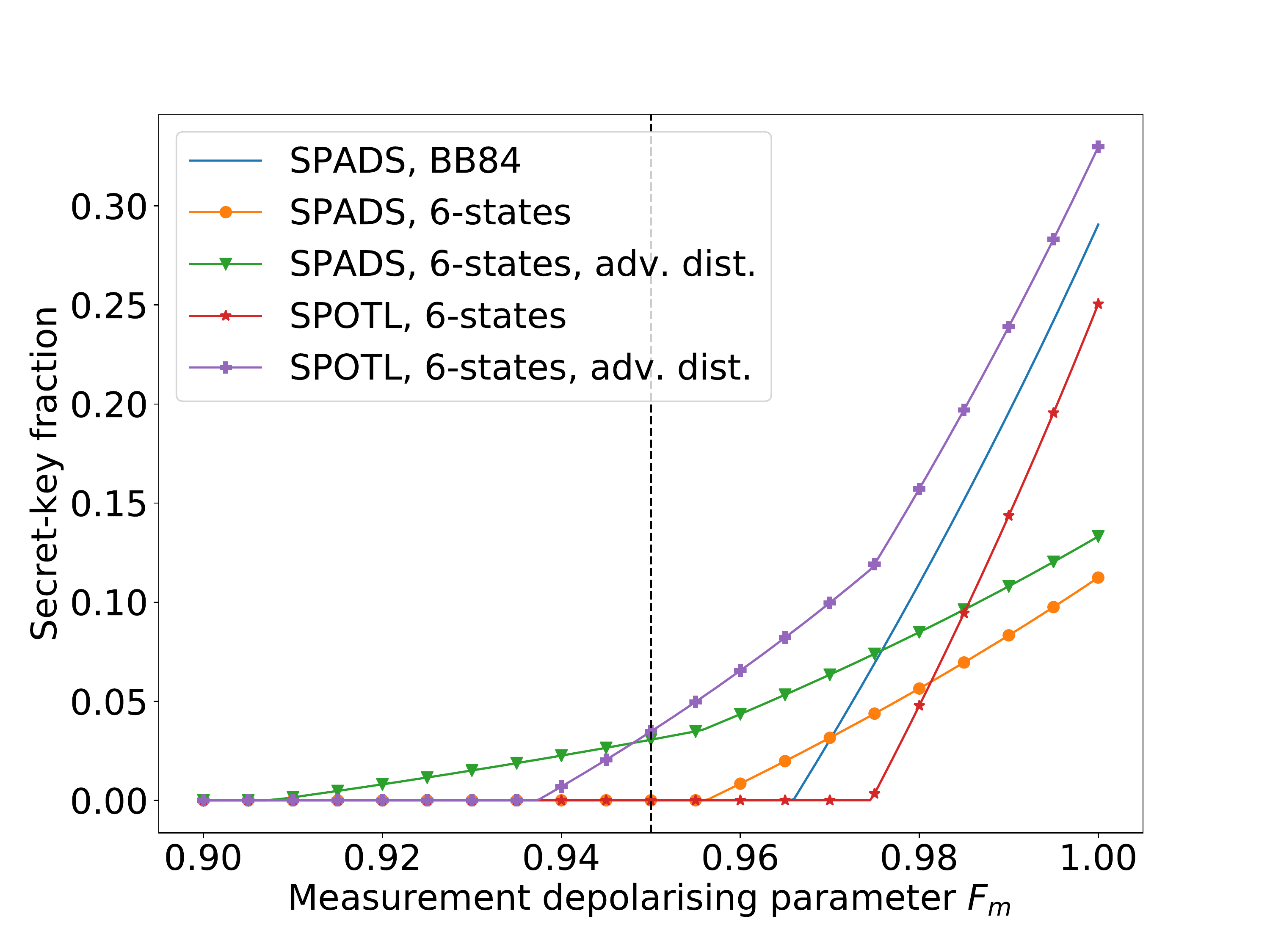}
\caption{Secret-key fraction as a function of the depolarizing parameter due to noisy measurement $F_m$ for the total distance of $12.5L_0$. We see that for the current experimental value of $F_m =0.95$ (marked with a dashed black vertical line) both schemes can generate key only if the advantage distillation post-processing is used. As $F_m$ increases, the protocols that do not utilize advantage distillation also start generating key. We also see that the curves can be divided into two groups in terms of their slope in the regime where they generate nonzero amount of key. Those two groups correspond to the scenarios where a fully asymmetric (bigger slope) or a symmetric (smaller slope) protocol is used. For all the plotted protocols, the cutoff $n^{*}$ is set to one and $t_{\textrm{w}}=5$ ns (the smallest detection time window we use) to maximize the secret-key fraction. Moreover, for each value of $F_m$, we optimize the secret-key fraction over the angle $\theta$. For the SPOTL scheme we assume $\theta_A = \theta_B$. For the SPADS scheme, we position the repeater node 2/3 of the total distance away from Alice and in the middle between Alice and Bob for the SPOTL scheme.}
\label{fig:optprot}
\end{figure}


\subsection{Optimal settings}
\label{sec:optimal}
We see that the above described repeater schemes include several tunable parameters. These parameters are the cut-off $\nstar$ for Bob's number of attempts until restart, the angle $\theta$ in the single-photon scheme, and the positioning of the repeater. These parameters can be optimized to maximize the secret-key rate. Here we will approach this optimization in a consistent way: We gradually restrict the parameter space by making specific observations based on numerical evidence.

The first claim that we will make is in relation to the \emph{optimal positioning of the repeater.} In Ref.~\cite{parameterregimes}, we have conjectured that for the SiSQuaRe scheme the middle positioning of the repeater is optimal. For the single-photon scheme, we want the probability of transmitting the photons from each of the two nodes to the beam-splitter heralding station to be equal. This effectively sets the target state between the electron spins to be the maximally entangled state. Hence, if we restrict ourselves to the case where the emission angles $\theta$ of both Alice and Bob are the same, then it is natural to position the heralding station symmetrically in the middle between them. Hence, the only nonobvious optimal positioning is for the SPADS and SPOTL scheme. 

\begin{figure}
\centering
\includegraphics[width=0.53\textwidth, trim= 0cm 0cm 0cm 0cm, clip]{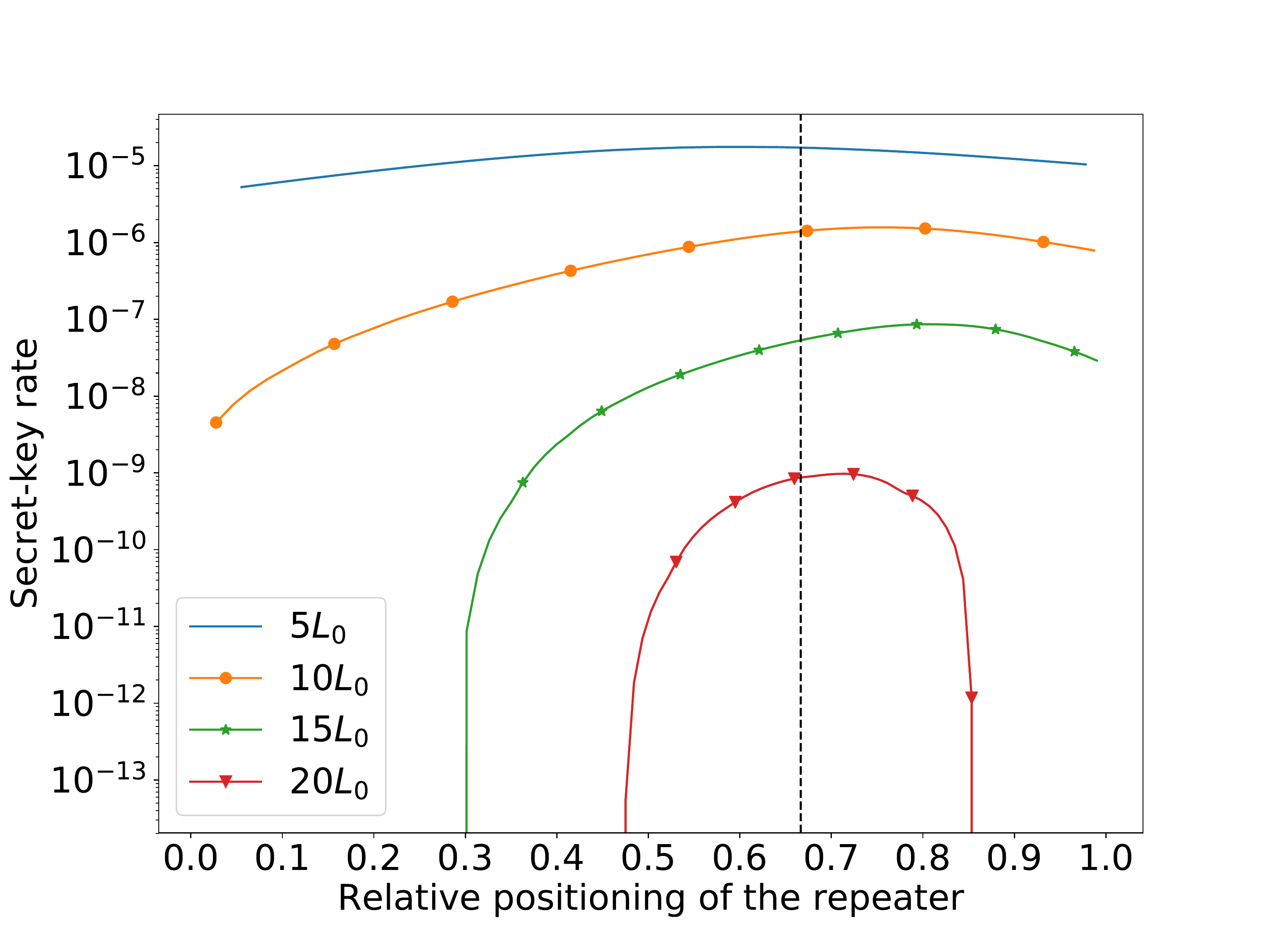}
\caption{Secret-key rate as a function of the relative positioning of the repeater for few different total distances for the SPADS scheme. The total distances are expressed in terms of the fiber attenuation length $L_0 = 0.542$ km. We see that positioning the repeater two-thirds of the distance away from Alice (marked by the vertical black dashed line) is a good positioning for all the distances. For each total distance considered and each positioning, the secret-key rate is optimized over the cutoff $n^{*}$, the angle $\theta$, and the time window of the detector $t_{\textrm{w}}$.}
\label{fig:position_2_5}
\end{figure}

For the SPADS scheme, positioning the repeater at two-thirds of the relative distance away from Alice could intuitively be expected to be optimal. This is because the single-photon scheme runs on two segments: Alice--beam-splitter, beam-splitter--repeater, while the one half of the SiSQuaRe scheme runs only over a single segment between the repeater and Bob. By segment, we mean here a distance over which we need to be able to independently transmit a photon. In Fig.~\ref{fig:position_2_5}, we show the secret-key rate as a function of the relative positioning of the repeater for a set of different total distances. We see there that despite the fact that positioning the repeater at two-thirds is not always optimal, it is a good enough positioning for all distances for our purposes. For each data point on the plot, we independently optimize over the cut-off $n^{*}$, the angle $\theta$ of the single-photon subscheme, and the duration of the detector time window $t_{\textrm{w}}$.

\begin{figure}
\centering
\includegraphics[width=0.53\textwidth, trim= 0cm 0cm 0cm 0cm, clip]{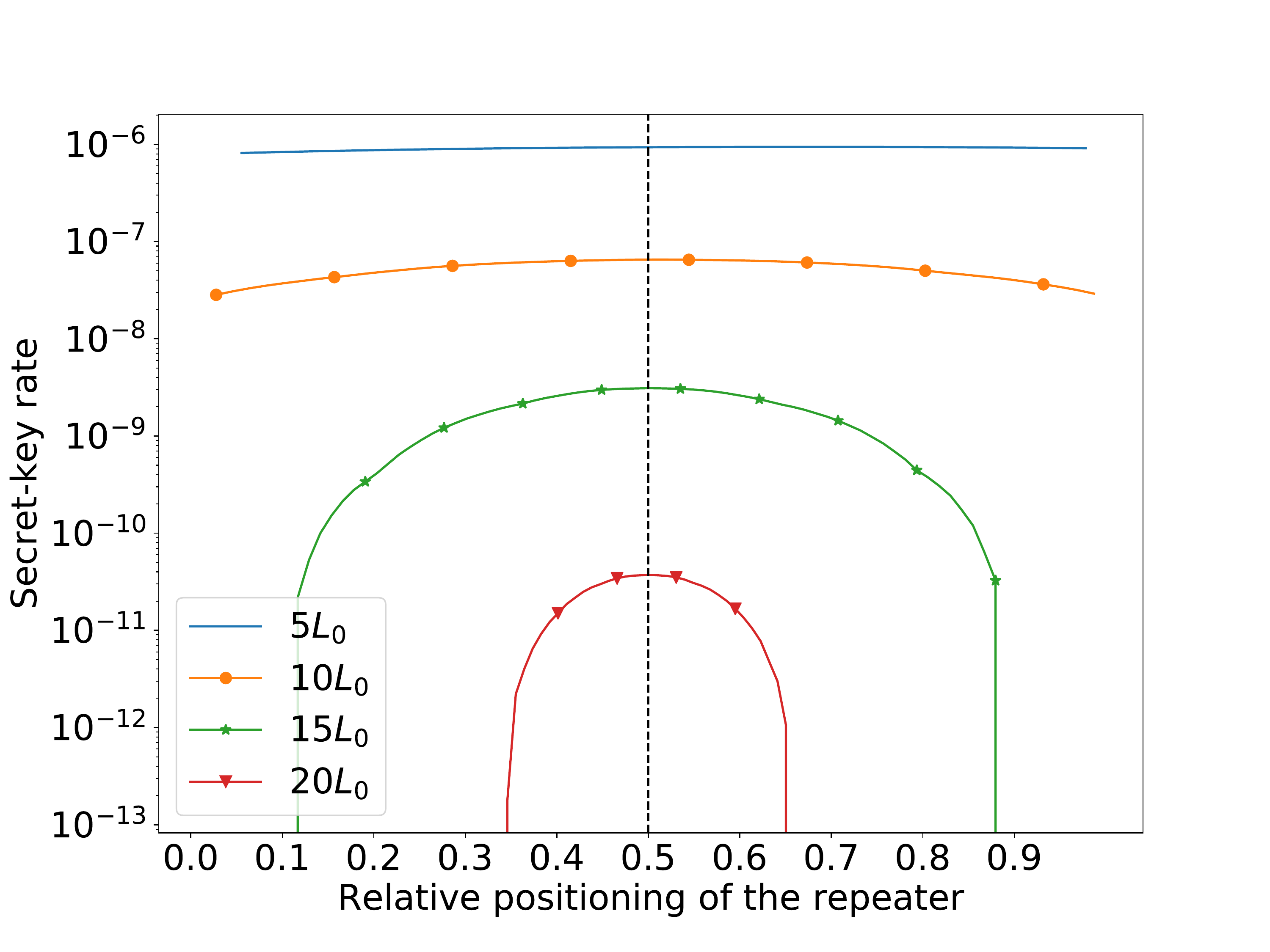}
\caption{Secret-key rate as a function of the relative positioning of the repeater for few different total distances for the SPOTL scheme. The total distances are expressed in terms of the fiber attenuation length $L_0 = 0.542$ km. We see that positioning the repeater in the middle between Alice and Bob (marked by the vertical black dashed line) is a good positioning for all the distances. For each total distance considered and each positioning the secret-key rate is optimized over the cutoff $n^{*}$, the angles $\theta_A$ and $\theta_B$ and the time window of the detector $t_{\textrm{w}}$.}
\label{fig:position3}
\end{figure}

The SPOTL scheme has the same symmetry as the SiSQuaRe scheme, in the sense that the part of the scheme performed on Alice's side is exactly the same as on Bob's side. This symmetry is only broken by the sequential nature of the scheme. Since we have already observed that the middle positioning is optimal for the SiSQuaRe scheme, we expect to see the same behavior for the SPOTL scheme. Indeed, we confirm this expectation numerically in Fig.~\ref{fig:position3}. Here for each data point we independently optimize over the cut-off $\nstar$, the angle $\theta_A$ ($\theta_B$) of the single-photon subscheme on Alice's (Bob's) side, and the duration of the detection time window.

To conclude, we will always place the heralding station within the single-photon (sub)protocol exactly in the middle between the two corresponding memory nodes. Moreover, we will also always place the memory repeater node in the middle for the SPOTL scheme and two-thirds of the distance away from Alice for the SPADS scheme.

Having established the optimal positioning of the repeater, we look into the relation between $\theta_A$ and $\theta_B$ for the SPOTL scheme. We observe that the relative error resulting from optimizing the secret-key rate over a single angle $\theta_A = \theta_B$ rather than two independent ones is smaller than $1\%$ for all distances. Hence, from now on we will restrict ourselves to optimizing only over one angle $\theta$ for the SPOTL scheme.

Having resolved the issues of the optimal positioning of the repeater for all schemes and reducing the number of angles to optimize over for the SPOTL scheme to one, we now investigate how our secret-key rate depends on the remaining parameters. These parameters are the angle $\theta$, the cut-off $\nstar$, and the duration of the detection time window $t_{\textrm{w}}$. The optimal time window follows a simple behavior for all schemes: For short distances, the probability of getting a dark count $p_d$ is negligible compared to the probability of detecting the signal photon. Hence, for those distances we can use a time window of 30 ns to make sure that almost all the emitted photons which are not polluted by the photons from the optical excitation pulse arrive inside the detection time window. We always need to sacrifice the photons arriving within the time $t_{\textrm{w}}^{\textrm{offset}}$ after the optical pulse has been applied to filter out the photons from that pulse; see Appendix~\ref{appendix:lossesandnoise} for details. Then, for larger distances where $p_d$ starts to become comparable with the probability of detecting the signal photon, the duration of the time window is gradually reduced. This reduces the effect of dark counts at the expense of having more photons arriving outside of the time window. See Appendix~\ref{appendix:lossesandnoise} for the modeling of the losses resulting from photons arriving outside of the time window.

\begin{figure}
\centering
\includegraphics[width=0.53\textwidth, trim= 0cm 0cm 0cm 0cm, clip]{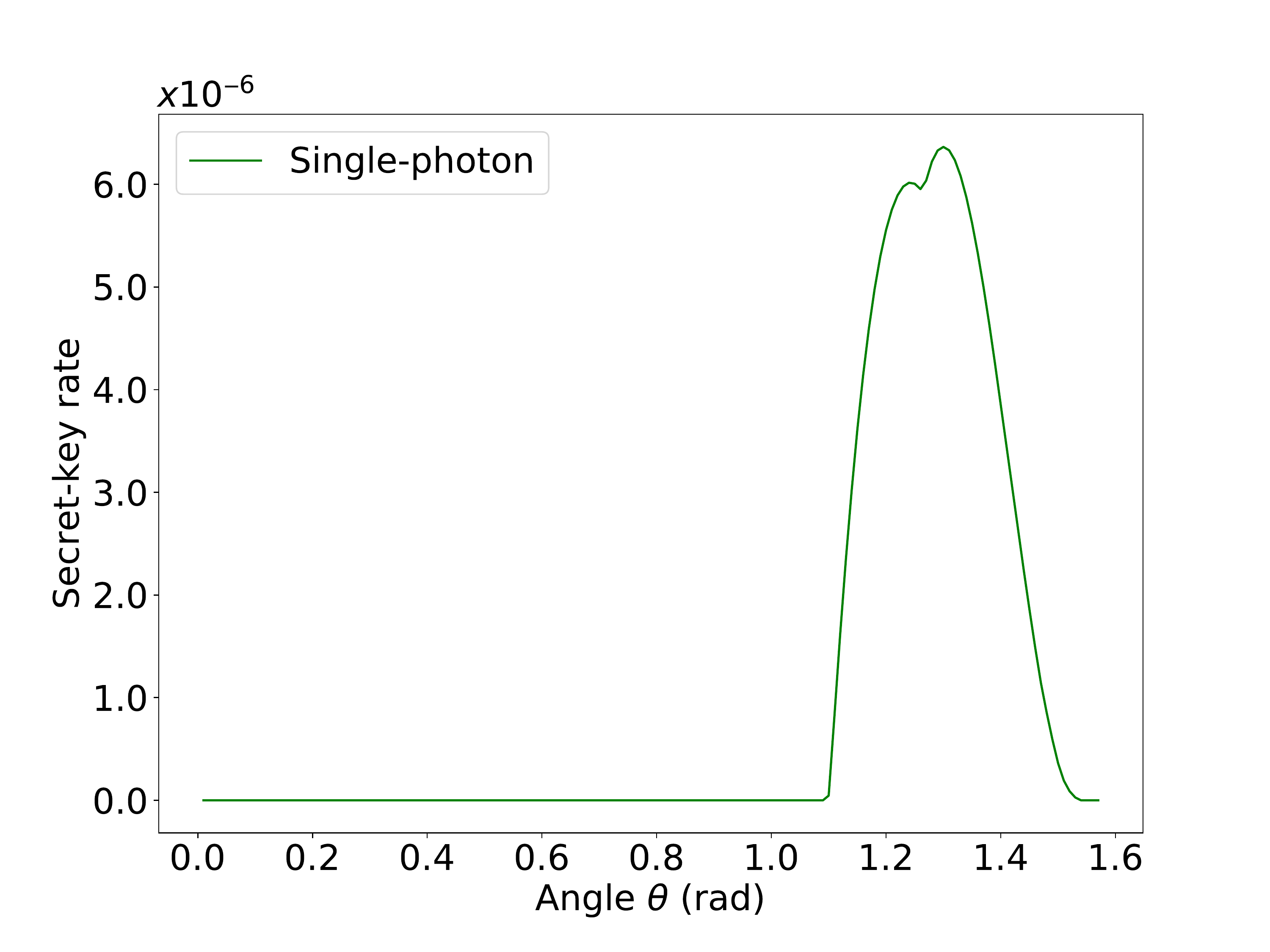}
\caption{Secret-key rate as a function of the $\theta$ angle for the single-photon scheme for the total distance of $12.5 L_0$, where $L_0 = 0.542$ km. We see that there is a relatively large range of angles for which nonzero amount of key can be generated. For each value of $\theta$, the secret-key rate is optimized over the time window $t_{\textrm{w}}$. The kink on the plot is a consequence of the fact that the six-state protocol with advantage distillation involves optimization over two subprotocols.}
\label{fig:theta_single_click}
\end{figure}

\begin{figure}
\centering
\includegraphics[width=0.53\textwidth, trim= 0cm 0cm 0cm 0cm, clip]{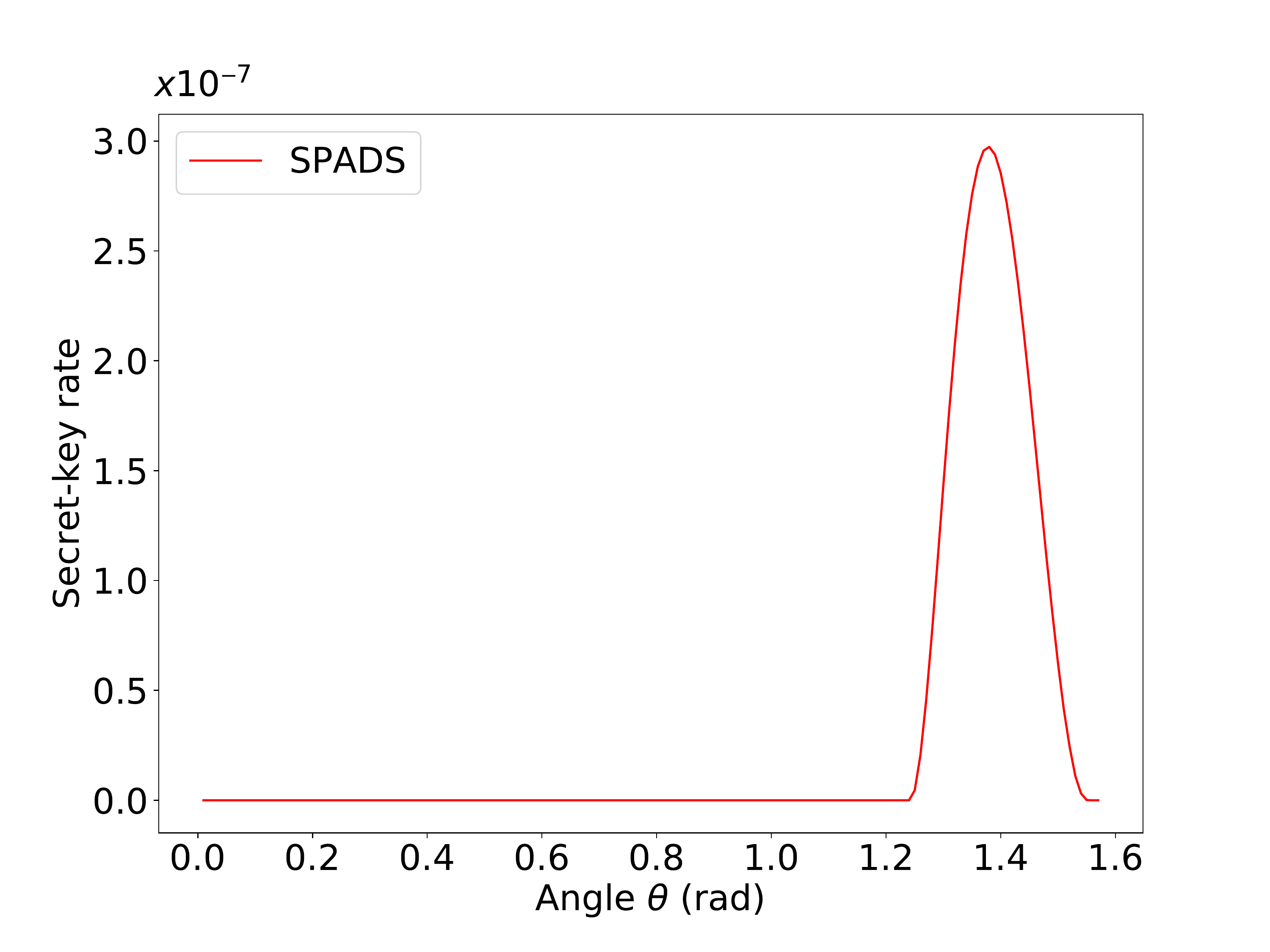}
\caption{Secret-key rate as a function of the $\theta$ angle for the SPADS scheme for the total distance of $12.5 L_0$, where $L_0 = 0.542$ km. We see that due to more noisy processes the range of $\theta$ that allows us to generate key is much more restricted than for the single-photon scheme. For each value of $\theta$, the secret-key rate is optimized over the cutoff $n^{*}$ and the time window $t_{\textrm{w}}$.}
\label{fig:theta_25}
\end{figure}

\begin{figure}
\centering
\includegraphics[width=0.53\textwidth, trim= 0cm 0cm 0cm 0cm, clip]{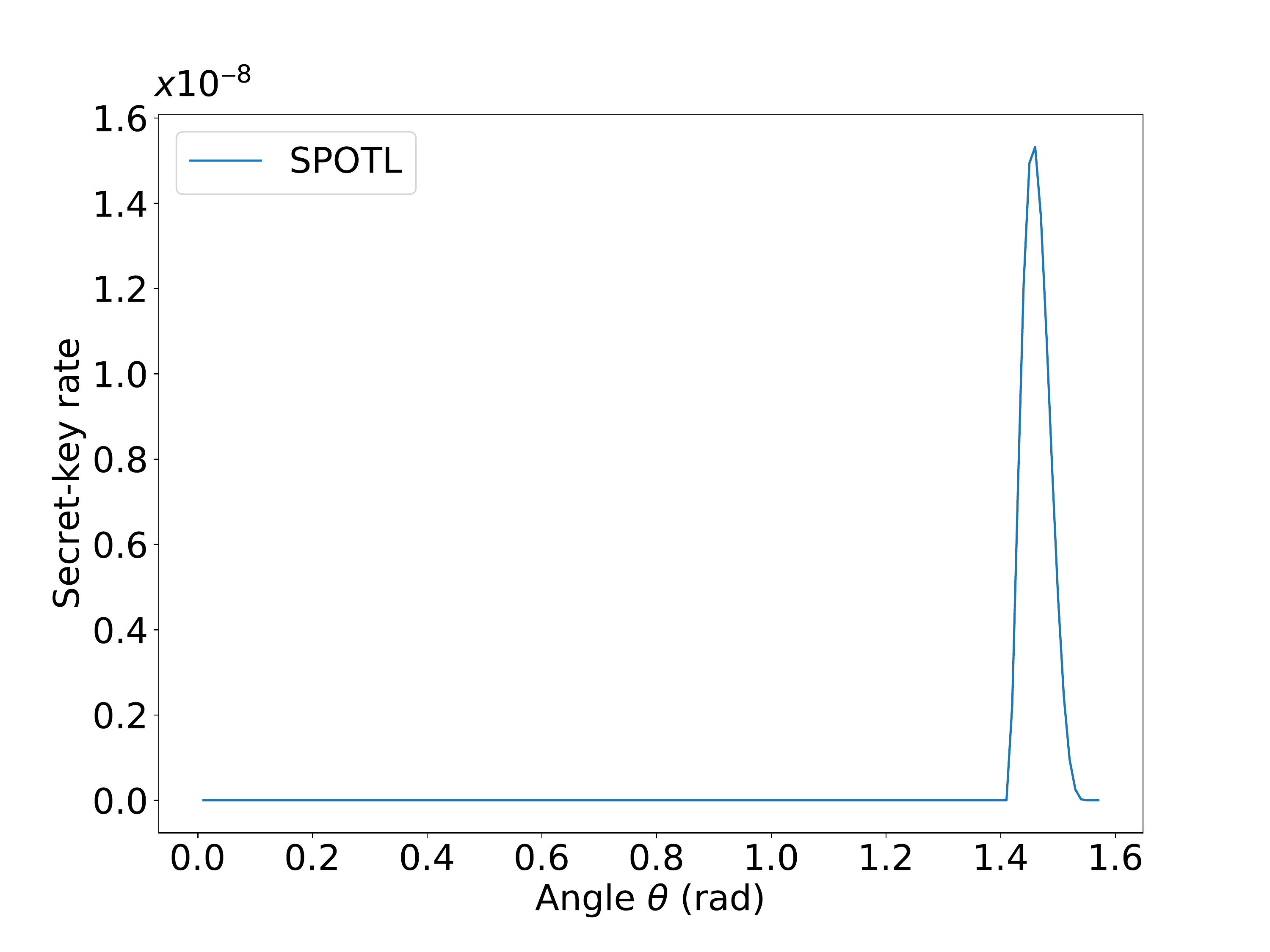}
\caption{Secret-key rate as a function of the angle $\theta = \theta_A = \theta_B$ for the SPOTL scheme for the total distance of $12.5 L_0$, where $L_0 = 0.542$ km. We see that, due to the increased amount of noisy processes, this scheme requires $\theta$ to be in a much narrower regime than for the single-photon and SPADS schemes, as can be seen by comparing the plot with the plots in Figs.~\ref{fig:theta_single_click} and~\ref{fig:theta_25}. This corresponds to the overwhelming dominance of the dark state of the spin (no emission of the photon) in order to avoid any extra noise coming from the photon loss. For each value of $\theta$, the secret-key rate is optimized over the cutoff $n^{*}$ and the time window $t_{\textrm{w}}$.}
\label{fig:theta_3_node}
\end{figure}

The dependence of the secret-key rate on the angle $\theta$, the tunable parameter that Alice and Bob choose in their starting state $\ket{\psi} = \sin\theta\ket{\downarrow}\ket{0} + \cos\theta\ket{\uparrow}\ket{1}$ in the single-photon scheme, is more complex. We observe that the optimal value of $\theta$ is closer to $\frac{\pi}{2}$ for schemes that involve more noisy processes. Informally, this means that Alice and Bob send `fewer' photons toward the beam splitter to overcome the noise coming from events in which both nodes emit a photon. At $\frac{\pi}{2}$ however, no photons are emitted and the rate drops down to zero. We illustrate this in Figs.~\ref{fig:theta_single_click},~\ref{fig:theta_25}, and~\ref{fig:theta_3_node}. We see that for the SPADS and SPOTL scheme, there is only a restricted regime of the angle $\theta$ for which one can generate nonzero amount of key. In particular, the SPOTL scheme requires a larger number of noisy operations, and therefore cannot tolerate much noise arising from the effect of photon loss in the single-photon subscheme. This means that there is only a small range of $\theta$ that allows for production of secret key. The single-photon scheme involves fewer operations and can tolerate more noise, and so lower values of the parameter $\theta$ still allow for the generation of key.

\begin{figure}
\centering
\includegraphics[width=0.53\textwidth, trim= 0cm 0cm 0cm 0cm, clip]{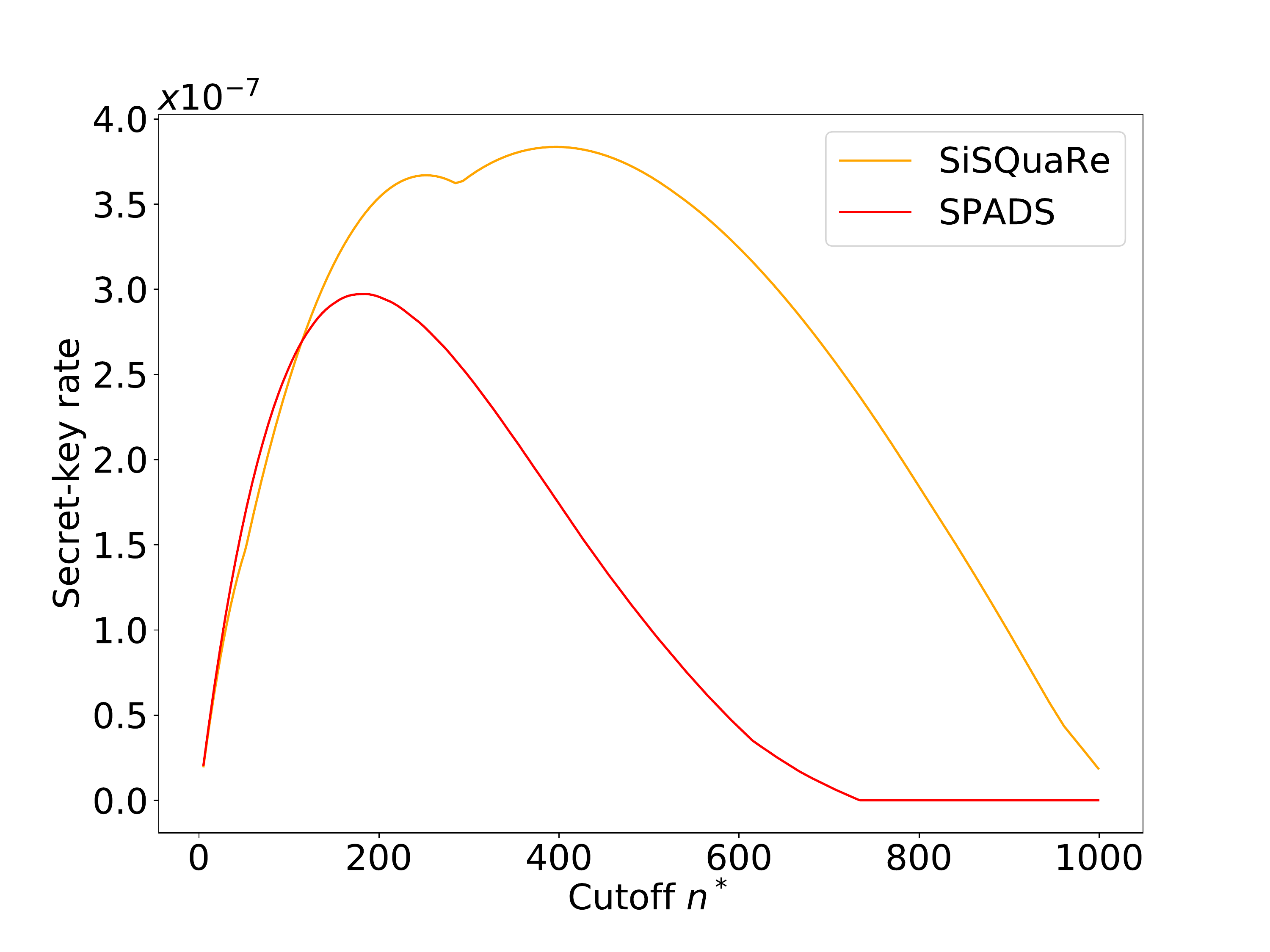}
\caption{Secret-key rate as a function of the cut-off for the SiSQuaRe and SPADS scheme for the total distance of $12.5 L_0$, where $L_0 = 0.542$ km. We see that the SPADS scheme requires a lower cut-off than the SiSQuaRe scheme because it involves more noisy operations. For each value of the cutoff $n^{*}$, we optimize the secret-key rate over the time window $t_{\textrm{w}}$ and for the SPADS scheme also over the $\theta$ angle. The kink for the SiSQuaRe scheme arises because of the optimization over the fully asymmetric one-way BB84 protocol and symmetric six-state protocol with advantage distillation, which itself involves optimization over two subprotocols.}
\label{fig:cut_off_depend_25_middle}
\end{figure}

\begin{figure}
\centering
\includegraphics[width=0.53\textwidth, trim= 0cm 0cm 0cm 0cm, clip]{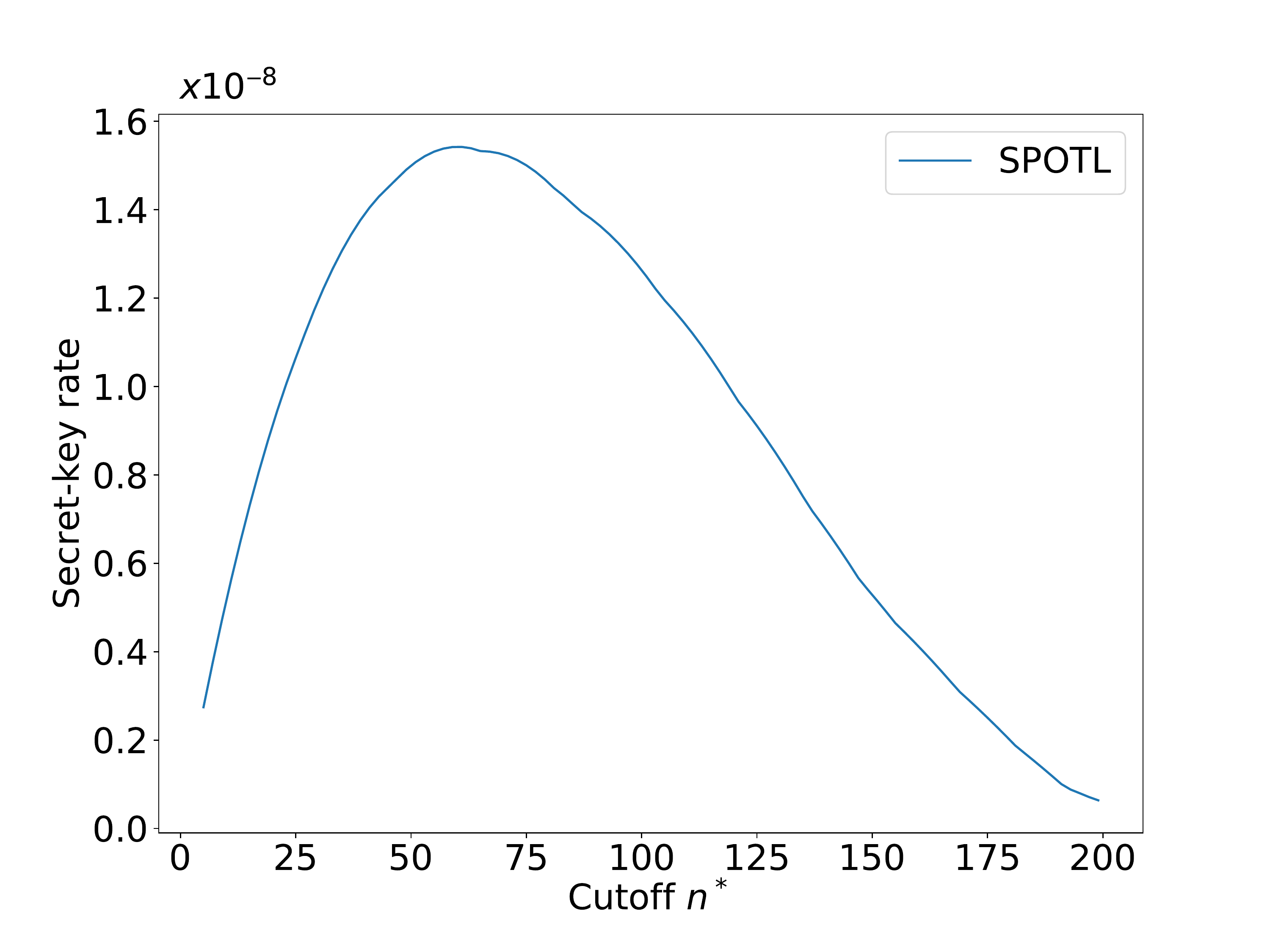}
\caption{Secret-key rate as a function of the cut-off for the SPOTL scheme for the total distance of $12.5 L_0$, where $L_0 = 0.542$ km. We see that due to the large number of noisy operations, this scheme requires a low cut-off in order to be able to generate key. For each value of the cutoff $n^{*}$ we optimize the secret-key rate over the time window $t_{\textrm{w}}$ and the $\theta$ angle.}
\label{fig:cut_off_depend_3_node}
\end{figure}

We also investigate the dependence of the rate on the cut-off. Both the SPADS and SPOTL scheme require a lower cut-off than the SiSQuaRe scheme; see Figs.~\ref{fig:cut_off_depend_25_middle} and~\ref{fig:cut_off_depend_3_node}. This is caused by the fact that each of them involves more noisy operations, and hence less noise tolerance is possible.

\subsection{Achieved secret-key rates of the quantum repeater proposals}
\label{sec:mainplots}
Now we are ready to present the main results, the secret-key rate for all the considered schemes as a function of the total distance when optimized over $\theta$, the cut-off $\nstar$, and the duration of the time window $t_{\textrm{w}}$. We compare the rates to the benchmarks from Sec.~\ref{sec:benchmarks}.

\begin{figure*}
\centering
\includegraphics[width=1.06\textwidth, trim= 0cm 0cm 0cm 0cm]{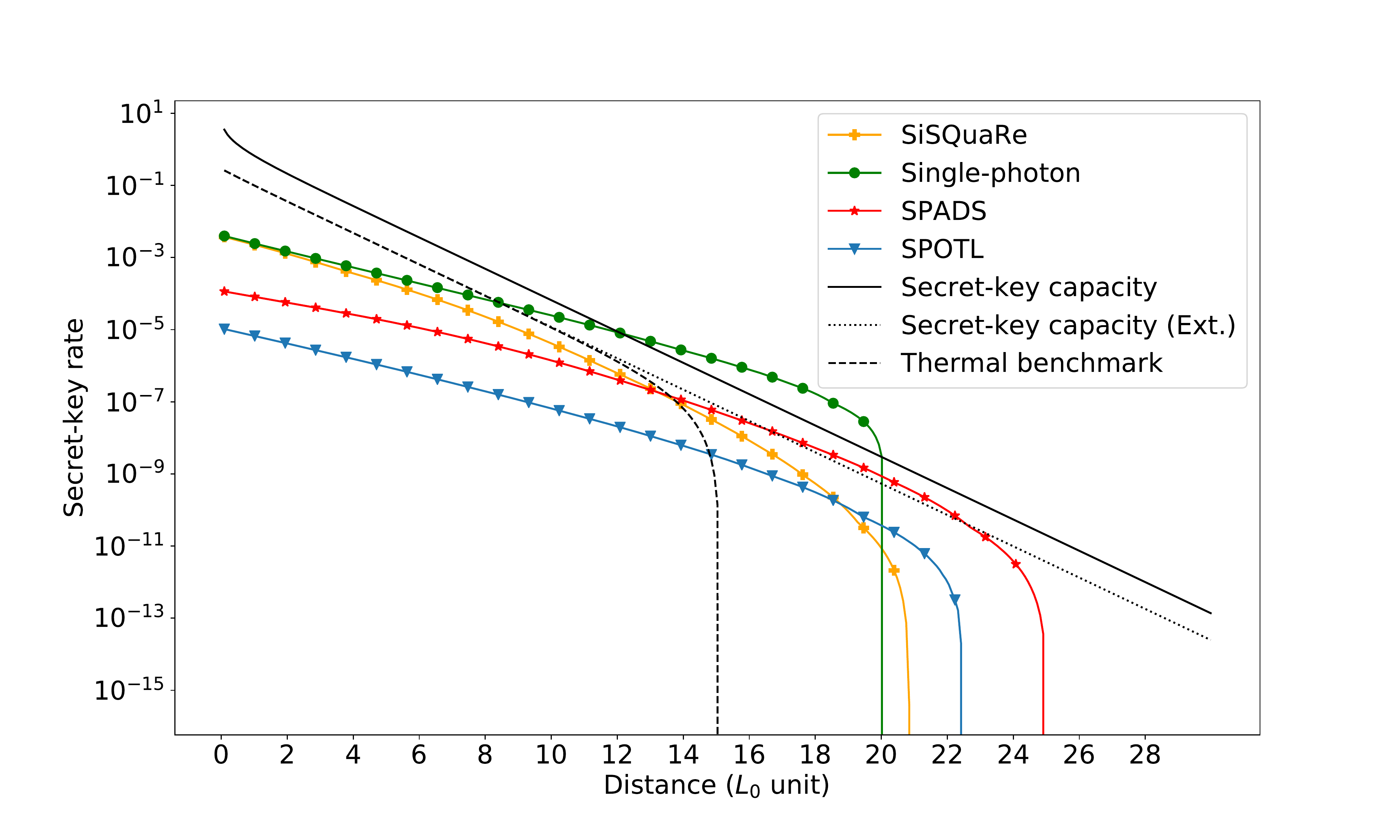}
\caption{Rate of all studied quantum repeater schemes as a function of the distance between Alice and Bob, expressed in the units of $L_0 = 0.542$ km. We also plot the different benchmarks from Sec.~\ref{sec:benchmarks}. We see that the single-photon scheme outperforms the secret-key capacity. For the achievable rates, the secret-key rate is optimized over the cutoff $n^{*}$, the angle $\theta$, and the time window $t_{\textrm{w}}$ independently for each distance.}
\label{fig:alloverdist}
\end{figure*}

In Fig.~\ref{fig:alloverdist}, we plot the rate of all four of the quantum repeater schemes as a function of the distance between Alice and Bob. We observe that already for realistic near-term parameters, the single-photon scheme can outperform the secret-key capacity of the pure-loss channel by a factor of 7 for a distance of $\approx 9.2$ km.

We have also investigated what improvements would need to be done in order for the SPADS and SPOTL schemes to also overcome the secret-key capacity. An example scenario in which the SPADS scheme outperforms this repeaterless bound includes better phase stabilization such that $\Delta \phi = \ang{5}$ and reduction of the decoherence effects in the carbon spin during subsequent entanglement generation attempts such that $a_0 = 1/8000$ and $b_0 = 1/20000$. Further improvement of these effective coherence times to $a_0 = 1/20000$ and $b_0 = 1/50000$ allows the SPOTL scheme to also overcome the secret-key capacity. We note that maintaining coherence of the carbon-spin memory qubit for such a large number of subsequent remote entanglement-generation attempts is expected to be possible using the method of decoherence-protected subspaces~\cite{reiserer2016robust,Kalb2018}. 

\begin{figure}
\centering
\includegraphics[width=0.53\textwidth, trim= 0cm 0cm 0cm 0cm, clip]{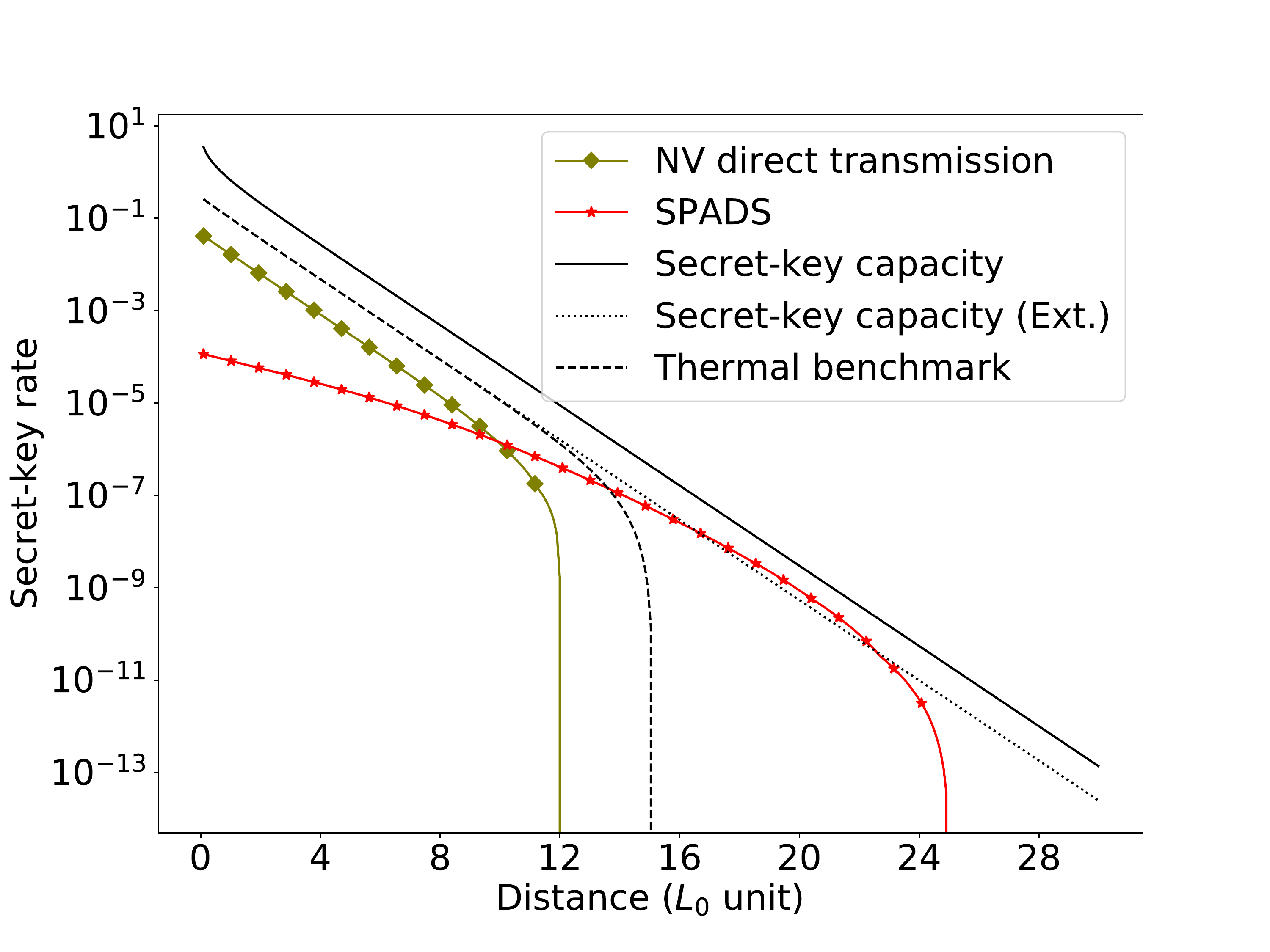}
\caption{Comparison of the SPADS scheme with the rate achievable using the direct transmission, with NV being the photon source. The secret-key rates for those schemes are plotted as a function of the distance between Alice and Bob, expressed in the units of $L_0 = 0.542$ km. We also plot the different benchmarks. We see that the SPADS scheme easily overcomes the direct transmission and the thermal benchmark (see Sec.~\ref{sec:benchmarks}). For the secret-key rate achievable with the SPADS scheme, we perform optimization over the cutoff $n^{*}$, the angle $\theta$, and the time window $t_{\textrm{w}}$ independently for each distance. Similarly, we also optimize the secret-key rate achievable with direct transmission over the time window $t_{\textrm{w}}$.}
\label{fig:Copmarison_2_5_direct_tr}
\end{figure}

As mentioned before, the SPADS scheme can be naturally compared against the benchmark of the direct transmission using NV as a source. The results are depicted in Fig.~\ref{fig:Copmarison_2_5_direct_tr}. We see that the SPADS scheme easily overcomes the NV-based direct transmission and the thermal benchmark for larger distances for which these benchmarks drop to zero.

\begin{figure*}
\centering
\includegraphics[width = 1.06\textwidth,trim= 0cm 0cm 0cm 0cm]{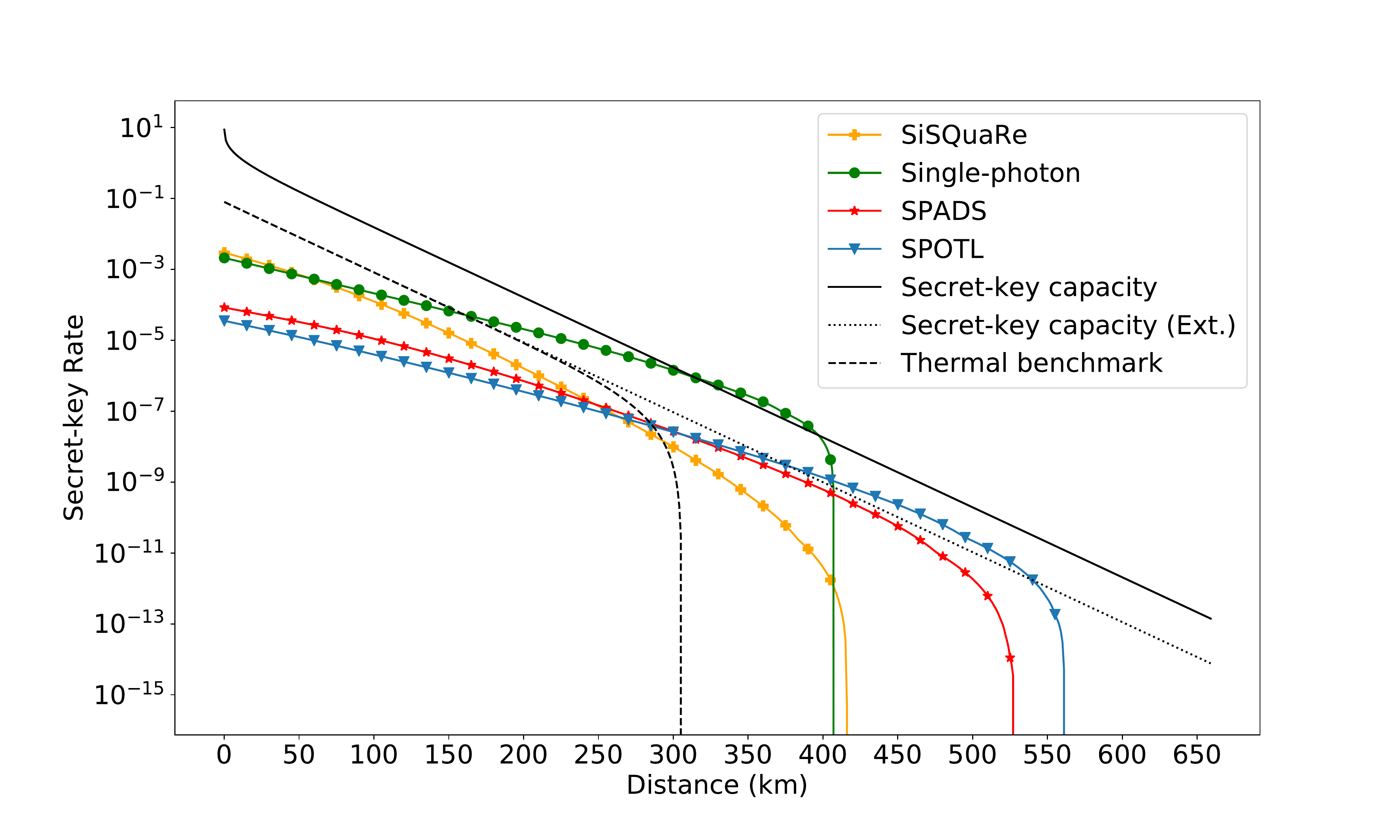}
\caption{Secret-key rate as a function of distance in units of km for transmission at telecom channel with $L_0 = 22$ km, along with the benchmarks from Sec.~\ref{sec:benchmarks}. We consider an improved measurement depolarizing parameter of $F_m =0.98$. The frequency conversion efficiency is assumed to be 0.3. We observe that the SPOTL scheme allows for the generation of secret-key over a distance of more than 550 km. For the achievable rates, the secret-key rate is optimized over the cutoff $n^{*}$, the angle $\theta$ and the time window $t_{\textrm{w}}$ independently for each distance.}
\label{fig:freq_conv}
\end{figure*}

In Fig.~\ref{fig:alloverdist}, we observe that for the SPOTL scheme, the total distance over which key can be generated is significantly smaller than for the SPADS scheme. This is despite the fact that the full distance is divided into four segments. The rather weak performance of this scheme is because it involves a larger number of noisy operations. As a result, the scheme can tolerate little noise from the single-photon subscheme, requiring the angle $\theta$ to be close to $\frac{\pi}{2}$, as can be seen in Fig.~\ref{fig:theta_3_node}. Hence, the probability of photon emission becomes greatly diminished and so the distance after which dark counts start becoming significant is much smaller than for the SPADS scheme. To overcome this problem, one would need to reduce the amount of noise in the system. One of the main sources of noise is the imperfect single-qubit measurement. Hence, we illustrate the achievable rates for the scenario with the boosted measurement depolarizing parameter $F_m=0.98$ in Fig.~\ref{fig:freq_conv}. Additionally, in this plot we also consider the application of probabilistic frequency conversion to the telecom wavelength at which $L_0=22$ km. Frequency conversion has already been achieved experimentally in the single-photon regime with success probability of $30\%$~\cite{zaske2012visible}. This is also the success probability that we consider here. The corresponding benchmarks have also been plotted for the new channel with $L_0=22$ km. We see in Fig.~\ref{fig:freq_conv} that with the improved measurement and using frequency conversion, the SPOTL scheme allows now to generate secret key over more than 550 km. We also see that under those conditions the single-photon scheme can also overcome the secret-key capacity of the telecom channel.

\subsection{Runtime of the experiment}
\label{sec:runtime}
While the theoretical capability of an experimental setup to surpass the secret-key capacity is a necessary requirement to claim a working quantum repeater, it does not necessarily mean that this can be experimentally verified in practice. Indeed, if a quantum repeater proposal only surpasses the secret-key capacity by a narrow margin at a large distance, the running time of an experiment could be too long for practical purposes. In this section, we will discuss an experiment which can validate a quantum repeater setup and calculate the running time of such an experiment, where we demonstrate that the single-photon scheme could be validated to be a quantum repeater within 12 hours.

A straightforward way of validating a quantum repeater would consist of first generating secret-key, calculating the achieved (finite-size) secret-key rate and then comparing the rate with the secret-key capacity. However, this requires a large number of raw bits to be generated, partially due to the loose bounds on finite-size secret-key generation. What we propose here is an experiment where the QBER and yield are separately estimated to be within a certain confidence interval. Then, if with the (worst-case) values of the yield and the QBER the corresponding asymptotic secret-key rate still confidently beats the benchmarks, one could claim that, in the asymptotic regime, the setup would qualify as a quantum repeater.

As we show in Appendix~\ref{section:runtime}, it is possible to run the single-photon scheme over a distance of $17L_0 \approx 9.2$ km for approximately 12 hours to find with high confidence ($\ge 1-1.5 \times {10}^{-4}$) that the scheme beats the capacity [see Eq.~\eqref{eq:benchmark1}] at that distance by a factor of at least 3.


\subsection{Discussion and future outlook}
\label{sec:discussion}
It is worth noting that our figure of merit --- the secret-key rate --- is weakly impacted by the latency of transmission, which grows linearly with distance for the SiSQuaRe, SPADS, and SPOTL schemes. Its only effect on the secret-key rate is the resulting decoherence time in the quantum memories while the memory nodes await the success or failure signals. This decoherence due to the waiting time is negligible in comparison to the noise due to interaction, arising from subsequent entanglement generation attempts. On the other hand, this latency would clearly be very visible in low throughput of these schemes. The single-photon scheme, on the other hand, has an advantage of the repetition rate being limited only by the local processing of the memory nodes, which would result in a higher throughput. We observe this fact in the modest expected duration of the experiment, even in the high-loss regime needed for overcoming the secret-key capacity. It is worth noting that while the single-photon scheme maintains constant latency for QKD, there exist schemes where such constant latency can be maintained also for remote entanglement generation; see e.g. Ref.~\cite{jones2016design}. It is hence clear that there are certain important properties of an efficient quantum repeater scheme that are not captured by the secret-key rate. However, achieving high throughputs for arbitrary distances would require almost all the components to be efficient in terms of rates and memories to be of high quality in terms of operational and long-storage fidelities. It is clear that demonstrating all these features together in a single experiment is still a future goal. The advantage of the secret-key rate is that overcoming the secret-key capacity would form a crucial step toward an implementation of an efficient and practical, long-distance quantum repeater architecture whose validity would carry an information-theoretic significance and will therefore be totally independent of any hardware-based reference scenario.

In our model, we have identified a significant amount of noise arising in the system. As a result, we find that it is not always beneficial to just divide the fixed distance into more elementary links. Hence, it is a natural question whether this noise could be eliminated e.g. using entanglement distillation. In fact, for the noise arising due to photon loss in the single-photon scheme, not only does there exist an efficient distillation procedure~\cite{campbell2008measurement, nickerson2014freely}, but it has also already been demonstrated in the NV platform~\cite{kalb2017entanglement}. Moreover, in the ideal case of noiseless operations and storage, a scheme based on generating two entangled states through the single-photon scheme and then distilling them as demonstrated in Ref.~\cite{kalb2017entanglement} should effectively also be able to overcome the secret-key capacity~\cite{vandam2017multiplexed} and provide a significant boost by completely removing the noise due to photon loss. Furthermore, an implementation of such a distillation-based remote entanglement generation scheme would alleviate the requirement of the optical phase stabilisation of the system. Therefore, this distillation-based scheme could be a natural fifth candidate for a proof of principle repeater. Nevertheless, we believe that the fidelities of quantum operations and the effective coherence times of the memories used in this paper might need to be improved before this distillation would prove useful.


\section{Conclusions}
\label{sec:conclusions}
We analyzed four experimentally relevant quantum repeater schemes on their ability to generate secret key. More specifically, the schemes were assessed by contrasting their achievable secret-key rate with the secret-key capacity of the channel corresponding to direct transmission. The secret-key rates have been estimated using near-term experimental parameters for the NV center platform. The majority of these parameters have already been demonstrated across multiple experiments. A remaining challenging element of our proposed schemes is the implementation of optical cavities. These cavities would enable the enhancement of both the photon emission probability into the zero-phonon line and the photon collection efficiency to the desired level.

With these near-term experimental parameters, our assessment shows the viability of one of the schemes, the single-photon scheme, for the first experimental demonstration of a quantum repeater. In fact, the single-photon scheme achieves a secret-key rate more than seven times greater than the secret-key capacity. We also estimated the duration of an experiment to conclude that a rate larger than the secret-key capacity is achievable. The duration of the experiment would be approximately 12 hours.

Finally, we show that a scheme based on concatenating the single-photon scheme twice (i.e., ~the SPOTL scheme), has the capability to generate secret-key at large distances. However, this requires converting the frequency of the emitted photons to the telecom wavelength and modestly improving the fidelity at which measurements can be performed.


\section*{Acknowledgements}
The authors would like to thank Koji Azuma, Tim Coopmans, Axel Dahlberg, Suzanne van Dam, Roeland ter Hoeven, Norbert Kalb, Victoria Lipinska, Marco Lucamarini, Gl\'aucia Murta, Matteo Pompili, and J\'er\'emy Ribeiro for helpful discussions and feedback. This work was supported
by the Dutch Technology Foundation (STW), the
Netherlands Organization for Scientific Research (NWO) through a VICI grant (RH), a VIDI grant (SW), the
European Research Council through a Starting Grant (RH and SW), the Ammodo KNAW award (RH) and the QIA project (funded by European Union's Horizon 2020, Grant Agreement No. 820445).


\bibliography{library}
\phantom{1mm}
\onecolumngrid
\phantom{1mm}

\appendix

\section{Losses and noise on the photonic qubits}
\label{appendix:lossesandnoise}
In this appendix, we describe how the losses and noise affect our photonic qubits. In particular, we first recall how the two types of encoding result in the losses acting as different quantum channels on the states. Then, we study the effects of a finite detector time window. More specifically, we first show that the arrival of a photon outside the time window is equivalent to all the other loss processes. Second, we calculate the probability of registering a dark count within the time window. We also show how to model the noise arising from those dark counts for the SiSQuaRe and SPADS schemes. Finally, we calculate the dephasing induced by the unknown phase shift for the single-photon scheme.


\subsection*{Effects of losses for the different encodings}
\label{appendix:lossestint}
 
The physical process of probabilistically losing photons corresponds to different quantum channels depending on the qubit encoding. In our repeater schemes we use two types of encoding: time-bin and presence or absence of a photon. For a time-bin encoded qubit in the ideal scenario of no loss we always expect to obtain a click in one of the detectors. Hence, loss of a photon resulting in a no-click event raises an erasure flag which carries the failure information. Therefore, it is clear that for this encoding the physical photon loss process corresponds to an erasure channel with the erasure probability given by one minus the corresponding transmissivity,
\begin{equation}
D(\rho) = \eta \rho + (1-\eta) \proj{\perp}\ .
\end{equation}
Here $\ket{\perp}$ is the loss flag, corresponding to the nondetection of a photon.
Since we are only interested in the quantum state of the system for the successful events when a detection event has occurred, we effectively post-select on the non-erasure events.
For presence or absence encoding, the situation is different since now there is no flag available that could explicitly tell us whether a photon got lost or not. In fact, for this encoding the photon loss results in an amplitude-damping channel applied to the photonic qubit. Here, the damping parameter equals one minus the transmissivity of the channel~\cite{chuang1997bosonic}.

\subsection*{Effects of the detector time window}
The detector only registers clicks that fall within a certain time window. It is \emph{a priori} not clear what kind of noisy or lossy channel should be used to model the loss of information due to nondetection of photons arriving outside of the time window. This is because in a typical loss process we have a probabilistic leakage of information to the environment. In the scenario considered here, the situation is slightly different as effectively no leakage occurs, but rather certain part of the incoming signal effectively gets discarded. Here we will show that despite this qualitative difference, within our model this process can effectively be modeled as any other loss process.

Now, let us provide a brief description of the physics of this process. First, the detection time window is chosen such that the probability of detecting a photon from the optical excitation pulse used to entangle the electron spin with the photonic qubit is negligible~\cite{hensen2015loophole}. For that reason, the detection time window is opened after a fixed offset $t^{\textrm{offset}}_{\textrm{w}}$ with respect to the beginning of the decay of the optical excited state of the electron spin. We note that for the considered enhancement of the ZPL-emission using the optical cavity, we predict the characteristic time of the NV emission $\tau$ to be approximately half of the corresponding value of $\tau$ if no cavity is used~\cite{hensen2015loophole,Fox2006,riedel2017deterministic}. Therefore, here we consider the scenario where the duration of the optical excitation pulse is made twice shorter with respect to the one used in Ref.~\cite{hensen2015loophole}. This will allow us to filter out the unwanted photons from the excitation pulse by setting $t^{\textrm{offset}}_{\textrm{w}}$ to half of the offset used in Ref.~\cite{hensen2015loophole}.

Second, we note that the detection time window cannot last too long; specifically, it needs to be chosen such that there is a good trade-off between detecting coherent and non-coherent (i.e.,~dark counts) photons. In this subsection, we will discuss the effects of photons arriving outside of this time window and the effects of registering dark counts within this time window. 


\subsubsection{Losses from the detector time window}
The NV center emits a photon through an exponential decay process with characteristic time $\tau$.
Therefore the probability of detecting a photon during a time window starting at $t^{\textrm{offset}}_{\textrm{w}}$ and lasting for $t_{\textrm{w}}$ is
\begin{equation}
p_{\textrm{in}}(t_{\textrm{w}}) =\frac{1}{\tau}\int_{t^{\textrm{offset}}_{\textrm{w}}}^{t^{\textrm{offset}}_{\textrm{w}}+t_{\textrm{w}}}dt\exp(-\frac{t}{\tau}) = \exp(-\frac{t^{\textrm{offset}}_{\textrm{w}}}{\tau})-\exp(-\frac{t^{\textrm{offset}}_{\textrm{w}} + t_{\textrm{w}}}{\tau})\ .
\end{equation}
Clearly the process of a photon arriving outside of the time window is qualitatively different from the loss process where the photons get lost to the environment. In the remainder of this section, we will now look at the difference between these two phenomena in more detail.

The emission process of the NV center is a coherent process over time. Consider a generic scenario in which we divide the emission time into two intervals, denoted by ``in'' and ``out'', respectively. Coherent emission then means that the state of the photon emitted by the electron spin in state $\ket{\uparrow}$ will be
\begin{equation}
\ket{\psi} = \sqrt{p_{\textrm{in}}}\ket{1}_{\textrm{in}}\ket{0}_{\textrm{out}} + \sqrt{1-p_{\textrm{in}}}\ket{0}_{\textrm{in}}\ket{1}_{\textrm{out}}\ .
\end{equation}
Now let us come back to our specific model, in which the ``in'' mode corresponds to the interval $\left[t^{\textrm{offset}}_{\textrm{w}},~t^{\textrm{offset}}_{\textrm{w}} + t_{\textrm{w}}\right]$ and the ``out'' mode to all the times $t \ge 0$ lying outside of this interval ($t=0$ is the earliest possible emission time). Here, the emission into the ``in'' mode occurs with probability $p_{\textrm{in}}(t_{\textrm{w}})$. Hence the spin-photon state resulting from the emission by the $\alpha \ket{\downarrow} + \beta \ket{\uparrow}$ spin state is
\begin{equation}
\ket{\psi} = \alpha \ket{\downarrow}\ket{0}_{\textrm{in}}\ket{0}_{\textrm{out}} + \beta \ket{\uparrow}\left(\sqrt{p_{\textrm{in}}(t_{\textrm{w}})}\ket{1}_{\textrm{in}}\ket{0}_{\textrm{out}} + \sqrt{1-p_{\textrm{in}}(t_{\textrm{w}})}\ket{0}_{\textrm{in}}\ket{1}_{\textrm{out}}\right)\ .
\end{equation}
If the presence or absence encoding is used, such a photonic qubit is then transmitted to the detector. Since only the spin and the ``in'' mode of the photon will be measured, we can now trace out the ``out'' mode:
\begin{equation}
\rho = \left(\abs{\alpha}^2 + \abs{\beta}^2 p_{\textrm{in}}(t_{\textrm{w}})\right)\proj{\phi} +  \abs{\beta}^2(1 - p_{\textrm{in}}(t_{\textrm{w}}))\proj{\uparrow}\otimes \proj{0}_{\textrm{in}}\ ,
\end{equation}
where
\begin{equation}
\ket{\phi} = \frac{1}{\sqrt{\abs{\alpha}^2 + \abs{\beta}^2 p_{\textrm{in}}(t_{\textrm{w}})}}\left( \alpha \ket{\downarrow}\ket{0}_{\textrm{in}} + \beta \sqrt{p_{\textrm{in}}(t_{\textrm{w}})} \ket{\uparrow}\ket{1}_{\textrm{in}} \right)\ .
\end{equation}
Note that this state can be obtained by passing the photonic qubit of the state
\begin{equation}
\ket{\psi} = \alpha \ket{\downarrow}\ket{0} + \beta \ket{\uparrow}\ket{1}\
\end{equation}
through the amplitude-damping channel with the damping parameter given by $1 - p_{\textrm{in}}(t_{\textrm{w}})$. Hence we can conclude that for the photon number encoding, the possibility of the photon arriving outside of the time window of the detector can be modeled in the same way as any other photon loss process, namely an amplitude-damping channel applied to that photonic qubit.

In the case of time-bin encoding we effectively have four photonic qubits, since now we have an ``in'' and ``out'' mode for both the early (denoted by ``e'') and the late (denoted by ``l'') time window. We assume here that the slots do not overlap. That is, a photon emitted in the ``out'' mode of the early time window is always distinct from any photon in the late time window. This can be achieved by making the time gap between the ``in'' modes of the early and late window long enough. In this case, the emission process results in a state
\begin{align}
\ket{\psi}	&= \alpha \ket{\downarrow}\left(\sqrt{p_{\textrm{in}}(t_{\textrm{w}})}\ket{1}_{e,\textrm{in}}\ket{0}_{e,\textrm{out}}\ket{0}_{l,\textrm{in}}\ket{0}_{l,\textrm{out}} + \sqrt{1-p_{\textrm{\textrm{in}}}(t_{\textrm{w}})}\ket{0}_{e,\textrm{in}}\ket{1}_{e,\textrm{out}}\ket{0}_{l,\textrm{in}}\ket{0}_{l,\textrm{out}}\right) \\
			&+ \beta \ket{\uparrow}\left(\sqrt{p_{\textrm{in}}(t_{\textrm{w}})}\ket{0}_{e,\textrm{in}}\ket{0}_{e,\textrm{out}}\ket{1}_{l,\textrm{in}}\ket{0}_{l,\textrm{out}} + \sqrt{1-p_{\textrm{in}}(t_{\textrm{w}})}\ket{0}_{e,\textrm{in}}\ket{0}_{e,\textrm{out}}\ket{0}_{l,\textrm{in}}\ket{1}_{l,\textrm{out}}\right)\ .
\end{align}
Again, tracing out the ``out'' modes results in a state
\begin{equation}
\rho = p_{\textrm{in}}(t_{\textrm{w}})\proj{\phi} + (1-p_{\textrm{in}}(t_{\textrm{w}})) \left(\abs{\alpha}^2\proj{\downarrow} + \abs{\beta}^2\proj{\uparrow}\right) \otimes \proj{00}_{e,l}\ ,
\end{equation}
where
\begin{equation}
\ket{\phi} = \alpha\ket{\downarrow}\ket{1}_e\ket{0}_l + \beta\ket{\uparrow}\ket{0}_e\ket{1}_l = \alpha\ket{\downarrow}\ket{e} + \beta\ket{\uparrow}\ket{l}\ .
\end{equation}
Here $\ket{00}_{e,l}$ corresponds to the loss flag from which we see that for the time-bin encoding the possible arrival of a photon outside of the time window results in an erasure channel with the erasure probability given by $\left[1-p_{\textrm{in}}(t_{\textrm{w}})\right]$. Hence this process can be also modeled as any other loss process for this encoding.

We have just shown that for both photon presence or absence and time-bin encodings the process of the photon arriving outside of the time window can be modeled by the source which prepares photons in a coherent superposition of the ``in'' and ``out'' modes and the detector tracing out (losing) the ``out'' modes. We have also shown that those two elements combined together result effectively in a loss process corresponding to the same channel as any other loss process for that encoding (amplitude damping for photon presence or absence and erasure channel for time-bin encoding).\\

However, between the source and the detector, there are other lossy or noisy components resulting in other quantum channels that need to be applied before the tracing out of the ``out'' mode at the detector. Now we show that for all loss and noise processes that occur in our model, the tracing out of the ``out'' mode can be mathematically commuted through all those additional noise and/or lossy processes. This means that the tracing out can be applied directly after the source, such that the above described reductions to amplitude-damping or erasure channel can be applied.

Consider the quantum channels acting on the photonic qubits of the form
\begin{equation}
\mathcal{N} = \sum_i p_i \mathcal{N}^i_{\textrm{in}} \otimes \mathcal{N}^i_{\textrm{out}}\ .
\label{eq:channel}
\end{equation}
Effectively, these are the channels that do not couple the ``in'' and ``out'' modes. Since in reality ``in'' and ``out'' modes correspond to different time modes, their coupling would require some kind of memory inside the channel. Hence we can think of the above defined channels as channels without memory.
Now it is clear that for a quantum state $\rho$ that among its registers includes both the ``in'' and the ``out'' mode, we have that
\begin{equation}
\tr_{\textrm{out}}[\mathcal{N}(\rho)] = \tr_{\textrm{out}}\left[\sum_i p_i \mathcal{N}^i_{\textrm{in}} \otimes \mathcal{N}^i_{\textrm{out}}(\rho)\right] = \sum_i p_i \mathcal{N}^i_{\textrm{in}}(\rho_{\textrm{in}})\ .
\end{equation}
Now, first tracing out the ``out'' modes and then applying the channel $\mathcal{N}$ (only the ``in'' part can be applied now) also results in $\sum_i p_i \mathcal{N}^i_{\textrm{in}}(\rho_{\textrm{in}})$ at the output. Hence, the tracing out of the ``out'' modes commutes with all the channels that are of the form \eqref{eq:channel}, which correspond to channels without memory. Clearly, the noise and/or loss processes that occur before the detection, such as photon loss or dephasing due to uncertainty in the optical phase of the photon, belong to this class of channels. In particular, this means that for photon presence or absence the amplitude damping due to photon loss in the channel and due to photon arrival outside of the time window can be both combined into one channel with the single damping parameter given by $1 - \eta p_{\textrm{in}}(t_{\textrm{w}})$ ($\eta$ denotes the transmissivity due to the loss process, e.g.,~the transmissivity of the fiber). The same applies to time-bin encoding where we now have a single erasure channel with erasure probability $1- \eta p_{\textrm{in}}(t_{\textrm{w}})$.

To conclude, the arrival of the photon outside of the time window can be modeled in the same way as any other loss process for both photon encodings used and therefore we can now redefine the detector efficiency $p'_{\textrm{det}} = p_{\textrm{det}} p_{\textrm{in}}(t_{\textrm{w}})$ and the total apparatus efficiency $p'_{\textrm{app}} = p_{\textrm{ce}}p_{\textrm{zpl}}p'_{\textrm{det}}$. We can then define $\eta_{\textrm{total}} =p'_{\textrm{app}}\eta_{f}$ as the total transmissivity, with probability $\eta_{\textrm{total}}$ a photon will be successfully transmitted from the sender to the receiver.


\subsubsection{Dark counts within the detector time window}

\par{Photon detectors are imperfect, and due to thermal excitations, they will register clicks that do not correspond to any incoming photons. These undesired clicks are called dark counts and can effectively be seen as a source of noise. The magnitude of this noise depends on the ratio between the probability of detecting the signal photon and measuring a dark count. Clearly, dark counts become a dominant source of noise when the probability of detecting the signal photon becomes comparable to the probability of a dark count click. The probability $p_{d}$ of getting at least one dark count within the time window $t_{\textrm{w}}$ of awaiting the signal photon is given by $p_{d}=1-\exp(-t_{\textrm{w}}d)$, where $d$ is the dark count rate of the detector~\cite{parameterregimes}.}

In the SiSQuaRe scheme, Alice and Bob perform measurements on time-bin-encoded photons. The same applies to Bob in the SPADS scheme. Since at least two detectors are required to perform this measurement, the presence of dark counts means that the outcome may lie outside of the qubit space. Moreover, this measurement needs to be trusted. In consequence, a squashing map needs to be used to process the multi-click events in a secure way. Here, as an approximation, we consider the squashing map for the polarization encoding~\cite{gittsovich2014squashing} in the same way as described in Ref.~\cite{parameterregimes}. Hence, this measurement can also be modeled as a perfect measurement preceded by a depolarizing channel with parameter $\alpha$, which depends on whether the BB84 or six-state protocol is used. The parameter $\alpha$ is given by~\cite{parameterregimes}
\begin{align}
\alpha_{A/B,\textrm{ BB84}} = \frac{p'_{\textrm{app}} \eta_{B}(1-p_d)}{1-(1-p'_{\textrm{app}} \eta_{A/B})(1-p_d)^2}\ , \\
\alpha_{A/B, \textrm{ six-state}} = \frac{p'_{\textrm{app}} \eta_{A/B}(1-p_d)^5}{1-(1-p'_{\textrm{app}} \eta_{A/B})(1-p_d)^6}\ .
\end{align}
Here $\eta_{A/B}$ denotes the transmissivity of the fiber between the memory repeater node and Alice's (Bob's) detector setup. Finally we note that dark counts increase the probability of registering a successful measurement event. For the optical measurement schemes utilizing the squashing map, the probability of registering a click in at least one detector is given by~\cite{parameterregimes}
\begin{align}
p_{A/B,~BB84} = 1 - (1-p'_{\textrm{app}}\eta_{A/B})(1-p_d)^2\ , \label{eq:ps1} \\
p_{A/B,~\textrm{six-state}} = 1 - (1-p'_{\textrm{app}}\eta_{A/B})(1-p_d)^6\ .
\label{eq:ps2}
\end{align}
The effect of dark counts in the single-photon scheme, which carries over to the SPOTL scheme, is analyzed in Appendix~\ref{sec:SingleClick}.

\subsection*{Noise due to optical phase uncertainty}
Another important noise process affecting photonic qubits is related to the fact that for the photon presence or absence encoding the spin-photon entangled state will also depend on the optical phase of the apparatus used. Specifically, it will depend on the phase of the lasers used to generate the spin photon entanglement as well as the optical phase acquired by the photons during the transmission of the photonic qubit. Knowledge about this phase is crucial for being able to generate entanglement through the single-photon scheme. In any realistic setup, however, there would be a certain degree of the lack of knowledge about this phase acquired by the photons. Since in the end what matters is the knowledge about the relative phase between the two photons, we can model this source of noise as the lack of knowledge of the phase on only one of the incoming photonic qubits. This noise process can be effectively modeled as dephasing. In this section, we will show that the phase uncertainty induces dephasing with a parameter $\lambda$ equal to
\begin{align}
\lambda &= \frac{I_1\left(\frac{1}{(\Delta \phi)^2}\right)}{2 I_0\left(\frac{1}{(\Delta \phi)^2}\right)} + \frac{1}{2}\ ,
\end{align}
where $\Delta \phi$ is the uncertainty in the phase and $I_{0}$ ($I_{1}$) is the Bessel function of order $0$ ($1$).
Let us assume that for Alice, the local phase of the photonic qubit has a Gaussian-like distribution on a circle, with standard deviation $\Delta \phi$ as observed in Ref.~\cite{humphreys2017deterministic}. This motivates us to model the distribution as a von Mises distribution~\cite{jammalamadaka2001topics}. The von Mises distribution reads
\begin{equation}
f(\phi) = \frac{e^{\kappa \cos(\phi -\mu)}}{2 \pi I_0(\kappa)}\ .
\end{equation}
Here $\mu$ is the measure of location, i.e., it corresponds to the center of the distribution, $\kappa$ is a measure of concentration and can be effectively seen as the inverse of the variance, and $I_0$ is the modified Bessel function of the first kind of order 0. One can then show~\cite{jammalamadaka2001topics} that
\begin{equation}
\int_{-\pi}^{\pi} d\phi f(\phi) e^{ \pm i \phi} = \frac{I_1(\kappa)}{I_0(\kappa)} e^{\pm i \mu}\ .
\end{equation}
Since we are only interested in the noise arising from the lack of knowledge about the phase rather than the actual value of this phase, without loss of generality we can assume $\mu = 0$. Moreover, the experimental parameter that we use here is effectively the standard deviation of the distribution $\Delta \phi$ and therefore we can write $\kappa = \frac{1}{(\Delta \phi)^2}$. 

Hence, let us write the spin-photon entangled state that depends on the optical phase $\phi$:
\begin{equation}
\ket{\psi^\pm (\phi)} = \sin(\theta)\ket{\downarrow 0} \pm e^{i\phi} \cos(\theta) \ket{\uparrow 1}\ .
\end{equation}

Now, the lack of knowledge about this phase leads to a mixed state:

\begin{equation}
\begin{aligned}
\int_{-\pi}^{\pi} f(\phi) \proj{\psi^\pm (\phi)} d \phi &= \sin^2(\theta)\proj{\downarrow 0} + \cos^2(\theta) \proj{\uparrow 1} \\
&\pm \sin(\theta)\cos(\theta) \int_{-\pi}^{\pi}f(\phi) (e^{i\phi} \ket{\uparrow 1}\bra{\downarrow 0} + e^{-i\phi} \ket{\downarrow 0}\bra{\uparrow 1}) d\phi\ .
\end{aligned}
\end{equation}

Let us now try to map this state onto a dephased state:
\begin{equation}
\begin{aligned}
\lambda \proj{\psi^\pm(0)} + (1-\lambda) \proj{\psi^\mp(0)} &= \sin^2(\theta)\proj{\downarrow 0} + \cos^2(\theta) \proj{\uparrow 1} \\
&\pm \sin(\theta)\cos(\theta) (2 \lambda -1) (\ket{\uparrow 1}\bra{\downarrow 0} + \ket{\downarrow 0}\bra{\uparrow 1})\ ,
\label{eq:}
\end{aligned}
\end{equation}
Hence, we observe that
\begin{align}
2\lambda - 1 &= \frac{I_1\left(\frac{1}{(\Delta \phi)^2}\right)}{I_0\left(\frac{1}{(\Delta \phi)^2}\right)}\ .\\
\rightarrow \lambda &= \frac{I_1\left(\frac{1}{(\Delta \phi)^2}\right)}{2 I_0\left(\frac{1}{(\Delta \phi)^2}\right)} + \frac{1}{2}\ .
\end{align}


\section{Noisy processes in NV-based quantum memories}
\label{sec:noisemem}

\par{In our setups we use $^{13}$C nuclear spins in diamond as long-lived memory qubits next to a Nitrogen Vacancy (NV) electron spin taking the role of a communication qubit. In this appendix, we will detail our model of the noisy processes in the NV.

The electron spin can be manipulated via microwave pulses and an optical pulse is used to create and send a photon entangled with it. This operation is noisy and can be modeled as having a dephasing noise of parameter $F_{\textrm{prep}}$. This means that, if the desired generated target state between the photon and the electron spin was $\ket{\psi^{+}}$, we actually have a mixture $F_{\textrm{prep}}\proj{\psi^{+}} +(1-F_{\textrm{prep}})(\id \otimes Z)\proj{\psi^{+}}(\id \otimes Z)$}.
\par{Information can be stored via a swapping of the electron spin state to the long living nuclear $^{13}$C spin. Through this swap operation we also free the communication qubit to be used for consecutive remote entanglement generation attempts. Because of the interaction with its environment, a quantum state stored in a $^{13}$C spin quantum memory undergoes an evolution that we model with a dephasing and a depolarizing channel with noise parameters $\lambda_{1}=(1+e^{-an})/2$ and $\lambda_{2}=e^{-bn}$, respectively. The form of the parameters $a$ and $b$ in general depends on the scheme. For the SiSQuaRe, SPADS and SPOTL schemes, there are two distinct effects that cause this decoherence: one induced by the time it takes to generate entanglement between the middle node and Bob, and one induced by the always-on hyperfine coupling between the electron spin and the carbon spin inside the middle NV node. This coupling becomes an additional source of decoherence for the carbon spin during probabilistic attempts to generate remote entanglement using the electron spin~\cite{reiserer2016robust,Kalb2018}. We model the decoherence effect on the qubit stored in the carbon spin of the middle node by a dephasing channel with parameter $\lambda_1$,
\begin{equation}
\mathcal{D}^{\lambda_{1}}_{\textrm{dephase}}(\rho) = \lambda_{1}\rho + (1-\lambda_{1})Z\rho Z\ ,
\end{equation}
and depolarizing channel with parameter $\lambda_2$,
\begin{equation}
\mathcal{D}^{\lambda_{2}}_{\textrm{depol}}(\rho) = \lambda_{2}\rho + (1-\lambda_{2})\frac{\mathbb{I}}{D}\ ,
\end{equation}
where $\lambda_{1}$ and $\lambda_{2}$ quantify the noise. The parameters depend as follows on the number of attempts $n$,
\begin{align}
\lambda_1 &= F_{T_{2}}=\frac{1+e^{-an}}{2}\ , \\
\lambda_2 &= F_{T_{1}}=e^{-bn}\ ,
\end{align}
where $a$ and $b$ are given by
\begin{equation}
a=a_{0} + a_{1}\left(L_{s}\frac{n_{ri}}{c} + t_{\textrm{prep}}\right)\ , \\
b=b_{0} + b_{1}\left(L_{s}\frac{n_{ri}}{c} + t_{\textrm{prep}}\right)\ .
\end{equation}
Here $n_{ri}$ is the refractive index of the fiber, $c$ is the speed of light in vacuum, $t_{\textrm{prep}}$ is the time it takes to prepare for the emission of an entangled photon, and $L_s$ is the distance the signal needs to travel before the repeater receives the information about failure or success of the attempt. Let $L_B$ denote the distance between the memory repeater node and Bob. Then for the SiSQuaRe and SPADS schemes $L_s = 2L_B$, since in each attempt first the quantum signal needs to travel to Bob, who then sends back to the middle node the classical information about success or failure. For the SPOTL scheme $L_s = L_B$, since in this case both the quantum and the classical signals need to travel only half of the distance between the middle node and Bob since the signals are exchanged with the heralding station, which is located halfway between the middle memory node and Bob. The parameters $a_{0}$ and $b_{0}$ quantify the noise due to a single attempt at generating an entangled spin-photon, induced by stochastic electron spin reset operations, quasi static noise, and microwave control infidelities. The parameters $a_{1}$ and $b_{1}$ quantify the noise during storage per second.

\par{Gates and measurements in the quantum memory are also imperfect. We model those imperfections via two depolarizing channels. The first one acts on a single qubit with depolarizing parameter $\lambda_2 = F_{m}$ corresponding to the measurement of the electron spin. The second one acts on two qubits with depolarizing parameter $\lambda_2 = F_{g}$ corresponding to applying a two-qubit gate to both the electron spin and the $^{13}$C spin. This means that every time a measurement is done on a $e^{-}$ qubit of a quantum state $\rho$, it is actually done on $\mathcal{D}^{F_{m}}_{\textrm{depol}}(\rho)$. Also a swapping operation between the $e^{-}$ spin and the nuclear spin (done experimentally via two two-qubit gates; see main text) leads to an error modeled by a depolarizing channel of parameter $F_{\textrm{swap}}=F_{g}^{2}$. Following the same logic, a Bell state measurement will cause the state to undergo an evolution given by a depolarizing channel. Specifically, following the decomposition of the Bell measurement into elementary gates for the NV implementation as described in Sec.~\ref{sec:modeling}, this evolution will consist of a depolarizing channel with parameter $F_{g}^{2}$ acting on both of the measured qubits and the depolarizing channel with parameter $F_{m}^{2}$ acting only on the electron spin qubit.}

\section{Expectation of the number of channel uses with a cut-off}
\label{sec:expnumbsum}
In this appendix, we derive an analytical formula for the expectation value of the number of channel uses between Alice and Bob needed to generate one bit of raw key for the SiSQuaRe, SPADS, and SPOTL schemes,
\begin{align}
\mathbb{E}[N] =  \frac{1}{p_A\cdot \left(1-(1-p_{B})^{n^{*}}\right)}+\frac{1}{p_{B}}\ .
\label{eq:channeluses}
\end{align}
For these three schemes, we implement a cut-off which is used to prevent decoherence. Each time the number of channel uses between the repeater node and Bob reaches the cut-off $n^{*}$, the entire protocol restarts from the beginning. Here, we take a conservative view and define the number of channel uses $N$ between Alice and Bob as the sum $N_A + N_B$, where $N_A$ $(N_B)$ corresponds to the number of channel uses between Alice (Bob) and the middle node. 
From the linearity of the expectation value, we have that
\begin{equation}
\mathbb{E}[N_{A}+N_{B}]= \mathbb{E}[N_{A}] + \mathbb{E}[N_{B}]\ .
\end{equation}
\par{We denote by $p_{A}$ and $p_{B}$ the probability of a successful attempt on Alice's and Bob's side respectively. Bob's number of channel uses follows a geometric distribution with parameter $p=p_{B}$, so that $\mathbb{E}[N_{B}] = \frac{1}{p_{B}}$. Without the cut-off, Alice's number of channel uses would follow a geometric distribution with parameter $p = p_{A}$. However, the cut-off parameter adds additional channel uses on Alice side. Since the probability that Bob succeeds within $n^{*}$ trials is $p_{\textrm{succ}} = 1-(1-p_{B})^{n^{*}}$, we in fact have that Alice's number of channel uses follows a geometric distribution with parameter $p'_{A}=p_{A}p_{\textrm{succ}}$. Hence, it is straightforward to see that
\begin{align}
\mathbb{E}[N_{A} + N_{B}] &= \frac{1}{p'_{A}} + \frac{1}{p_{B}}\\
&= \frac{1}{p_A\left(1-(1-p_{B})^{n^{*}}\right)}+\frac{1}{p_{B}}\ .
\end{align}}

\section{SiSQuaRe scheme analysis}
\label{sec:SiSQuaReChanges}
The analysis of the SiSQuare scheme has been performed in~\cite{parameterregimes}. In this work we use the estimates of the yield and QBER as derived in~\cite{parameterregimes} with the following modifications:
\begin{enumerate}[label={(\arabic*)}]
\item For the calculation of the yield, we now adopt a conservative perspective and calculate the number of channel uses as $\mathbb{E}[N_{A}+N_{B}]$, as derived in Appendix~\ref{sec:expnumbsum}, rather than $\mathbb{E}[\textrm{max}(N_{A},N_{B})]$. Note that $\mathbb{E}[\textrm{max}(N_{A},N_{B})] \leq \mathbb{E}[N_{A}+N_{B}] \leq 2 \mathbb{E}[\textrm{max}(N_{A},N_{B})]$.
\item The total depolarizing parameter for gates and measurements $F_{\textrm{gm}}$ defined in Ref.~\cite{parameterregimes} is now decomposed into individual operations as described in Appendix~\ref{sec:noisemem}. That is, in this work depolarization due to imperfect operations on the memories is expressed in terms of depolarizing parameter due to imperfect measurement, $F_m$, and imperfect two-qubit gate, $F_g$. Since in the analysis of the SiSQuaRe scheme we only deal with Bell diagonal states, the overall noise due to imperfect swap gate and the Bell measurement leads to $F_{\textrm{gm}} = F_g^4 F_m^2$.
\item In Ref.~\cite{parameterregimes}, we have assumed the duration of the detection time window to be fixed to 30 ns and assumed that all the emitted photons will fall into that time window. Here, similarly as for other schemes, we perform a more refined analysis in which we include the trade-off between the duration of the time window and the dark count probability as described in Appendix~\ref{appendix:lossesandnoise}. 
\end{enumerate}

\section{Single-photon scheme analysis}
\label{sec:SingleClick}
In this appendix, we provide a detailed analysis of the single-photon scheme between two remote NV-center nodes. This section is structured as follows. First, we describe the creation of the spin-photon entangled state followed by the action of the lossy channel on the photonic part of this state, including the noise due to the uncertainty in the phase of the state induced by the fiber. Second, we apply the optical Bell measurement. Then we evaluate the effect of dark counts, which introduce additional errors to the generated state. Finally, we calculate the yield of this scheme and extract the QBER from the resulting state.


\subsection*{Spin-photon entanglement and action of a lossy fiber on the photonic qubit}
First, both Alice and Bob generate spin-photon entangled states, parameterized by $\theta$. As we will later see, this parameter allows for trading off the quality of the final entangled state of the two spins with the yield of the generation process. The ideal spin-photon state would then be described as
\begin{equation}
\ket{\psi^+} = \sin\left(\theta\right)\ket{\downarrow}\ket{0} + \cos\left(\theta\right)\ket{\uparrow}\ket{1}\ .
\label{eq:spinphoton}
\end{equation}
The preparation of the spin-photon entangled state is not ideal. That is, the spin-photon entangled state is not actually as described above, but rather of the form (see Appendix~\ref{sec:noisemem})
\begin{equation}
\rho = F_{\textrm{prep}} \proj{\psi^+} + (1-F_{\textrm{prep}}) (\id \otimes Z) \proj{\psi^+} (\id \otimes Z) = F_{\textrm{prep}} \proj{\psi^+} + (1-F_{\textrm{prep}}) \proj{\psi^-}\ .
\label{eq:prepstate}
\end{equation}
Here,
\begin{equation}
\ket{\psi^-} =  \sin\left(\theta\right)\ket{\downarrow}\ket{0} - \cos\left(\theta\right)\ket{\uparrow}\ket{1}\ .
\end{equation}

For the next step, we need to consider two additional noise processes that affect the photonic qubits before the optical Bell measurement is performed. The first one is the loss of the photonic qubit. This can happen at the emission, while filtering the photons that are not of the required ZPL frequency, in the lossy fiber, in the imperfect detectors, or due to the arrival outside of the time window in which detectors expect a click. All these losses can be combined into a single loss parameter
\begin{equation}
 \eta = \eta_{\textrm{total}} = p_{\textrm{ce}}p_{\textrm{zpl}}\sqrt{\eta_{f}}p'_{\textrm{det}}\ ,
\end{equation} 
with $\eta_{f}=\exp(-\frac{L}{L_{0}})$, where $L$ is the distance between the two remote NV-center nodes in the scheme (see Fig. \ref{fig:photonlosses} and Appendix A). Hence, a photon is successfully transmitted through the fiber and detected in the middle heralding station with probability $\eta$.}
Now we note that the action of the pure-loss channel on the qubit encoded in the presence or absence of a photon corresponds to the action of the amplitude-damping channel with the damping parameter $1-\eta$~\cite{chuang1997bosonic}.

The second process that effectively happens at the same time as loss is the dephasing noise arising from the optical instability of the apparatus as described in Appendix~\ref{appendix:lossesandnoise}. We note that the amplitude-damping and dephasing channel commute; hence, it does not matter in which order we apply the two noise processes corresponding to the loss of the photonic qubit and unknown drifts of the phase of the photonic qubit in our model. Here, we first apply the dephasing due to the lack of knowledge of the phase on Alice's photon and then amplitude damping on both photons due to all the loss processes.

Following the model in Appendix~\ref{appendix:lossesandnoise}, the lack of knowledge about the optical phase will effectively transform Alice's state to
\begin{equation}
\rho_A = \left(F_{\textrm{prep}}\lambda + (1-F_{\textrm{prep}})(1-\lambda)\right) \proj{\psi^+} + \left((1-F_{\textrm{prep}})\lambda + F_{\textrm{prep}}(1-\lambda)\right) \proj{\psi^-}\ .
\end{equation}
where
\begin{equation}
\lambda = \frac{I_1\left(\frac{1}{(\Delta \phi)^2}\right)}{2 I_0\left(\frac{1}{(\Delta \phi)^2}\right)} + \frac{1}{2}.
\end{equation}

Now we can apply all the transmission losses modeled as the amplitude-damping channel.
The action of this channel on the photonic part of the state $\rho$ results in the state that we can describe as follows. First, let us introduce two new states:
\begin{equation}
\ket{\psi_{\eta}^{\pm}} = \frac{1}{\sqrt{\sin^2(\theta) + \eta \cos^2(\theta)}}(\sin\left(\theta\right)\ket{\downarrow}\ket{0} \pm \sqrt{\eta} \cos\left(\theta\right)\ket{\uparrow}\ket{1})\ .
\end{equation}
Then, after the losses and before the Bell measurement, the state of Alice can be written as
\begin{align}
\begin{split}
\rho'_A &= \left(\sin^2(\theta) + \eta \cos^2(\theta)\right)\left(\left(F_{\textrm{prep}}\lambda + (1-F_{\textrm{prep}})(1-\lambda)\right) \proj{\psi_{\eta}^{+}} + \left((1-F_{\textrm{prep}})\lambda + F_{\textrm{prep}}(1-\lambda)\right) \proj{\psi_{\eta}^{-}} \right) \\
&+(1-\eta) \cos^2(\theta) \proj{\uparrow}\proj{0} \ ,
\end{split}
\end{align}
and for Bob
\begin{equation}
\rho'_B = \left(\sin^2(\theta) + \eta \cos^2(\theta)\right)\left(F_{\textrm{prep}} \proj{\psi_{\eta}^{+}} + (1-F_{\textrm{prep}}) \proj{\psi_{\eta}^{-}} \right) +(1-\eta) \cos^2(\theta) \proj{\uparrow}\proj{0}\ .
\end{equation}


\subsection*{States after the Bell measurement}

Now we need to perform a Bell measurement on the photonic qubits within the states $\rho'_A$ and $\rho'_B$. Here we consider the scenario with non-photon-number-resolving detectors. Assuming for the moment the scenario without dark counts, we have at most two photons in the system. Hence we can consider three possible outcomes of our optical measurement: left detector clicked, right detector clicked, and none of the detectors clicked.  The measurement operators can be easily derived by noting that in our scenario without dark counts, each of the detectors can be triggered either by one or two photons and no cross clicks between detectors are possible due to the photon-bunching effect. Then we can apply the reverse of the beam splitter mode transformations to the projectors on the events with one or two photons in each of the detectors to obtain these projectors in terms of the input modes. Finally, we truncate the resulting projectors to the qubit space since in our scenario it is not possible for more than one photon to be present in each of the input modes of the beam splitter. In this way we obtain the following measurement operators:
\begin{align}
\begin{split}
A_0 &= \proj{\Psi^+} + \frac{1}{\sqrt{2}}\proj{11}\ , \\
A_1 &= \proj{\Psi^-} + \frac{1}{\sqrt{2}}\proj{11}\ , \\
A_2 &= \proj{00}\ .
\end{split}
\label{eq:opticalBSM_POVM}
\end{align}
These outcomes occur with the following probabilities:
\begin{align}
p_0 &= p_1 = \eta \cos^2(\theta) \left(1-\frac{\eta}{2}\cos^2(\theta)\right)\ , \\
p_2 &= (1-\eta \cos^2(\theta))^2\ .
\end{align}
The post-measurement state of the two spins for the outcome $A_0$ is
\begin{equation}
\rho_0 = \frac{2 \sin^2(\theta)}{2-\eta \cos^2(\theta)}\left( a \proj{\Psi^+} + b \proj{\Psi^-}\right) + \frac{\cos^2(\theta) (2- \eta)}{2-\eta \cos^2(\theta)}\proj{\uparrow \uparrow}\ .
\end{equation}
Here,
\begin{align}
\ket{\Psi^{\pm}} &= \frac{1}{\sqrt{2}}\left(\ket{\downarrow \uparrow} \pm \ket{\uparrow \downarrow}\right)\ ,\\
a &= \lambda (F_{\textrm{prep}}^2 + (1-F_{\textrm{prep}})^2) + 2F_{\textrm{prep}} (1-F_{\textrm{prep}})(1- \lambda)\ , \\
b &= (1 - \lambda) (F_{\textrm{prep}}^2 + (1-F_{\textrm{prep}})^2) + 2F_{\textrm{prep}} (1-F_{\textrm{prep}})\lambda\ .
\end{align}
For the outcome $A_1$, the postmeasurement state of the spins is the same up to a local $Z$ gate which Bob can apply following the trigger of the $A_1$ outcome.
The postmeasurement state of the spins for the outcome $A_2$, that is, when none of the detector clicked, is
\begin{equation}
\rho_2 = \frac{1}{\left[1 - \eta \cos^2(\theta)\right]^2} \left(\sin^4(\theta) \proj{\downarrow \downarrow} + (1 - \eta) \cos^2(\theta) \sin^2(\theta) \left(\proj{\downarrow \uparrow} + \proj{\uparrow \downarrow}\right) + (1- \eta)^2 \cos^4(\theta) \proj{\uparrow \uparrow}\right)\ .
\end{equation}
This is a separable state and so events corresponding to outcome $A_2$ (that is, no click in any of the detectors) will be discarded as failure. However, dark counts on our detectors can make us draw wrong conclusions about which of the three outcomes we actually obtained.

The effect of dark counts can be seen as follows
\begin{enumerate}[label={(\arabic*)}]
\item We measured $A_2$ (no actual detection) but one of the detectors had a dark count. This event will happen with probability $2 p_2 p_d (1-p_d)$ and will make us accept the state $\rho_2$. Note that this is a classical state so application of the $Z$ correction by Bob does not affect this state at all.
\item We measured $A_1$ or $A_2$ but we also got a dark count in the other detector. This event will happen with probability $(p_0 + p_1) p_d$. This will effectively lead us to rejection of the desired state $\rho_0$. Hence, effectively $\rho_0$ will only be accepted if we measured $A_1$ or $A_2$ but the other detector did not have a dark count, which will happen with probability $(p_0 + p_1) (1 - p_d)$.
\end{enumerate}


\subsection*{The yield and QBER}
Taking dark counts into account, we see that the yield of the single-photon scheme, which is just the probability of registering a click in only one of the detectors, will be
\begin{equation}
Y = (p_0 + p_1)(1 - p_d) + 2 p_2 p_d (1-p_d) =  2(1-p_d) \left[\eta \cos^2(\theta) \left(1-\frac{\eta}{2}\cos^2(\theta)\right) + (1-\eta \cos^2(\theta))^2 p_d\right]\ .
\label{eq:yieldSingleClick}
\end{equation}
The effective accepted state after a click in one of the detectors will then be
\begin{equation}
\rho_{\textrm{out}} = \frac{1}{Y}\left((p_0 + p_1)(1 - p_d) \rho_0 + 2 p_2p_d (1-p_d) \rho_2 \right)\ .
\label{eq:rhoSingleClick}
\end{equation}

Note that both Alice and Bob perform a measurement on their electron spins immediately after each of the spin-photon entanglement generation events. This measurement causes an error modeled as a depolarizing channel of parameter $F_{m}$ on each qubit, which means that after a successful run of the single-photon protocol, the effective state shared by Alice and Bob including the noise of their measurements will be given by
\begin{align}
\rho_{AB} = F_m^2 \rho_{\textrm{out}} + (1-F_m)F_m \left[\frac{\mathbb{I}_{2,A}}{2} \otimes \tr_A[\rho_{\textrm{out}}] + \tr_B[\rho_{\textrm{out}}]\otimes\frac{\mathbb{I}_{2,B}}{2}\right]
+ (1-F_m)^2~\frac{\mathbb{I}_{4,AB}}{4}\ .
\end{align}
One can then extract the QBER for this state in all the three bases using the appropriate correlated/anti-correlated projectors such that:
\begin{align}
e_{z} &= \Tr[(\outerproduct{00} +\outerproduct{11})\rho_{AB}]\ , \\
e_{xy} &= \Tr((\outerproduct{+-} +\outerproduct{-+})\rho_{AB}) = \Tr((\outerproduct{0_y 1_y} +\outerproduct{1_y 0_y})\rho_{AB})\ .
\end{align}
Here $\ket{+}$ and $\ket{-}$ denote the two eigenstates of $X$ and $\ket{0_y}$ and $\ket{1_y}$ denote the two eigenstates of $Y$. We note that for our model of the single-photon scheme the QBER in $X$ and $Y$ bases are the same and therefore we denote both by a single symbol $e_{xy}$.

\section{SPADS and SPOTL schemes analysis}
\label{appendix:QBERsingleclicktimes2}
\par{In order to compute the quantum bit error rate (QBER) of the Single-Photon with Additional Detection Setup (SPADS) scheme and the Single-Photon Over Two Links (SPOTL) scheme, we derive step by step the quantum state shared between Alice and Bob. The following results have been found using \texttt{Mathematica}. Finally, we also calculate the yield of the SPADS and SPOTL schemes. }


\subsection*{Generation of elementary links}
\subsubsection*{Single-photon scheme on Alice side}

The application of the single-photon scheme on Alice's side leads Alice and the quantum repeater to share a state given in Eq.~\eqref{eq:rhoSingleClick}. This state can be rewritten as
\begin{gather}
\rho_{\textrm{A-$\textrm{QR}^{e}$}} = A_{1}\outerproduct{\Psi^{+}} + B_{1}\outerproduct{\Psi^{-}} + C_{1}\left(\outerproduct{10} + \outerproduct{01}\right) + D_{1}\outerproduct{11} + E_{1}\outerproduct{00}\ ,
\end{gather}
with $A_1 = A(\theta_A, Y_A)$, $B_1 = B(\theta_A, Y_A)$, $C_1 = C(\theta_A, Y_A)$, $D_1 = D(\theta_A, Y_A)$ and $E_1 = E(\theta_A, Y_A)$. Here, we have that
\begin{equation}
A(\theta, Y) = \frac{1}{Y}2\cos^{2}(\theta)\sin^{2}(\theta)\eta (1-p_d)\left[(F_{\textrm{prep}}^{2} + (1-F_{\textrm{prep}})^{2})\lambda + 2F_{\textrm{prep}}(1-F_{\textrm{prep}})(1-\lambda)\right] \ ,\\
B(\theta, Y) = \frac{1}{Y}2\cos^{2}(\theta)\sin^{2}(\theta)\eta (1-p_d)\left[(F_{\textrm{prep}}^{2} + (1-F_{\textrm{prep}})^{2})(1-\lambda) + 2F_{\textrm{prep}}(1-F_{\textrm{prep}})\lambda\right]\ ,\\
C(\theta, Y) = \frac{2}{Y}\cos^{2}(\theta)\sin^{2}(\theta) p_d (1-p_d)(1-\eta)\ ,\\
D(\theta, Y) = \frac{1}{Y}\cos^{4}(\theta)\left(2(1-\eta)\eta (1-p_d) + \eta^2(1-p_d) + 2(1-\eta)^2 p_d (1-p_d)\right)\ ,\\
E(\theta, Y) = \frac{2}{Y}\sin^{4}(\theta) p_d (1-p_d)\ .
\end{equation}
In the above $Y$ denotes the yield or the probability of success of the single-photon scheme and is given by Eq.~\eqref{eq:yieldSingleClick}. Subscript $A$ indicates that in that expression for the yield and for each of the above defined coefficients we use $\theta = \theta_A$. Moreover, we have made here the following change of notation with respect to the Appendix~\ref{sec:SingleClick}, $\ket{\downarrow} \rightarrow \ket{0}$ and $\ket{\uparrow} \rightarrow \ket{1}$.
\subsubsection*{SWAP gate in the middle node}
In the next step, a SWAP gate is applied in the middle node to transfer the electron state to the nuclear spin of the NV center. This causes a depolarizing noise of parameter $F_{\textrm{swap}}=F_{g}^{2}$ (see Appendix \ref{appendix:lossesandnoise}). The resulting state can then be written as
\begin{equation}
\rho_{\textrm{A-$\textrm{QR}^{C}$}} = F_{\textrm{swap}} \rho_{\textrm{A-$\textrm{QR}^{e}$}} + (1 - F_{\textrm{swap}}) \tr_{\textrm{QR}}[\rho_{\textrm{A-$\textrm{QR}^{e}$}}] \otimes \frac{\mathbb{I}_{\textrm{2,QR}}}{2}\ .
\end{equation}

\subsubsection*{The procedure on Bob's side}
We now use the electron spin of the quantum repeater to generate the second quantum state. Here, the procedures for the SPADS and SPOTL schemes diverge.

In the procedure for the SPADS scheme, the quantum repeater generates a spin-photon entangled state where the photonic qubit is encoded in the time-bin degree of freedom. Since the spin-photon entangled state is imperfect, the electron and the photon share a state
\begin{equation}
\rho_{\textrm{$\textrm{QR}^{e}$}-B} = F_{\textrm{prep}}\outerproduct{\Psi^{+}} + (1-F_{\textrm{prep}})\outerproduct{\Psi^{-}}\ .
\end{equation}
Here we use the following labeling for time-bin encoded early and late modes of the photon: $\ket{e} = \ket{1}, \, \ket{l} = \ket{0}$. This photon is then sent toward Bob's detector. The lossy channel acts on such a time-bin encoded qubit as an erasure channel and so the quantum spin-photon state of the successful events in which the photonic qubit successfully arrives at the detector is unaffected by the lossy channel.

For the SPOTL scheme, the repeater's electron spin and Bob's quantum memory generate a second state of the form given in Eq.~\eqref{eq:rhoSingleClick}. We can rewrite this state as
\begin{gather}
\rho_{\textrm{$\textrm{QR}^{e}$}-B} = A_2\outerproduct{\Psi^{+}} + B_2\outerproduct{\Psi^{-}} + C_2\left(\outerproduct{10} + \outerproduct{01}\right) + D_2\outerproduct{11} + E_2\outerproduct{00}\ ,
\end{gather}
with $A_2 = A(\theta_B, Y_B)$, $B_2 = B(\theta_B, Y_B)$, $C_2 = C(\theta_B, Y_B)$, $D_2 = D(\theta_B, Y_B)$, and $E_2 = E(\theta_B, Y_B)$.


\subsubsection*{Decoherence in the quantum memories}
Decoherence of the carbon spin in the middle node can be modeled identically for both the SPADS and SPOTL scheme.
 \par{During the $n<n^{*}$ attempts to generate the state $\rho_{\textrm{$\textrm{QR}^{e}$-B}}$, the carbon spin in the middle node holding half of the state $\rho_{\textrm{A-$\textrm{QR}^{C}$}}$ will decohere. Using the decoherence model discussed in Appendix~\ref{sec:noisemem}, decoherence of the carbon spin will thus give us }
\begin{equation}
\rho'_{\textrm{A-$\textrm{QR}^{C}$}}= F_{T_{1}}[F_{T_{2}}\rho_{\textrm{A-$\textrm{QR}^{C}$}} + (1-F_{T_{2}})(\mathbb{I}_{2}\otimes Z)\rho_{\textrm{A-$\textrm{QR}^{C}$}}(\mathbb{I}_{2}\otimes Z)^{\dagger}] + (1-F_{T_{1}})\tr_{\textrm{QR}}[\rho_{\textrm{A-$\textrm{QR}^{C}$}}] \otimes \frac{\mathbb{I}_{\textrm{2,QR}}}{2}\ .
\end{equation}
For key generation, Alice (SPADS and SPOTL schemes) and Bob (SPOTL scheme) can actually measure their electron spin(s) immediately after the generation of spin photon entanglement, preventing the effect of decoherence on these qubit(s).

\subsection*{Noise due to measurements}
\subsubsection*{Measurement of the qubits of Alice and Bob}
In the SPADS scheme, Alice performs a measurement on her electron spin immediately after each of the spin-photon entanglement generation events to prevent any decoherence with time of this qubit. This measurement causes an error modeled as a depolarizing channel of parameter $F_{m}$. Bob, on the other hand, performs a measurement on a photonic qubit that is encoded in the time-bin degree of freedom. His measurement utilizes the squashing map so that we can model the noise arising from this measurement as a depolarizing channel with parameter $\alpha_B$ as described in Appendix~\ref{appendix:lossesandnoise}. 
Hence, the total state just before the Bell measurement is given by 
\begin{align}
\rho_{A-QR-B} &= F_m \alpha_{B} \rho'_{\textrm{A-$\textrm{QR}^{C}$}}\otimes\rho_{\textrm{$\textrm{QR}^{e}$}-B} + (1-F_m)\alpha_{B} \frac{\mathbb{I}_{2,A}}{2} \otimes \tr_A[\rho'_{\textrm{A-$\textrm{QR}^{C}$}}]\otimes\rho_{\textrm{$\textrm{QR}^{e}$}-B} \nonumber\\
&+ (1-\alpha_{B})F_m\rho'_{\textrm{A-$\textrm{QR}^{C}$}} \otimes \tr_B[\rho_{\textrm{$\textrm{QR}^{e}$}-B}]\otimes\frac{\mathbb{I}_{2,B}}{2} + (1-F_m)(1-\alpha_{B})  \tr_{AB}[\rho'_{\textrm{A-$\textrm{QR}^{C}$}}\otimes \rho_{\textrm{$\textrm{QR}^{e}$}-B}]\otimes\frac{\mathbb{I}_{4,AB}}{4}\ .
\end{align}
For the SPOTL scheme, both
Alice and Bob perform a measurement on their electron spins immediately after each of the spin-photon entanglement generation events. This measurement causes an error modeled as a depolarizing channel of parameter $F_{m}$ on each qubit, which means that after both Alice and Bob succeeded in performing the single-photon scheme with the repeater, the total, four-qubit state just before the Bell-measurement and including the noise of the measurements of Alice and Bob will be given by
\begin{align}
\rho_{A-QR-B} &= F_m^2 \rho'_{\textrm{A-$\textrm{QR}^{C}$}}\otimes\rho_{\textrm{$\textrm{QR}^{e}$}-B} + (1-F_m)F_m \left[\frac{\mathbb{I}_{2,A}}{2} \otimes \tr_A[\rho'_{\textrm{A-$\textrm{QR}^{C}$}}]\otimes\rho_{\textrm{$\textrm{QR}^{e}$}-B} + \rho'_{\textrm{A-$\textrm{QR}^{C}$}} \otimes \tr_B[\rho_{\textrm{$\textrm{QR}^{e}$}-B}]\otimes\frac{\mathbb{I}_{2,B}}{2}\right] \nonumber\\
&+ (1-F_m)^2 \tr_{AB}[\rho'_{\textrm{A-$\textrm{QR}^{C}$}}\otimes \rho_{\textrm{$\textrm{QR}^{e}$}-B}]\otimes\frac{\mathbb{I}_{4,AB}}{4}\ .
\end{align}


\subsubsection*{Bell state measurement}
Before the entanglement swapping, we have a total state $\rho_{A-QR-B}$. We now perform a Bell state measurement on the two qubits in the middle node. The error coming from this measurement is modeled by concatenation of depolarizing channels (see Appendix~\ref{appendix:lossesandnoise}), which means that the measurement is actually performed on 
\begin{equation}
\rho_{\textrm{fin}} = F^2_{g}F^2_{m}\rho_{A-QR-B} + F^2_g(1-F^2_{m})\tr_{QR^e}[\rho_{A-QR-B}]\otimes \frac{\mathbb{I}_{2,QR^e}}{2} + (1-F^2_g)\tr_{QR}[\rho_{A-QR-B}]\otimes \frac{\mathbb{I}_{4,QR}}{4}\ .
\end{equation}
While $\rho'_{\textrm{A-$\textrm{QR}^{C}$}}$ is not Bell diagonal for the SPADS scheme, $\rho_{\textrm{$\textrm{QR}^{e}$}-B}$ is, and so we find that taking into account the classical correction (which will be performed on the measured bit-value by Alice and Bob) the four cases corresponding to different measurement outcomes are equivalent. This means that if we model the correction to be applied to the quantum state rather than the classical bit, then the four post-measurement bipartite states shared between Alice and Bob are exactly the same.

For the SPOTL scheme, both $\rho'_{\textrm{A-$\textrm{QR}^{C}$}}$ and $\rho_{\textrm{$\textrm{QR}^{e}$}-B}$ are not Bell diagonal which means that the resulting state of qubits of Alice and Bob after the Bell state measurement depends on the outcome of this Bell measurement and those four corresponding states are not equivalent under local unitary corrections. In fact, the two states corresponding to the $\Phi^{\pm}$ outcomes and the two states corresponding to the $\Psi^{\pm}$ outcomes are pairwise equivalent under local Pauli corrections. Hence, we will derive two different QBER corresponding to the following different resulting states shared between Alice and Bob,
\begin{align}
\rho_{\Phi,AB} &= (\mathbb{I}_A \otimes U_{\Phi^{\pm},B})\Tr_{QR}\left[\frac{(\mathbb{I}\otimes\outerproduct{\Phi^{\pm}}\otimes\mathbb{I})\rho_{\textrm{fin}}(\mathbb{I}\otimes\outerproduct{\Phi^{\pm}}\otimes\mathbb{I})^{\dagger}}{\Tr(\rho_{\textrm{fin}}(\mathbb{I}\otimes\outerproduct{\Phi^{\pm}}\otimes\mathbb{I}))}\right](\mathbb{I} \otimes U_{\Phi^{\pm},B})^\dag\ , \\
\rho_{\Psi,AB} &= (\mathbb{I}_A \otimes U_{\Psi^{\pm},B})\Tr_{QR}\left[\frac{(\mathbb{I}\otimes\outerproduct{\Psi^{\pm}}\otimes\mathbb{I})\rho_{\textrm{fin}}(\mathbb{I}\otimes\outerproduct{\Psi^{\pm}}\otimes\mathbb{I})^{\dagger}}{\Tr(\rho_{\textrm{fin}}(\mathbb{I}\otimes\outerproduct{\Psi^{\pm}}\otimes\mathbb{I}))}\right](\mathbb{I} \otimes U_{\Psi^{\pm},B})^\dag\ .
\end{align}
Here $U_{\Phi^{\pm},B}$ and $U_{\Psi^{\pm},B}$ denote the four Pauli corrections implemented by Bob after the corresponding outcome of the Bell measurement. Note that for the SPADS scheme $\rho_{\Phi,AB} = \rho_{\Psi,AB}$.


\subsection*{The yield and QBER}
\subsubsection{Yield}
For both the SPADS and SPOTL scheme, we calculate the yield as the inverse of the number of channel uses required to generate one bit of raw key, $Y = 1/\mathbb{E}[N]$, where $\mathbb{E}[N]$ is given by Eq.~\eqref{eq:channeluses}. For the SPOTL scheme in that formula we use $p_{A} = Y_{A}$ ($p_{B} = Y_{B}$), where $Y_{A}$ ($Y_{B}$) denotes the yield of the single-photon scheme on Alice's (Bob's) side given by Eq.~\eqref{eq:yieldSingleClick}. For the SPADS scheme, $p_A$ takes the same form as for the SPOTL scheme (but is now calculated for two thirds of the total distance between Alice and Bob rather than half), while $p_B$ is the probability of registering a click in Bob's optical detection setup as in the SiSQuaRe scheme. 


\subsubsection*{Extraction of the qubit error rates}
By projecting these final corrected states onto the correct subspaces, we can obtain the qubit error rates $e_{z}$ and $e_{xy}$ (with our model we find that for both SPADS and SPOTL schemes the error rates in $X$ and $Y$ bases are the same). The state shared between Alice and Bob after the Pauli correction will always be the same for the SPADS scheme. Thus, there is only a single QBER $e_{z}$ and $e_{xy}$ independently of the outcome of the Bell measurement. For the SPOTL scheme that is not the case, there will be two sets of QBER corresponding to the states $\rho_{\Phi,AB}$ and $\rho_{\Psi,AB}$:
\begin{align}
e_{z,\Phi} &= \Tr[(\outerproduct{00} +\outerproduct{11})\rho_{\Phi}]\ , \\
e_{z,\Psi} &= \Tr[(\outerproduct{00} +\outerproduct{11})\rho_{\Psi}]\ , \\
e_{xy,\Phi} &= \Tr[(\outerproduct{+-} +\outerproduct{-+})\rho_{\Phi}] = \Tr((\outerproduct{0_y 1_y} +\outerproduct{1_y 0_y})\rho_{\Phi})\ , \\
e_{xy,\Psi} &= \Tr[(\outerproduct{+-} +\outerproduct{-+})\rho_{\Psi}] = \Tr[(\outerproduct{0_y 1_y} +\outerproduct{1_y 0_y})\rho_{\Psi}]\ .
\end{align}
Again, for the SPADS scheme $e_{z,\Phi} = e_{z,\Psi} = e_{z}$ and $e_{xy,\Phi} = e_{xy,\Psi} = e_{xy}$.


\subsubsection*{Averaging the qubit error rates}
We have now derived the qubit error rates as a function of the experimental parameters. For the SPOTL scheme, we now average the QBER over the two outcomes to get the final average QBER,
\begin{align}
\langle e_{z}\rangle &= \langle p_{\Psi} e_{z,\Psi} + p_{\Phi}  e_{z,\Phi}\rangle \ ,\\
\langle e_{xy}\rangle &= \langle p_{\Psi}  e_{xy,\Psi} + p_{\Phi} e_{xy,\Phi}\rangle\ ,
\end{align}
where $p_{\Psi}$ ($p_{\Phi}$) is the probability of measuring one of the $\ket{\Psi}$ ($\ket{\Phi}$) states in the Bell measurement and $\langle \cdots \rangle$ is found by averaging the expression over the number of Bob's attempts $n$ with the geometric distribution within the first $n^{*}$ trials. For the SPADS scheme $\langle e_{z}\rangle$ and $\langle e_{xy}\rangle$ can be averaged directly. The dependence on $n$ arises from the decoherence terms $F_{T_{1}}$ and $F_{T_{2}}$. Indeed, those terms correspond to the decoherence in the middle node during the attempts on Bob's side. Denoting by $p_{B}$ the probability that in a single attempt Bob generates entanglement with the quantum repeater using the single-photon scheme for the SPOTL scheme and using direct transmission of the time-bin encoded qubit from the repeater to Bob for the SPADS scheme, we have that the exponentials in those expressions can be averaged as follows~\cite{parameterregimes}
\begin{equation}
\langle e^{-cn}\rangle = \frac{p_{B}e^{-c}}{1-(1-p_B)^{n^{*}}}\frac{1-(1-p)^{n^{*}}e^{-cn^{*}}}{1-(1-p_{B})e^{-c}}\ .
\end{equation}

\section{Secret-key fraction and advantage distillation}
\label{sec:secretkeyfracderiv}
In this section, we review the formulas for the secret-key fraction for the QKD protocols used in our model as a function of the QBER.


\subsection*{One-way BB84 protocol}
For the fully asymmetric BB84 protocol with standard one-way post-processing, the secret-key fraction is given by~\cite{scarani2009security, lo2005efficient}
\begin{equation}
r = 1- h(e_x) - h(e_z)\ ,
\end{equation}
where $h(x)$ is the binary entropy function. Note that this formula is symmetric under the exchange of $e_x$ and $e_z$; that is, the secret-key fraction is the same independent of whether we extract the key in the $Z$ or $X$ basis. As we will see later in this section, this is not the case for the six-state protocol with advantage distillation.


\subsection*{Six-state protocol with advantage distillation}
Now we shall examine the six-state protocol with advantage distillation of Ref.~\cite{watanabe2007key}. For the purpose of this section, following the notation of~Ref.\cite{watanabe2007key}, we shall denote the four Bell states as
\begin{eqnarray}
\ket{\psi(\san{x},\san{z})} =
\frac{1}{\sqrt{2}}(
\ket{0}\ket{0 + \san{x}} + (-1)^{\san{z}}
\ket{1} \ket{1 + \san{x}\,\,(\textrm{mod} \, 2)}
),
\end{eqnarray}
for $\san{x},\san{z} \in \{0,1\}$.
We then write the Bell-diagonal state as
\begin{eqnarray}
\rho_{AB} = \sum_{\mathclap{\san{x},\san{z} \in \{0,1\} }}\; p_{\san{xz}}
\ket{\psi(\san{x},\san{z})}\bra{\psi(\san{x},\san{z})}\ .
\label{eq:belldiag}
\end{eqnarray}
The considered advantage distillation protocol is described in Ref.~\cite{watanabe2007key}. It is shown there that if the key is extracted in the $Z$ basis, then the secret-key fraction for the fully asymmetric six-state protocol supplemented with this two-way postprocessing technique is given by
\begin{eqnarray} 
r_{\text{six-state}} = \max \left\lbrace1 - H(P_{\san{XZ}}) + \frac{P_{\bar{\san{X}}}(1)}{2}
h\left( \frac{p_{00} p_{10} + p_{01} p_{11}}{
(p_{00} + p_{01})(p_{10} + p_{11})}\right), 
\frac{P_{\bar{\san{X}}}(0)}{2} [
1 - H(P_{\san{XZ}}^\prime) ] \right\rbrace ,
\label{eq-key-rate-vs-error-rate}
\end{eqnarray} 
where
\begin{eqnarray}
P_{\bar{\san{X}}}(0) &=& (p_{00} + p_{01})^2
+ (p_{10} + p_{11})^2\ , \\
P_{\bar{\san{X}}}(1) &=& 2 (p_{00} + p_{01})
(p_{10} + p_{11})\ , \\
p_{\san{00}}^\prime &=&
\frac{p_{00}^2 + p_{01}^2}{(p_{00} + p_{01})^2 + (p_{10}+p_{11})^2}, \\
p_{\san{10}}^\prime &=& 
\frac{2 p_{00} p_{01}}{(p_{00} + p_{01})^2 + (p_{10}+p_{11})^2}, \\
p_{\san{01}}^\prime &=&
\frac{p_{10}^2 + p_{11}^2}{(p_{00} + p_{01})^2 + (p_{10}+p_{11})^2}, \\
p_{\san{11}}^\prime &=&
\frac{2 p_{10}p_{11} }{(p_{00} + p_{01})^2 + (p_{10}+p_{11})^2} \ ,
\end{eqnarray}
$P_{\san{XZ}}$ ($P_{\san{XZ}}^\prime$) is the probability distribution over the coefficients $p_{\san{xz}}$ ($p_{\san{xz}}^\prime$) and $H(P_{\san{XZ}})$ ($H(P_{\san{XZ}}^\prime)$) is the Shannon entropy of this distribution.

Now let us have a look at how to link the Bell coefficients $p_{\san{xz}}$ with our QBER $e_z$ and $e_{xy}$ (for all our schemes, the estimated QBER in the $X$ basis is the same as in the $Y$ basis). In this section, we assume the target state that Alice and Bob want to generate is $\ket{\psi(\san{0},\san{0})}$. Note that in the analysis in Appendixes~\ref{sec:SingleClick} and~\ref{appendix:QBERsingleclicktimes2} it is the state $\ket{\psi(\san{1},\san{0})}$ that is a target, but of course the secret-key fraction analysis is independent of which Bell state is a target state as they are all the same up to local Pauli rotations. Hence, the relation between the Bell-diagonal coefficients and the QBER is

\begin{align}
p_{10}+p_{11} &= e_z\ , \\
p_{01}+p_{11} &= e_{xy}\ , \\
p_{01}+p_{10} &= e_{xy}\ , \\
p_{00} + p_{01} + p_{10} + p_{11} &= 1\ .
\end{align}
Therefore,
\begin{align}
\begin{split}
p_{00} &= 1 - \frac{e_z}{2} - e_{xy}\ ,\\
p_{01} &=  e_{xy} - \frac{e_z}{2}\ ,\\
p_{10} &= p_{11} = \frac{e_z}{2}\ .
\end{split}
\label{eq:bellcoeffsQBERmap}
\end{align}
And so,
\begin{align}
P_{\bar{\san{X}}}(0) &= 1-2e_z + 2e_z^2\ ,\\
P_{\bar{\san{X}}}(1) &= 2(1-e_z)e_z\ .
\end{align}

It is important to note that for the above described advantage distillation, the amount of generated secret key depends on the basis in which it is extracted, as has been shown in Ref.~\cite{murta2018}. Let us now have a look at the amount of key that can be extracted in the $X$ and $Y$ bases.
As has been shown in Ref.~\cite{murta2018}, the secret-key fraction in these cases is also given by Eq.~\eqref{eq-key-rate-vs-error-rate} but now the Bell coefficients depend on QBER in the following way:

\begin{align}
\begin{split}
p_{00} &= 1 - \frac{e_z}{2} - e_{xy}\ ,\\
p_{10} &=  e_{xy} - \frac{e_z}{2}\ ,\\
p_{01} &= p_{11} = \frac{e_z}{2}\ .
\end{split}
\label{eq:bellcoeffsQBERx}
\end{align}
And so,
\begin{align}
\begin{split}
P_{\bar{\san{X}}}(0) &= 1-2e_{xy} + 2e_{xy}^2\ ,\\
P_{\bar{\san{X}}}(1) &= 2(1-e_{xy})e_{xy}\ .
\end{split}
\label{eq:ADprobx}
\end{align}

We note that we have assumed here that in the case of key extraction in $Y$ basis, either Alice or Bob applies a local bit flip in the $Y$ basis to the shared state, as the target state $\ket{\psi(0,0)}$ is anticorrelated in that basis.

In Ref.~\cite{murta2018} it has been also observed that in the considered case of having the QBER in the $X$ and $Y$ bases being equal, the six-state protocol with advantage distillation allows us to extract more key if it is extracted in the basis with higher QBER. This observation determines the basis that we use for extracting key for the single-photon and the SPOTL schemes that use fully asymmetric six-state protocol with advantage distillation. Specifically, for the single-photon scheme we observe higher QBER in the $Z$ basis, while for the SPOTL scheme the QBER is higher in the $X$ and $Y$ bases. Therefore, these are the bases that we choose to use for extracting key for those schemes. 

For the SiSQuaRe and SPADS schemes, the symmetric six-state protocol is used. Hence, for those schemes we group the raw bits into three groups corresponding to three different key-extraction bases and we extract the key separately for each of these bases. Finally, to obtain the final secret-key fraction, we note that for the symmetric six-state protocol we also need to include sifting; that is, only one third of all the raw bits were obtained by Alice and Bob measuring in the same basis (the raw bits for the protocol runs in which they measured in different bases are discarded). Hence, if we denote by $r_i$ the secret-key fraction obtained from the group of raw bits in which both Alice and Bob measured in the basis $i$, the final secret-key fraction for the six-state protocol for those schemes is given by
\begin{equation}
r= \frac{1}{3}\left(\frac{1}{3}r_x + \frac{1}{3}r_{y} + \frac{1}{3}r_z\right)\ .
\end{equation}
Clearly, in our case we have $r_x = r_y = r_{xy}$.


\subsection*{One-way six-state protocol}
In Fig.~\ref{fig:optprot} we have also plotted the secret-key fraction for the one-way six-state protocol. For the fully asymmetric protocol and the case in which the key is extracted in the $Z$ basis, it is given by~\cite{scarani2009security}
\begin{equation}
r = 1 - e_z h\left(\frac{1+(e_x - e_y)/e_z}{2}\right) - (1-e_z)h\left(\frac{1-(e_x + e_y + e_z)/2}{1-e_z}\right) - h(e_z)\ .
\label{eq:onewaysixstate}
\end{equation}
Although this formula does not appear to be symmetric under the permutation of $e_x, \, e_y, \, e_z$, it is in fact invariant under this permutation~\cite{renner2005eth}. This means that for the symmetric one-way six-state protocol, in our case the final secret-key fraction is given by the expression in Eq.~\eqref{eq:onewaysixstate} multiplied by the sifting efficiency of one-third.

\section{Runtime of the experiment}
\label{section:runtime}
In this section, we will detail how to perform an experiment that will be able to establish that a setup can surpass the capacity of a quantum channel modeling losses in a fiber (see Eq.~\eqref{eq:benchmark1}). This experiment can validate a setup to qualify as a quantum repeater, without explicitly having to generate secret-key. We show then that, for the listed parameters in the main text, the single-photon scheme can be certified to be a quantum repeater within approximately 12 hours.

The experiment is based on estimating the yield of the scheme and the individual QBER of the generated states. More specifically, here we will calculate the probability that, assuming our model is accurate and each individual run is independent and identically distributed, the observed estimates of the yield and the individual QBER are larger and smaller, respectively, than some fixed threshold values. If, with these threshold values for the yield and QBER, the calculated asymptotic secret-key rate still surpasses the capacity, we can claim a working quantum repeater. The experiment consists of first performing $n$ attempts at generating a state between Alice and Bob, from which the yield can be estimated by calculating the ratio of the successful attempts and $n$. Then, the QBER in each basis is estimated by Alice and Bob measuring in the same basis in each of the successful attempts.\\

Central to our calculation is the fact that, for $n$ instances of a Bernoulli random variable with probability $p$, the probability that the number of observed successes $S(n)$ is smaller or equal than some value $k$ is equal to
\begin{align}
P\left(S(n)\leq k \right) = \sum_{i=0}^k\binom{n}{i}p^i\left(1-p\right)^{n-i}\ .
\end{align}

Assuming the outcomes of our experiment are independent and identically distributed, the observed yield $\bar{Y}$ satisfies

\begin{align}
P\left(\bar{Y}\leq \left(Y-t_Y\right) \right) = P\left(n\bar{Y}\leq n\left(Y-t_Y \right)\right) = \sum_{i=0}^{\left\lfloor n\left(Y-t_Y \right)\right\rfloor}\binom{n}{i}Y^i\left(1-Y\right)^{n-i}\ ,
\end{align}
where $Y-t_Y$ is the lower threshold.
Let us make this more concrete with a specific calculation. For a distance of $17L_0$ the yield is equal to $\approx 5.6\times 10^{-6}$. Setting the maximum deviation in the yield to $\bar{Y} = Y-t_Y$ with $t_Y = 2.0\times 10^{-7}$ and the number of attempts to $n = 5\times 10^9$ (which corresponds to approximately a runtime of 12 hours assuming a single attempt takes $8.5\times 10^{-6}$ s, corresponding to $t_{\textrm{prep}}$ and a single-shot readout lasting $2.5\times 10^{-6}$ s), we find that 

\begin{align}
P\left(\bar{Y}\leq \left(Y-t_Y\right) \right) \leq 9.2\cdot 10^{-10}\ .
\end{align}

Similarly, for the individual errors $\lbrace e_k \rbrace_{k \in \lbrace x, y, z\rbrace}$ in the three bases we have that 
\begin{align}
P\left(\bar{e}_k\geq \left(e_k+t_k\right) \right) = P\left(m\cdot \bar{e}_k\geq m\left(e_k+t_k\right) \right)  = \sum_{i=\left\lceil m\left(e_k+t_k \right)\right\rceil}^{m}\binom{m}{i}\left(e_k\right)^i\left(1-e_k\right)^{m-i}\ .
\end{align}
Here we set $m = \left\lfloor \frac{n}{3}\left(Y-t_Y\right) \right \rfloor$, which is an estimate for the number of raw bits that Alice and Bob obtain from measurements in each of the three bases, for the total $n$ attempts of the protocol. All the raw bits from those three sets are then compared to estimate the QBER in each of the three bases. Note that we gather the same amount of samples for each basis, even when an asymmetric protocol would be performed. Setting $t_i = t = 0.015,~\forall i \in \lbrace x, y, z \rbrace$ and, as before, $n = 5\times 10^9$, we find, at a distance of $17L_0$ where $e_z \approx 0.171$ and $e_y = e_x \approx 0.141$, that 
\begin{align}
P\left(\bar{e}_z\geq \left(e_z+t\right) \right) \leq 9.0 \times 10^{-5}\ ,\\
P\left(\bar{e}_y\geq \left(e_y+t\right) \right) = P\left(\bar{e}_x\geq \left(e_x+t\right) \right) &\leq 2.7 \times 10^{-5}\ .
\end{align}

Then, with probability at least
\begin{equation}
\begin{gathered}
\left(1-P\left(\bar{e}_x\geq \left(e_x+t\right) \right)\right) \left(1-P\left(\bar{e}_y\geq \left(e_y+t\right) \right)\right) \left(1-P\left(\bar{e}_z\geq \left(e_z+t\right) \right)\right)\left(1- P\left(\bar{Y}\leq \left(Y-t_Y\right) \right)\right)\\
\geq 1-1.5\times 10^{-4}\ ,
\end{gathered}
\end{equation}
none of the observed QBER and yield exceed their threshold conditions. The corresponding lowest secret-key rate for these parameters (with a yield of $Y-t_Y$ and QBER of $e_x+t_x,~e_y+t_y,~e_z+t_z$) is $\approx 1.97\times 10^{-7}$, which we observe is greater than the secret-key capacity by a factor $\approx 3.29$ (see Eq.\eqref{eq:benchmark1}) at a distance of $17L_0$, since the secret-key capacity equals $-\log_2\left(1-e^{-17}\right)\lesssim 5.97\times 10^{-8}$.\\

Thus, with high probability we can establish that the single-photon scheme achieves a secret-key rate significantly greater than the corresponding secret-key capacity for a distance of $17L_0\approx 9.2$ km within approximately 12 hours.

\section{MDI QKD}
\label{sec:MDI}
We note here that the single-photon scheme for generating key is closely linked to the measurement-device-independent (MDI) QKD protocol~\cite{lo2012measurement}. In particular, it is an entanglement-based version of a scheme in which Alice and Bob prepare and send specific photonic qubit states to the heralding station in the middle, where the qubits are encoded in the presence or absence of the photon. We note that in the ideal case of the single-photon scheme, the spin-photon state is given in Eq.~\eqref{eq:spinphoton}. For the six-state protocol the spin part of this state is then measured in the $X$, $Y$, or $Z$ basis at random according to a fixed probability distribution (this probability distribution dictates whether we use symmetric or asymmetric protocol). Considering the probabilities of the individual measurement outcomes, this is equivalent to the scenario in which Alice and Bob choose one of the three set of states at random according to the same probability distribution and prepare each of the two states from that set with the probability equal to the corresponding measurement outcome probability. These sets do not form bases, as the two states within each set are not orthogonal. We will therefore refer to these sets here as ``pseudobases''. Depending on the chosen pseudobasis, they prepare one of the six states encoding the bit value of ``0'' or ``1'' in that pseudobasis. These states and the corresponding preparation probabilities are the following:
\begin{enumerate}[label={(\arabic*)}]
\item pseudo-basis 1:  $\{\ket{0}, \ket{1}\}$ with probabilities $\{\sin^2\theta,~\cos^2\theta\}$,
\item pseudo-basis 2: $\{\sin\theta \ket{0} + \cos\theta \ket{1},~\sin\theta \ket{0} - \cos\theta \ket{1}\}$ with probabilities $\{\frac{1}{2}, \frac{1}{2}\}$,
\item pseudo-basis 3: $\{\sin\theta \ket{0} + i\cos\theta \ket{1},~\sin\theta \ket{0} - i\cos\theta \ket{1}\}$ with probabilities $\{\frac{1}{2}, \frac{1}{2}\}$.  
\end{enumerate}

These states are then sent toward the beam-splitter station. The station performs the standard photonic Bell-state measurement and sends the outcome to both Alice and Bob. Alice and Bob discard all the runs for which the beam-splitter station measured $A_2$ (recall the measurement operators in Eq.~\eqref{eq:opticalBSM_POVM}). They then exchange the classical information about their pseudo-basis choice and keep only the data for the runs in which they both used the same basis. For those data they apply the following post-processing in order to obtain correlated raw bits
\begin{enumerate}[label={(\arabic*)}]
\item Pseudo-basis 1: For both outcomes $A_0$ and $A_1$, Bob flips the value of his bit.
\item Pseudo-basis 2: For the outcome $A_0$, they do nothing; for the outcome $A_1$, Bob flips the value of his bit.
\item Pseudo-basis 3: For the outcome $A_0$, they do nothing; for the outcome $A_1$, Bob flips the value of his bit.
\end{enumerate}
In this way, Alice and Bob have generated their strings of raw bits.

We note here that the direct preparation of the six states from the three pseudo-bases described above in the photonic presence or absence degree of freedom is experimentally hard. This is related to the fact that linear optics does not allow to easily perform the single qubit rotations necessary to prepare these states. The use of memory-based NV centers offers a great advantage here, as in these schemes the rotations that allow us to obtain the required amplitudes of the photonic states are performed on the electron spins rather than the photons themselves. There has also been proposed an alternative scheme that also benefits from single-photon detection events in which Alice and Bob send coherent pulses to the heralding station~\cite{lucamarini2018overcoming,tamaki2018information}.  

\end{document}